\documentclass[11pt]{article}
\usepackage{amsmath,amssymb,latexsym,tabu,array}
\usepackage{epsfig}
\usepackage{graphicx}
\usepackage{color}
\usepackage{soul}
\usepackage{verbatim}
\usepackage[normalem] {ulem}
\allowdisplaybreaks

\newcommand{\simg}{\stackrel{\sim}{>}}
\newcommand{\be}{\begin{equation}}
\newcommand{\ee}{\end{equation}}

\begin{document}
\title{An alternative approach to Michaelis-Menten kinetics that is based on 
the Renormalization Group: Comparison with the perturbation expansion beyond 
the sQSSA.}
\author{Barbara Coluzzi$^{a}$\!\!
\footnote{Corresponding author. E-mail: \tt barbara.coluzzi@sbai.uniroma1.it} ,
 Alberto M. Bersani$^{b}$, and Enrico Bersani$^{c}$}
\maketitle
\begin{center}

a) {\em Dipartimento di Scienze di Base ed Applicate per l'Ingegneria}, 
Sapienza University, via A. Scarpa 14, 00161 Rome, Italy. \\

b) {\em Dipartimento di Ingegneria Meccanica e Aerospaziale}, 
Sapienza University, via Eudossiana 18, 00184 Rome, Italy. \\

c) {\em Laboratorio di Strutture e Materiali Intelligenti}, Sapienza 
University, Palazzo Caetani, 
via San Pasquale s.n.c., 04012 Cisterna di Latina, Latina, Italy. 

\begin{abstract}
\noindent
We recall the perturbation expansion for Michaelis-Menten kinetics, beyond 
the standard quasi-steady-state approximation (sQSSA). Against
this background, we are able to appropriately apply the  
alternative approach to the study of singularly perturbed 
differential equations that is based on  the renormalization group (SPDERG), by 
clarifying similarities and differences. In the present demanding 
situation, we directly renormalize  the bare initial condition value 
for the substrate. Our main results are: i) the 2nd order SPDERG 
uniform approximations to the correct solutions contain, up to 
1st order,  the same outer components  as the known perturbation expansion 
ones; ii) the differential equation to be solved for the derivation of the 
1st order outer substrate component is simpler within the SPDERG approach;  
iii) the approximations better reproduce the numerical solutions of the 
original problem in a region encompassing the matching one, because of the 
2nd order terms in the inner components, calculated here for the first time to 
our knowledge: iv) the refined SPDERG uniform approximations, that we propose, 
give the correct  asymptotically vanishing solutions, too, and allow to obtain 
results nearly indistinguishable from the solutions of the original problem in 
a large part of the whole relevant time window, even in the studied 
unfavourable kinetic constant  case, for an expansion parameter value as large 
as $\varepsilon=0.5$.
\end{abstract}
\end{center}

\topmargin=-2.5cm

\section{Introduction}
\noindent
Michaelis-Menten (MM) kinetics, that characterizes enzymatic reactions
\cite{MiMe,BeBeDAPe,MaMo,LiSe}, is a well known example in biomathematics 
\cite{MaMo,LiSe,Mu,EdKe} of a system of ordinary differential equations 
(ODEs) characterized by two definitely different time scales. In fact, usually 
the complex reaches a {\em quasi-steady-state} of equilibrium with the 
substrate at the very beginning, whereas the generally experimentally observed 
part of the reaction happens on a time scale that can be as larger as several 
orders of magnitude \cite{Anetal}. Correspondingly, the standard 
quasi-steady-state approximation (sQSSA) \cite{MiMe,BeBeDAPe,MaMo,LiSe}, 
in which the independent variables are chosen to be the substrate and the
complex, with the time derivative of the complex taken to be zero, is 
the routinely considered starting point for investigating the system's dynamics.
Nevertheless, different starting points are considered in the literature. 
 
Actually, from the theoretical point of view \cite{BeBeDAPe,MaMo,LiSe,BeOr}, 
MM kinetics is an example of boundary layer problem. Therefore, in the 
standard methods the solutions are approximated by the perturbation expansion 
(PE), in an appropriate parameter $\varepsilon$, of both the inner and the 
outer components of the two chosen independent variables, {\em i.e.}, 
of the solutions of systems of two ODEs with regular and singular 
perturbations, respectively, with the further imposition of the appropriate 
matching conditions (MCs), at each order in $\varepsilon$. Within this
framework, both the sQSSA and the {\em total} quasi-steady-state approximation
(tQSSA) \cite{BeBeDAPe,BoBoSe,TzEd,DABe,La,Sw} represent the 0th order terms
of the outer solutions. Noticeably, the PE beyond the tQSSA, in which the 
independent variables are chosen to be the total substrate ({\em i.e.}, the 
sum of the substrate and of the complex) and the complex, has the advantage 
that the value of the expansion parameter is lower than 1 (indeed, one has 
$\varepsilon \le 1/4$ in this case), for whatever kinetic constants and for 
whatever initial condition values (ICVs)
\cite{BeBeDAPe,BoBoSe,TzEd,DABe}. 

In the present work, we are interested in testing the correctness, in the case 
of MM kinetics, of the renormalization group approach to singularly perturbed 
differential equations (SPDERG) proposed by Chen, Goldenfeld and Oono in 
\cite{ChGoOo1,ChGoOo2}. Therefore, we focus on the simpler case of the PE
beyond the sQSSA, by moreover making the usual choice for the expansion 
parameter ({\em i.e.}, $\varepsilon=e_0/s_0$ with $e_0$ the initial enzyme 
concentration and $s_0$ the initial substrate one) \cite{BeBeDAPe,MaMo}.  
It can be shown that in this case the approximated solution in the 
substrate / complex phase space converges rapidly to the exact one, for 
reasonably small $\varepsilon$ values \cite{MaMo}, and it was indeed proved 
that this PE (for $\varepsilon < 1$) converges uniformly for all times  
$t \in [0, \infty)$  \cite{LiSe}. 

On the other hand, the problem is not obvious to be solved, because both of 
the practical difficulty in explicitly finding the outer components, and of 
the peculiarity in the MCs to be imposed, whose number of terms increases with 
the considered order in the expansion \cite{LiSe}. These difficulties are 
shared by the other standard PE methods for MM kinetics considered in
the literature, such as in particular the one beyond the tQSSA 
\cite{BeBeDAPe,BoBoSe,TzEd,DABe}.
 
Even more because of these difficulties, after recalling the sQSSA and the main 
known results of the PE beyond it, we apply to MM kinetics the SPDERG 
alternative approach proposed in \cite{ChGoOo1,ChGoOo2}, that is based on the 
renormalization group \cite{ZJ}. Noticeably, a preliminary attempt to apply 
this approach in the present case, in the even more demanding case of the 
tQSSA framework, has been presented in \cite{Ra}, but we follow here the 
different way of directly renormalizing the bare ICVs (for the sake of 
precision, the one of the substrate). 

In fact, though this approach appears more generally applicable, it was 
in particular already shown successful for obtaining the leading terms of the 
uniform approximations (UAs) of the corresponding PE in other cases of 
dynamics characterized by the presence of a boundary layer 
\cite{ChGoOo1,ChGoOo2}. From the point of view of the present application, 
in which we calculate in detail the solutions up to 2nd order, we anticipate 
that the work could appear quite technical and cumbersome, nevertheless the 
calculations do not imply particular difficulties. On the other hand, this 
application to the demanding situation of MM kinetics, within the sQSSA
framework, appears to allow to better understand both the working principles 
and the advantages / limits of the approach in similar cases, besides  
highlighting the analogies and the differences with the standard PE. 
Indeed, the study appears to make possible to get some insights  
on the working principles of the PE itself, too. Moreover,
we obtain a more detailed knowledge of MM dynamics within the sQSSA framework, 
even only from the point of view of the 2nd order contributions to the inner 
solutions, that are presented here for the first time to our knowledge. From 
this last perspective, it is to be noticed that the calculation of the 2nd order
inner contribution is not merely a technical exercise, but is fundamental
for an in-depth understanding of the SPDERG scheme in the present case,
as we will show in the paper.

The paper substantially consists of a first more introductive part, and of a 
second part in which we present and discuss our results. In detail, in the 
first part, we recall: in Section \ref{review1}, the basis of MM kinetics and 
the sQSSA, by moreover introducing the particular values of the kinetic 
constants and the two ICV sets that we consider; in Section \ref{review2}, the 
results of the standard PE beyond the sQSSA, up to the known 1st order; in 
Section \ref{alter}, the basis of the alternative SPDERG approach 
proposed in \cite{ChGoOo1,ChGoOo2}, with attention to the case of boundary 
layer problems. In the second part, we present and discuss: in Section 
\ref{rdI}, the derivation of the 1st order SPDERG UAs; in Section \ref{rdII}, 
the calculation of the 2nd order contributions to the SPDERG UAs; 
in Section \ref{rdIIref}, the refined 2nd order SPDERG UAs that can be 
proposed; in Section \ref{rdcomp}, the comparison between the different 
best UAs that we considered. Finally, in Section  \ref{conc}, we present
our conclusions. The paper is moreover completed by three Appendices:
in Appendix A and in Appendix B we report the 1st order inner solution 
for the complex, and the 2nd order ones both for the substrate and the complex 
for partially general ICVs, respectively; in Appendix C we verify that the ODE 
obtained at the 2nd order from the study of the substrate is indeed the same 
ODE that one finds from the study of the complex.

\section{The sQSSA}
\label{review1}
\noindent
The  sQSSA represents a milestone in the mathematical modelling of enzymatic 
reactions \cite{MiMe,BeBeDAPe,MaMo,LiSe,Mu,EdKe,He,BrHa}. Here we 
just remind that the original paper by Michaelis and Menten dates back to more 
than one century ago \cite{MiMe}, that the idea was already present in the 
previous paper by Henry \cite{He}, and that the approach was further developed 
in particular by Briggs and Haldane \cite{BrHa}. Schematically 
\cite{BeBeDAPe,MaMo}, one is modelling the reaction between the enzyme $E$, 
the substrate $S$, the complex $C$, and the product $P$:
\begin{equation}
E+S 
\left.\right.^{\textstyle \stackrel{k_1}{\longrightarrow}}_{ 
\stackrel{\textstyle \longleftarrow}{ k_{-1}}} 
C \stackrel{k_2}{\longrightarrow} E+P,
\end{equation}
that is reversible in the first part and irreversible in the second one,
with associated kinetic constants $k_1$, $k_{-1}$, and $k_2$. 

When introducing the concentrations $e$, $s$, $c$ and $p$, respectively,
by using the mass action law, we arrive to describe the process by means 
of a system of four 1st order ODEs. Then, within the standard
framework \cite{BeBeDAPe,MaMo}, we start by using the conservation law
$e+c=e_0+c_0$, that implies that the enzyme concentration $e$ does only depend 
on the complex one $c$. Moreover, we observe that the product 
concentration $p$ can be obtained from the complex concentration $c$ by 
integrating (equivalently, one can use the other conservation law, 
$s+c+p=s_0+c_0+p_0$). Finally, we assume that the concentrations of the complex
and of the product are zero at the beginning for simplicity ({\em i.e.},
$c_0=p_0=0$). 

Hence, we end up with the well known system of two 1st order ODEs that are to 
be obeyed by the variables $s$ and $c$ (with ICVs $s(0)=s_0$ and $c(0)=0$):
\begin{eqnarray}
\left \{
\begin{array}{lcl}
\dot{s}(t)&=&k_1 [s(t)+K_D] \left [c(t) - \frac{\textstyle e_0 s(t)}
{\textstyle s(t)+K_D} \right  ]\\
\dot{c}(t)&=&-k_1[s(t)+K_M] \left [c(t) - \frac{\textstyle e_0 s(t)}
{\textstyle s(t)+K_M} 
\right ],\\
\end{array}
\right.
\label{mmineq}
\end{eqnarray}
where the dot means the time derivative. Here $K_D=k_{-1}/k_1$ is the so-called
dissociation constant, whereas $K_M=(k_{-1}+k_2)/k_1$ is the parameter that is 
generally known as Michaelis constant. It can be further noticed that 
$K_M-K_D=k_2/k_1=K$ is the usual Van Slyke-Cullen 
constant \cite{VSCu}. Though the original kinetic constants, $k_1$,
$k_{-1}$ and $k_2$, are the key physical parameters for the studied
system, $K_D$, $K_M$ and $K$ turn out to be the experimentally
measurable ones, usually in particular on the basis of the sQSSA. 

\begin{figure}[t]
\begin{center}
\leavevmode
\epsfig{figure=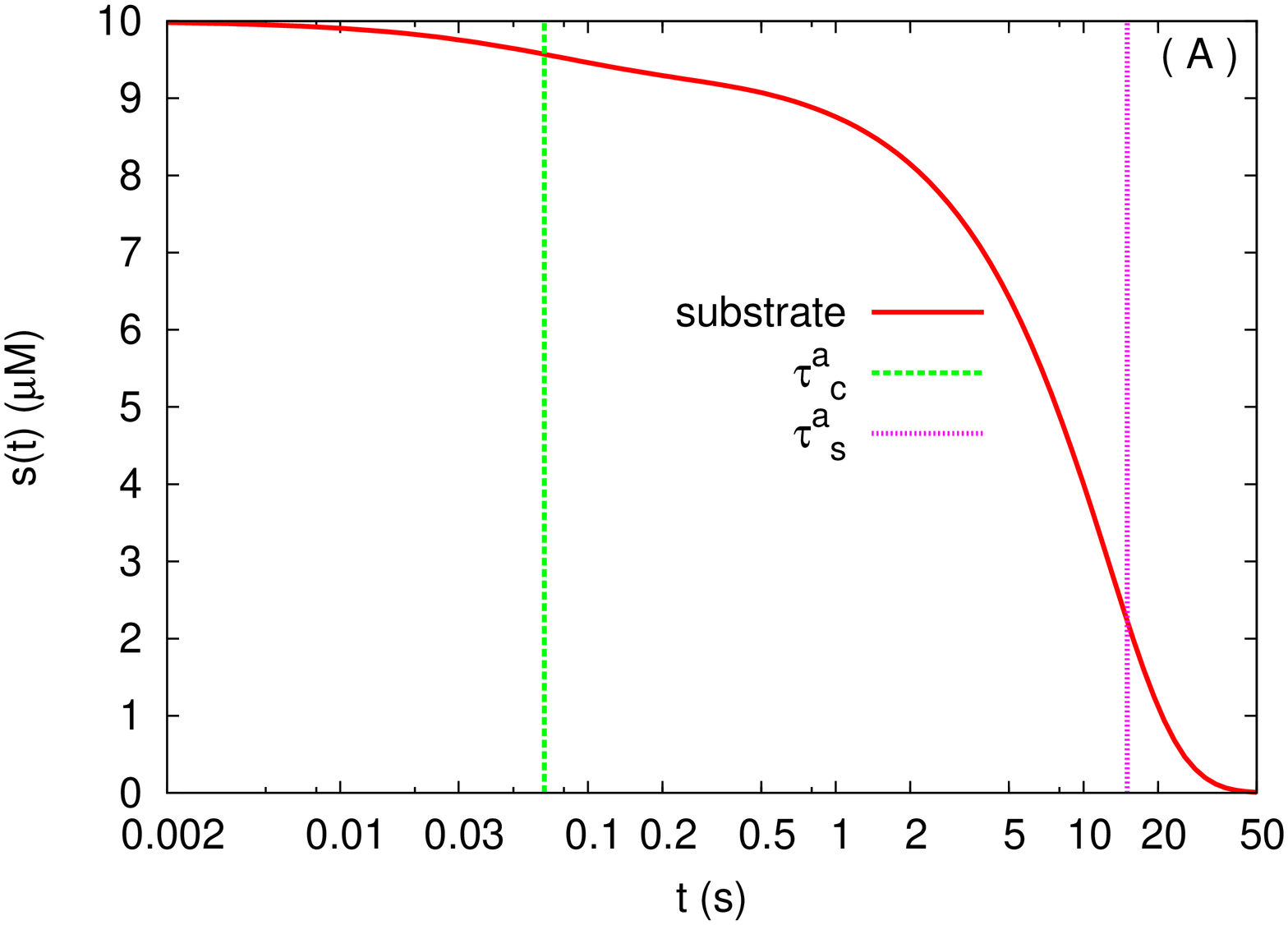,angle=270,width=8cm}
\epsfig{figure=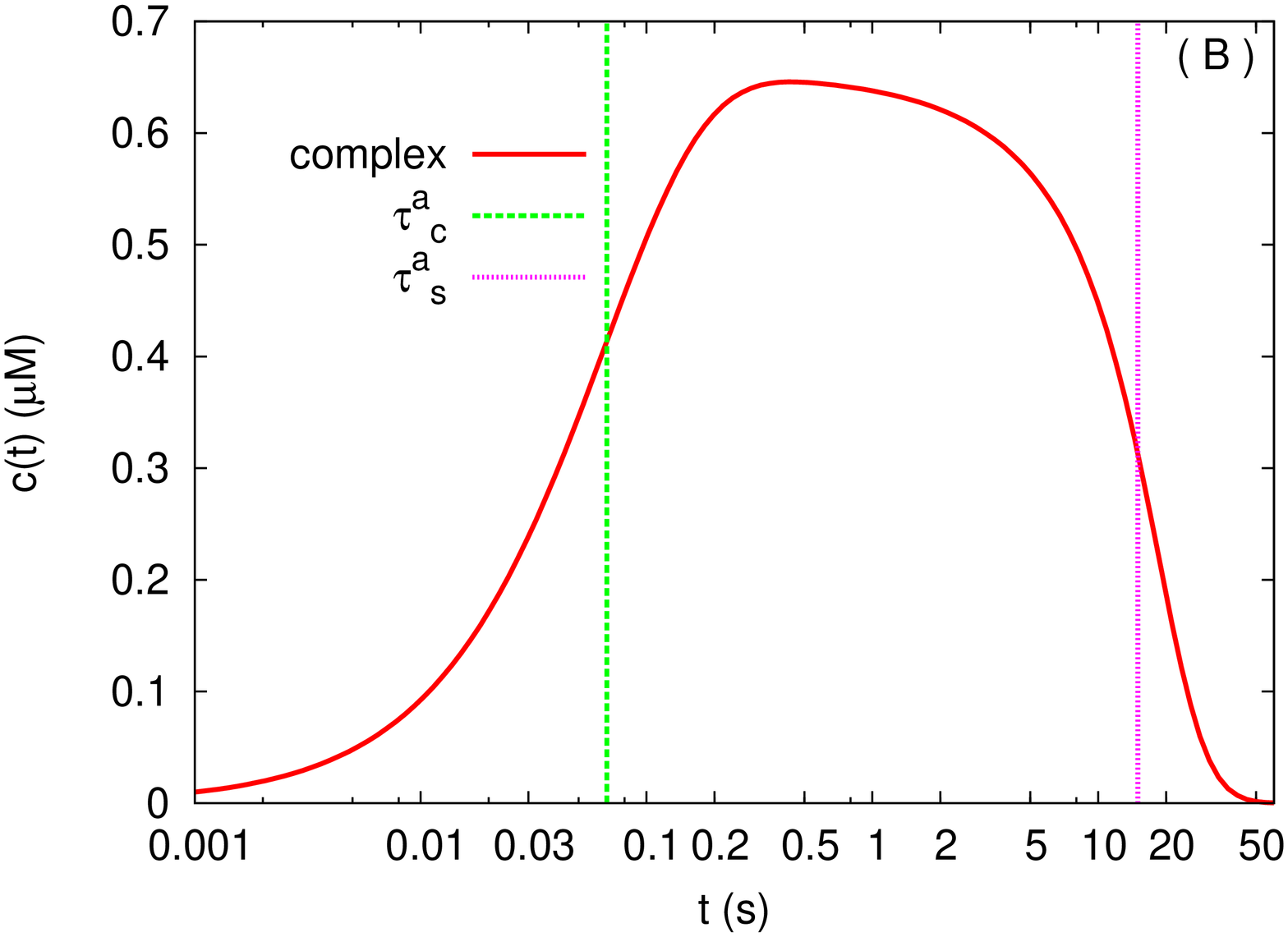,angle=270,width=8cm}
\epsfig{figure=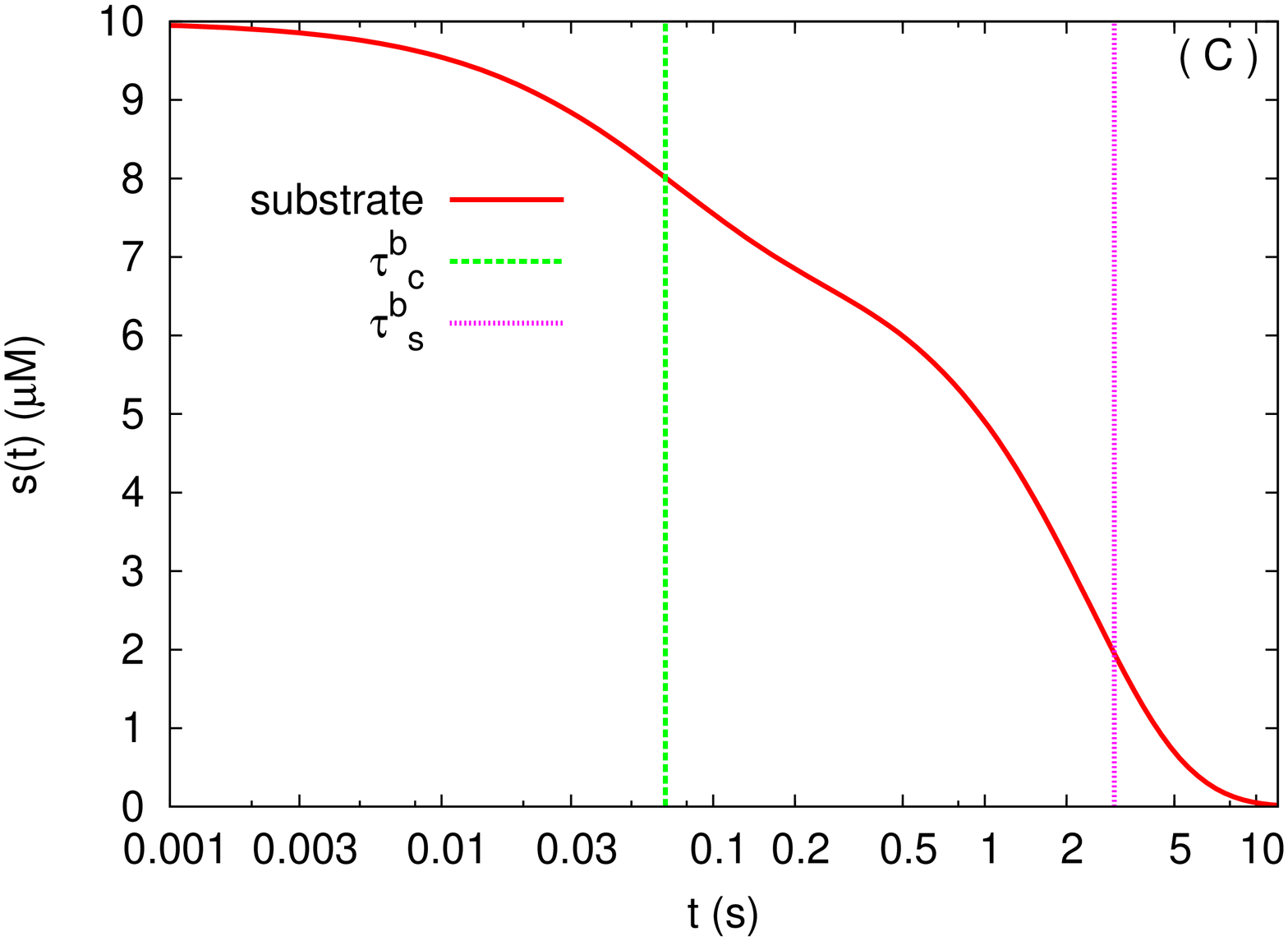,angle=270,width=8cm}
\epsfig{figure=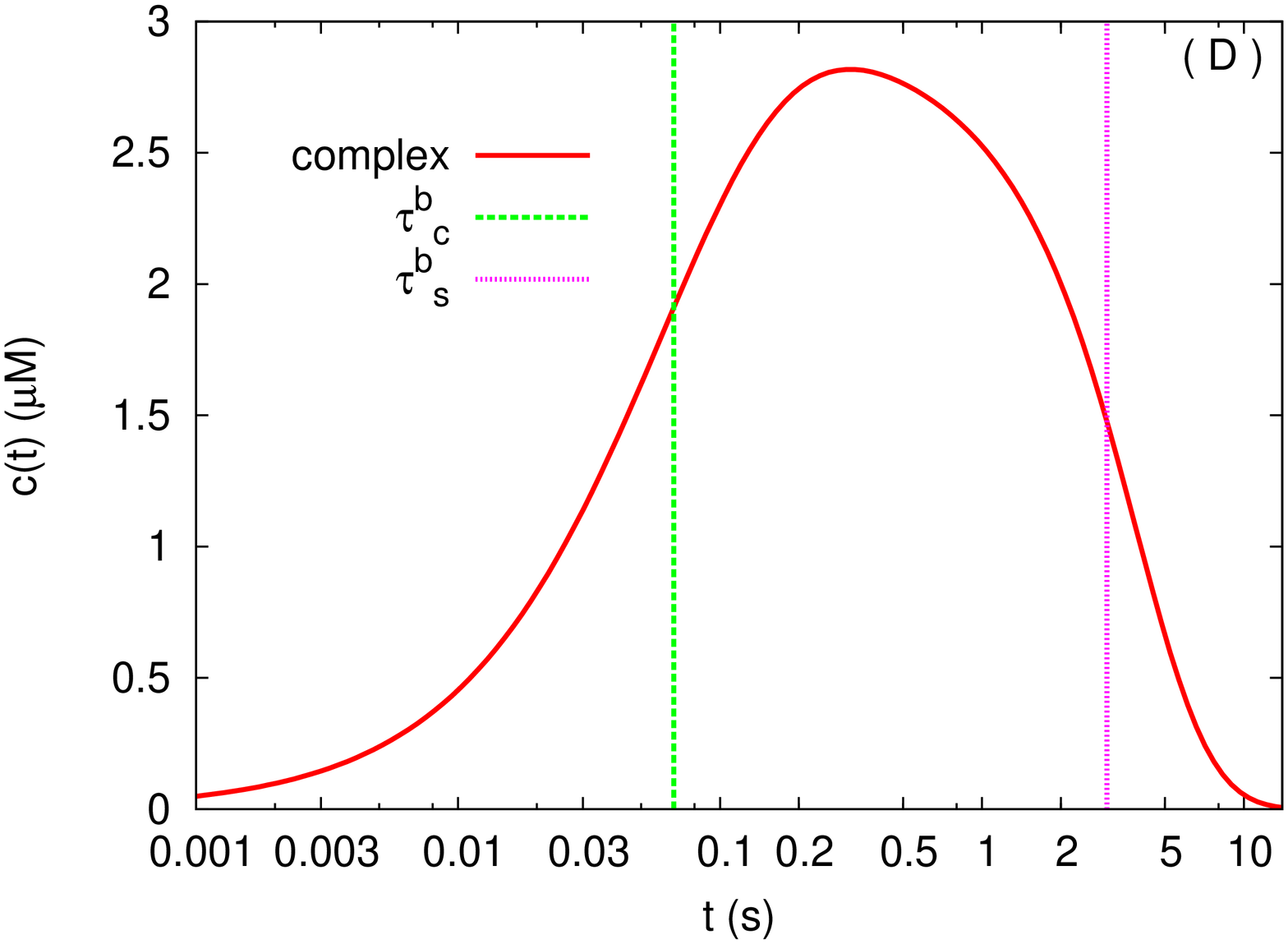,angle=270,width=8cm}
\caption{{\footnotesize In $A)$ and in $C)$ we present the behaviour of 
the concentrations of the substrate $s(t)$, whereas in $B)$ and in $D)$ we 
present the ones of the complex $c(t)$, solutions of Eqs.~(\ref{mmineq}), for 
the $a$ and $b$ sets of ICVs in (\ref{initialvalues}), respectively. 
Notice that the time is in logarithmic scale. We plot the corresponding rough 
evaluations the two different time scales involved, too, with $\tau_s$ 
describing the substrate decay time and $\tau_c$ the complex saturation
time (see the text for details).}}
\label{fig1}
\end{center}
\end{figure}

For the sake of clarity, a set of values for the kinetic constants can be 
chosen to be \cite{BeBeDAPe}:
\begin{equation}
k_1=1 \mu M^{-1} s^{-1}; \hspace{.2cm}
k_{-1}=4 s^{-1}; \hspace{.2cm}
k_2=1 s^{-1};
\label{kineticconstants}
\end{equation}
As we will discuss in detail in the following, this choice already corresponds
to a demanding case for applying the sQSSA, thus it is appropriate for the 
present analysis. In particular, these kinetic constants give 
${K_D}=4 \mu M$, ${K_M}=5 \mu M$ and $K=1 \mu M$. We will moreover consider 
two sets of ICVs:
\begin{equation}
e^a_0=1 \mu M; \hspace{.2cm} e^b_0=5 \mu M; \hspace{.2cm} 
s^a_0=s^b_0=10 \mu M;
\label{initialvalues}
\end{equation}
where we are taking different values of the initial enzyme concentration 
$e_0$, with $e^b_0 > e^a_0$, by labelling $a$ and $b$ the two corresponding 
sets. We plot in [Fig. \ref{fig1}] the solutions of Eqs.~(\ref{mmineq}),
obtained by numerically integrating the system, for these two sets of 
values, respectively.

The logarithmic scale in the figures enhances the typical presence of two
definitely different time scales, with the concentration of the complex $c$ 
that evolves very rapidly at the beginning whereas it turns out to be 
in a {\em quasi-steady-state} or {\em pseudo quasi-equilibrium} in the second 
part, of definitely longer duration. Indeed, the presence of a {\em plateau}, 
in which the complex is really roughly constant, is well more evident in case 
$a$, and we will show in the following that this choice of the ICVs 
corresponds to a situation in which the sQSSA is expected to work better.

In fact, in the sQSSA one looks at the second part, by directly 
taking $\dot{c} \sim 0$ in Eqs.~(\ref{mmineq}) (which is obviously quite a 
coarse approximation, though it does not mean that the complex is assumed
to be constant, but only that it depends algebraically on the substrate). 
Therefore, the approximation is the more correct the 
more the time scale corresponding to the rapid transient phase for the 
complex, $\tau_c$, is small with respect to the time scale $\tau_s$, that 
rules the slow decay of the substrate. 

Though one can argue about more refined time scale evaluations \cite{SeSl}, 
the most intuitive choice in order to roughly get the $\tau_c$ order of 
magnitude corresponds to  assume a constant substrate concentration 
in the first part, {\em i.e.}, a complex concentration that approximatively 
approaches exponentially its {\em plateau} value at the beginning \cite{MaMo}, 
$\dot{c} \sim -(c-c_{eq})/\tau_c$, with $\tau_c=1/(k_1(s_0+K_M))$, that takes 
the same values $\tau_c^a=\tau_c^b=0.0667 s$ in our case, for the $a$ and $b$ 
ICV sets respectively. On the other hand, when instead assuming 
$\dot{c} \sim 0$ in the second part, one finds \cite{MaMo} 
$\dot{s} \sim -k_2e_0s/(s+K_M) \sim -s/\tau_s$, that approximatively describes 
the exponential decay to zero of the substrate in the end. Correspondingly, 
the substrate decay time is $\tau_s=(s_0+K_M)/(k_2e_0)$, that takes  
instead the two different values $\tau_s^a=15 s$ and $\tau_s^b=3 s$ in our 
case. Hence, the substrate decay time is well larger for the $a$ set of ICVs 
than for the $b$ one. 
 
Let us now look more in detail at the behaviours displayed by the present 
numerical solutions of the original problem (\ref{mmineq}) when 
plotted in logarithmic scale. In fact, in the case of the substrate, one 
observes, both in [Fig. \ref{fig1}A] and in [Fig. \ref{fig1}B], the presence 
of three inflection points. Indeed, this feature appears to depend on the 
present choices of the kinetic constants and of the ICVs. We checked 
in particular the case $k_1=1 \mu M^{-1} s^{-1}$, $k_{-1}=10 s^{-1}$ and 
$k_2=20 s^{-1}$ with the $a$ set of ICVs and the present case with 
a definitely smaller value of $e_0=0.1 \mu M$, and we found a (not
shown) simpler behaviour, with only one inflection point in the
curve for the substrate.

It is therefore even more interesting that, as can be seen from 
[Fig. \ref{fig1}], though the used approximations for evaluating 
the time scales can look questionable, in logarithmic scale 
the obtained estimation of $\tau_s$ captures quite accurately the last 
({\em i.e.}, the third) inflection point in 
the curve for the substrate, that therefore appears to be interpretable 
as the substrate decay time. On the other hand, the obtained estimation for 
$\tau_c$ captures quite accurately the first inflection
point in the curve for the complex, that therefore appears
interpretable as the complex saturation time. In fact, 
the position of the third inflection point in the curve of the
substrate roughly coincides with the one of the second inflection 
point in the curve for the complex. Correspondingly, both of the inflection 
points in the curve for the complex are captured by $\tau_c$ and $\tau_s$, 
respectively (this is more evident in the case of the $a$ set of ICV). 
Moreover, the complex saturation time $\tau_c$ turns out to be quite near 
to the first inflection point in the curve for the substrate.

From this point of view, we notice that, 
in the previously recalled simpler considered cases, in which the 
(not shown) curves for the substrates display single inflection points
in logarithmic scale, their positions are more accurately captured by the 
corresponding substrate decay times $\tau_s$, and both $\tau_c$ and $\tau_s$ 
more accurately capture the positions of the 
two inflection points in the complex curves.

With the aim of deepening the analysis, we are led to the problem of 
adimensionalising the system in Eqs.~(\ref{mmineq}). This involves the 
choice of the $\varepsilon$ variable as the one giving the
sQSSA condition $\dot{c} \sim 0$ for $\varepsilon=0$. In fact, this 
$\varepsilon$ variable is clearly also the candidate for PEs 
that make possible to go beyond the sQSSA.
 
At this point, we remind that there are different approaches, in which for 
instance one chooses a different expansion parameter $\varepsilon$ 
\cite{BeBeDAPe,MaMo,SeSl}, or different independent variables at the beginning, 
such as in the already recalled tQSSA \cite{BeBeDAPe,BoBoSe,TzEd,DABe,La,Sw}, 
that turns out to make possible to describe a larger range of 
experimental situations and that could be therefore particularly useful in 
many cases.

In the present work, in order to test the SPDERG approach to MM kinetics, we 
focus on the sQSSA. In detail, within the two known possible 
adimensionalization schemes and the two corresponding different choices of the 
$\varepsilon$ variable \cite{BeBeDAPe,MaMo,SeSl} in this case, we study 
the one that is more largely considered in the literature 
\cite{BeBeDAPe,MaMo,Mu,EdKe,LiSe,HeTsAr}, {\em i.e.}, $\varepsilon=e_0/s_0$. 

In detail, one introduces the adimensional variables $m=K_D/s_0$,
$M=K_M/s_0$ and one scales the time of a factor $\delta$, 
$t \rightarrow \delta t$ with $\delta=k_1 e_0$. The substrate concentration is 
made adimensional by taking $\tilde{s}(t)=s(t)/s_0$. Moreover, in the
presently considered scheme, the complex concentration is made adimensional by 
taking $\tilde{c}(t)=c(t)/e_0$ (which is a possible correct choice since, 
thanks to the first of the recalled conservation laws, $c(t)\le e_0$). 

Correspondingly, one obtains the (singular with respect to $\varepsilon$) 
system of ODEs:
\begin{eqnarray}
\left\{
\begin{array}{lcl}
\dot{\tilde{s}}^{out}(t)&=& [\tilde{s}^{out}(t)+m] \left [\tilde{c}^{out}(t) - 
\frac{\textstyle  \tilde{s}^{out}(t)}
{\textstyle \tilde{s}^{out}(t)+m} \right  ]\\
\varepsilon \dot{\tilde{c}}^{out}(t)&=&-[\tilde{s}^{out}(t)+M] 
\left [\tilde{c}^{out}(t) - \frac{\textstyle \tilde{s}^{out}(t)}
{\textstyle \tilde{s}^{out}(t)+M} 
\right ].\\
\end{array}
\right.
\label{outer}
\end{eqnarray}
Here the label {\em out} refers to the fact that these are the ODEs that
capture the long time behaviours, {\em i.e.}, the ones to be obeyed by the 
{\em outer} solution. With the present kinetic constant choice 
(\ref{kineticconstants}), we have $m=0.4$ and $M=0.5$, respectively. 
On the other hand, the two sets of ICVs (\ref{initialvalues}) yield two
different values for $\delta$ and for the expansion parameter $\varepsilon$:
\begin{eqnarray}
\begin{array}{lclclcl}
\delta^a&=&1 s^{-1}; &\hspace{.2cm}& \delta^b&=&5 s^{-1};  \\
\varepsilon^a&=&0.1; & \hspace{.2cm} &  \varepsilon^b&=&0.5; 
\end{array}
\label{parameters2}
\end{eqnarray}
Thus, though the basic condition $\varepsilon < 1$ is 
verified in both of the cases, we are in the situation 
$\varepsilon^b > \varepsilon^a$, 
that is expected to be the most appropriate for
highlighting differences in the considered approximations. 

The sQSSA corresponds clearly to take $\varepsilon=0$. In fact 
\cite{BeBeDAPe,MaMo,LiSe}, it can be rigorously interpreted as the 0th order 
term of an asymptotic expansion in $\varepsilon$ of this example of 
singular perturbation, in which the original system is reduced to one ODE 
and one algebraic relation. For $\varepsilon=0$, one has: 
\begin{eqnarray}
\left \{
\begin{array}{lcl}
\dot{\tilde{s}}^{out}_{0}(t)&=&-\frac{\textstyle M-m}{\textstyle 
\tilde{s}^{out}_{0}(t)+M}
\tilde{s}^{out}_{0}(t) \\
{\tilde{c}}^{out}_{0}(t)&=&\frac{\textstyle \tilde{s}^{out}_{0}(t)}
{\textstyle \tilde{s}^{out}_{0}(t)+M}. \\
\end{array}
\right.
\label{QSSA}
\end{eqnarray}
It is to be noted that, in the numerator of the ODE to be obeyed by 
${\tilde{s}}^{out}_{0}(t)$, the quantity $M-m$ is just the adimensionalized
Van Slyke-Cullen constant $K/s_0$  \cite{VSCu}. 

The system is to be considered together with the ICV 
${\tilde{s}}^{out}_{0}(0)=1$, in fact implying that the other ICV   
is automatically fixed to ${\tilde{c}}^{out}_{0}(0)=1/(1+M)$.

The ODE for the adimensional substrate concentration can be solved explicitly 
\cite{BeBeDAPe,ScMe}, by means of the Lambert function $\omega(x)$ 
\cite{Coetal}, that verifies the equation $\omega(x)e^{\omega(x)}=x$:
\begin{equation}
\tilde{s}^{out}_{0}(t)=M \omega(e^{\textstyle -(M-m)t/M+1/M}/M).
\label{sout0}
\end{equation}
This solution does also satisfy the ICV  $\tilde{s}^{out}_{0}(0)=1$, since:
\begin{equation}
M \omega(e^{\textstyle 1/M}/M)=M\omega \left[ \omega^{-1}(1/M)\right].
\end{equation}
Correspondingly, one gets:
\begin{equation}
\tilde{c}^{out}_{0}(t)=\frac{\textstyle \omega(e^{\textstyle -(M-m)t/M+1/M}/M)}
{\textstyle \omega(e^{\textstyle -(M-m)t/M+1/M}/M)+1},
\label{cout0}
\end{equation}
with, in particular, ${\tilde{c}}^{out}_{0}(0)=1/(1+M)$, as expected.

\begin{figure}[tb]
\begin{center}
\leavevmode
\epsfig{figure=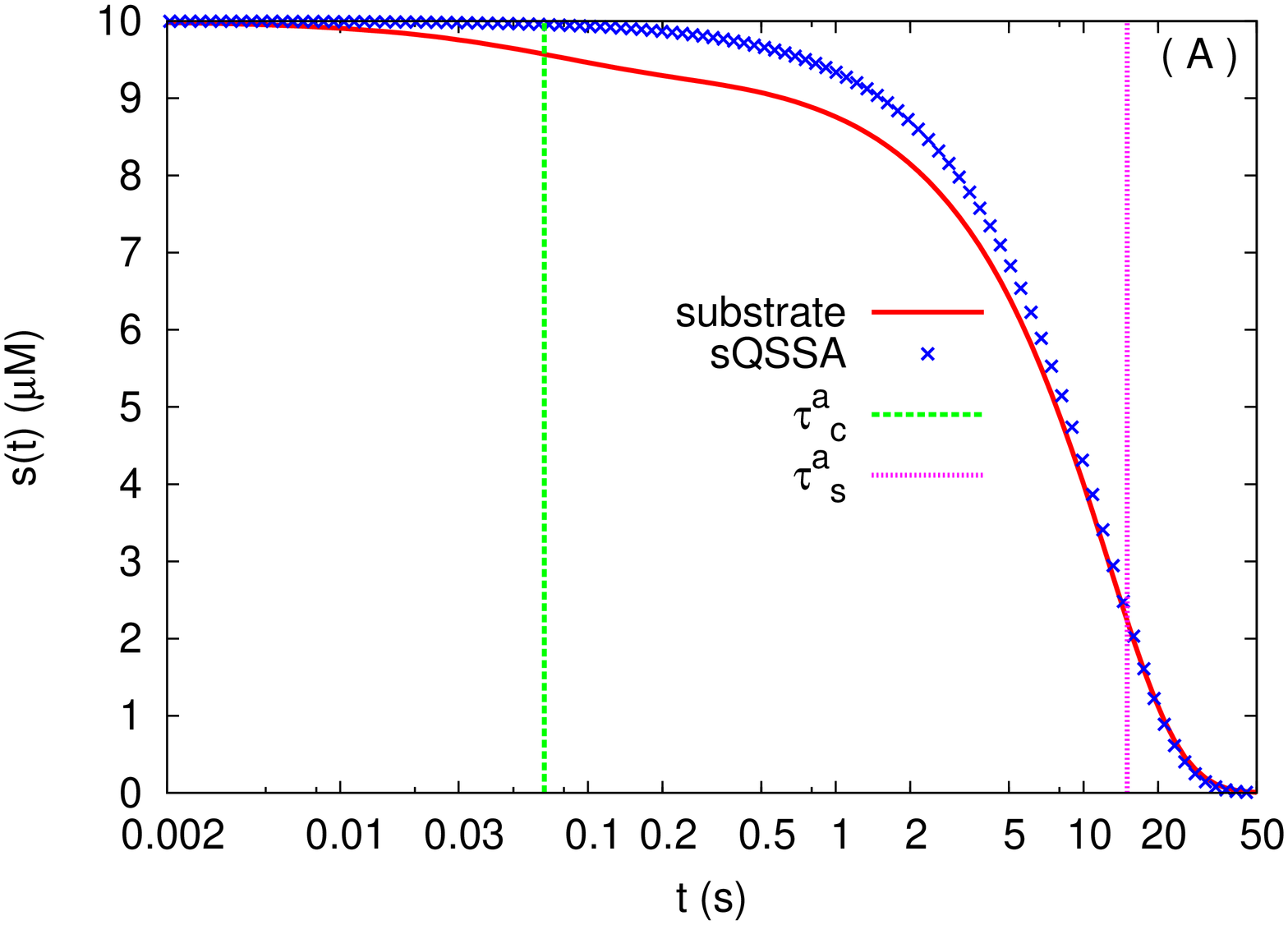,angle=270,width=8cm}
\epsfig{figure=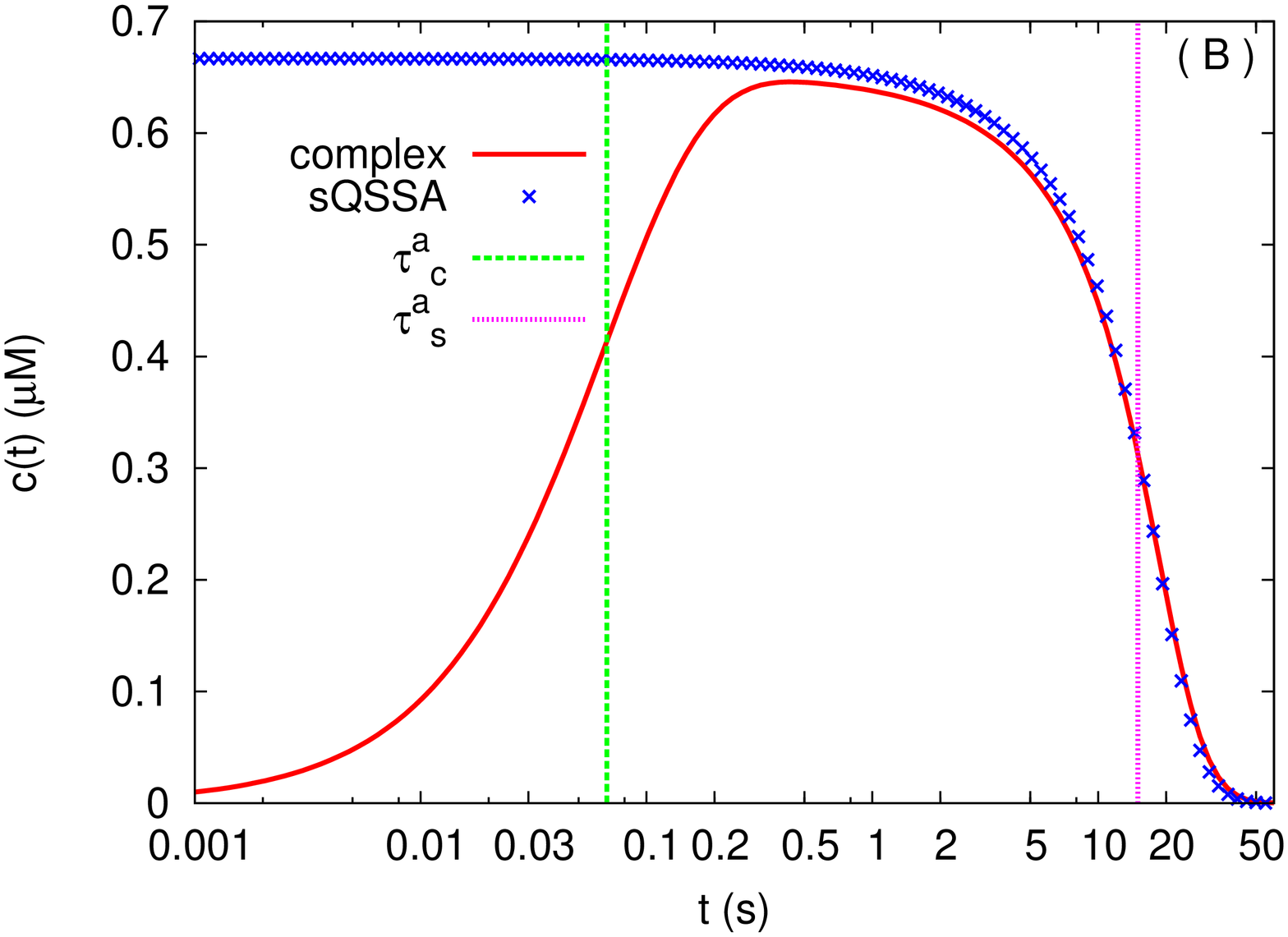,angle=270,width=8cm}
\epsfig{figure=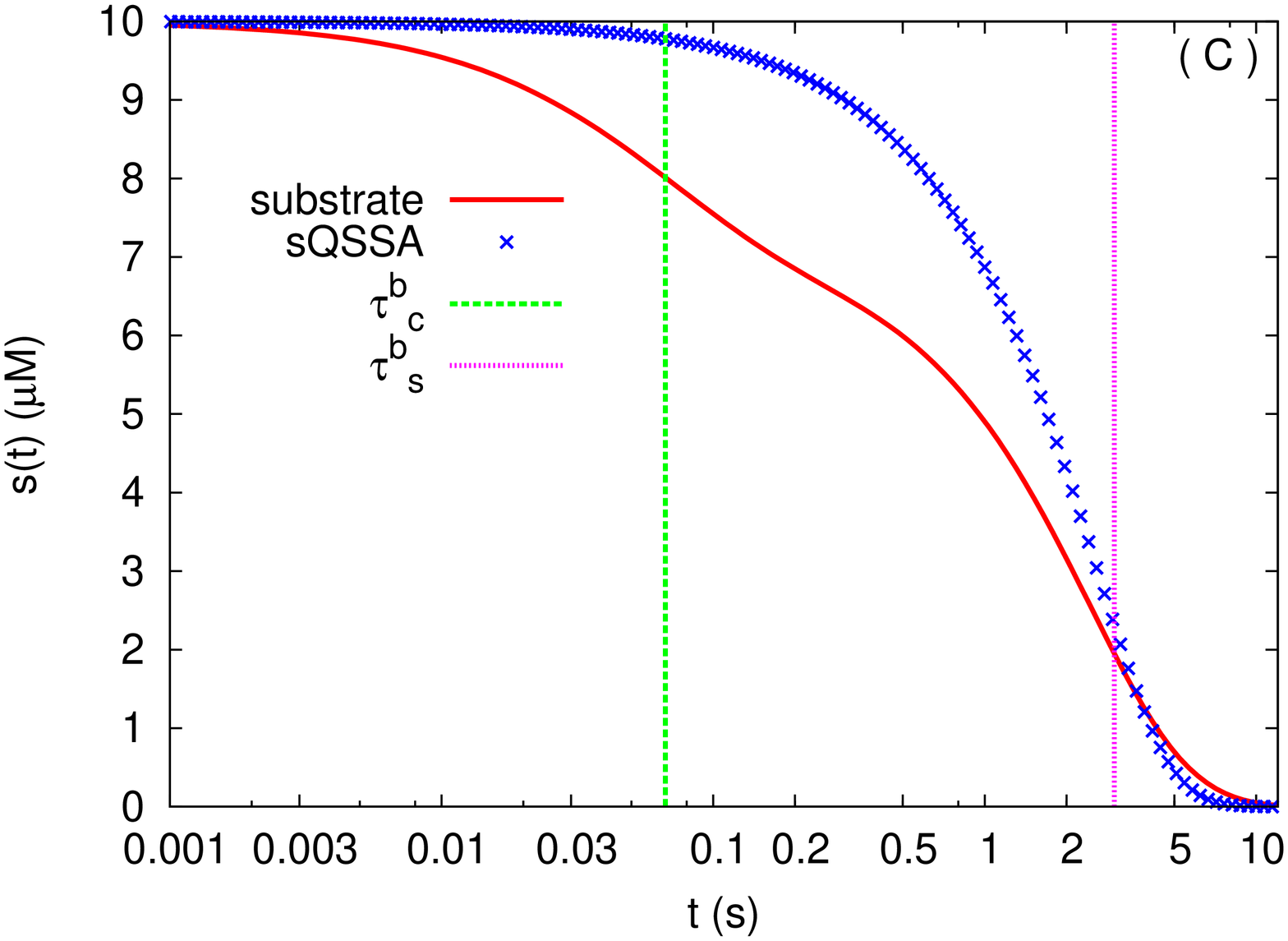,angle=270,width=8cm}
\epsfig{figure=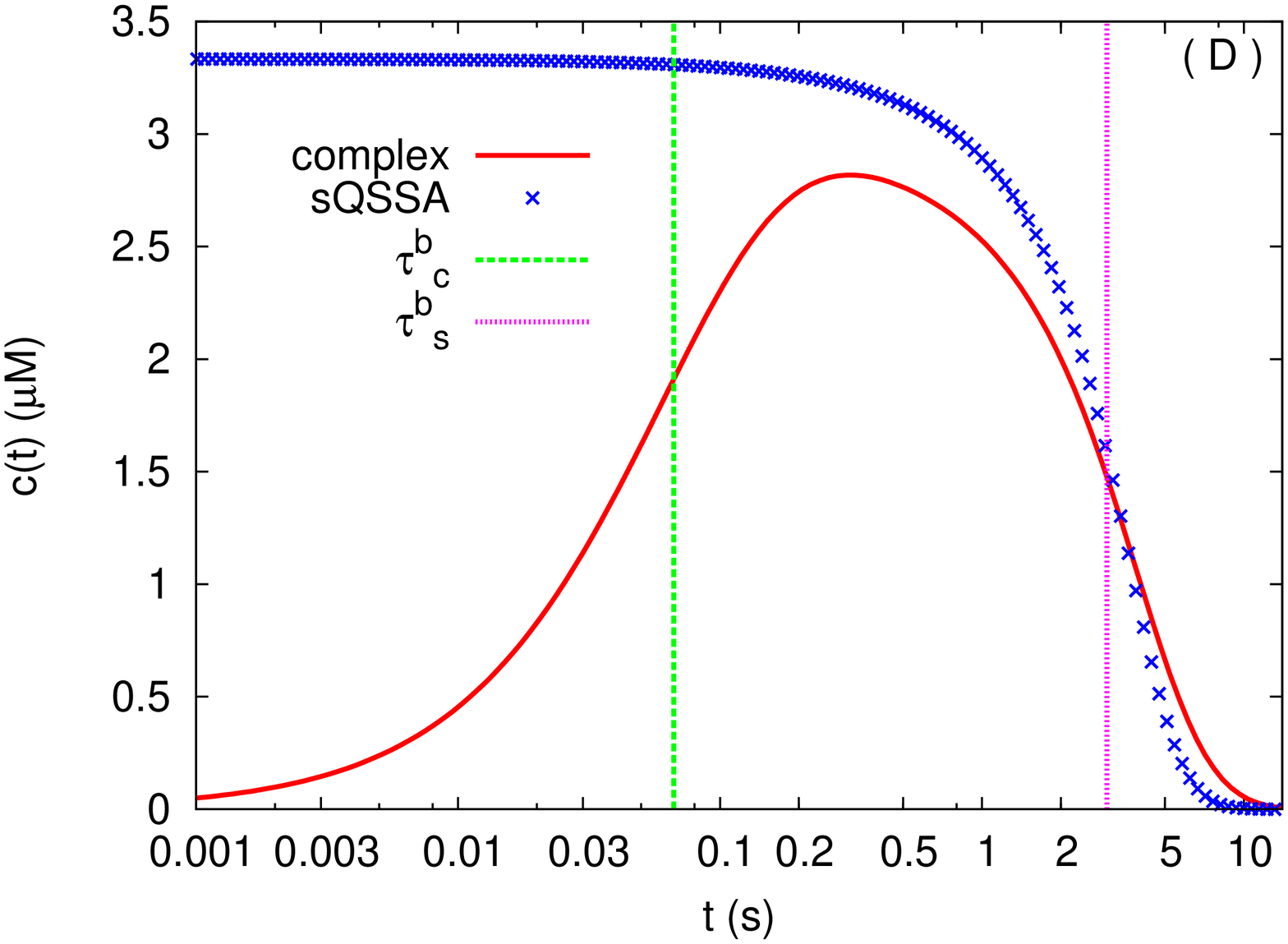,angle=270,width=8cm}
\caption{{\footnotesize In $A)$ and in $C)$ we present the behaviour of the 
concentrations of the substrate $s(t)$, whereas in $B)$ and in $D)$ we present 
the ones of the complex $c(t)$ for the $a$ and $b$ sets of ICVs 
given in (\ref{initialvalues}), respectively. Hence, in $A)$ and $B)$ we are 
in the case with $\varepsilon=\varepsilon^a=0.1$, whereas in $C)$ and $D)$ we 
are in the one with $\varepsilon=\varepsilon^b=0.5$. We plot both the 
numerical solutions of Eqs.~(\ref{mmineq}) already shown in the previous 
[Fig. \ref{fig1}], and the analytical solutions computed from the sQSSA ones 
(by using a standard numerical approximation for the Lambert function) as 
given in (\ref{sout0}) and (\ref{cout0}). We finally plot the corresponding 
rough evaluations of the two different time scales involved, too, 
with $\tau_s$ describing the substrate decay time and $\tau_c$ the complex 
saturation time. Notice that the time is in logarithmic scale.}}
\label{fig2}
\end{center}
\end{figure}

Hence, by using a standard numerical approximation for the Lambert
function, we obtain the behaviours of $\tilde{s}^{out}_{0}(t)$ and
$\tilde{c}^{out}_{0}(t)$, {\em i.e.}, the behaviours of the
substrate and complex concentrations within the sQSSA. We show in 
[Fig. \ref{fig2}] our results for the two considered sets of ICVs, in 
comparison with the numerical solution of the original problem 
(\ref{mmineq}) (the same curves as in [Fig. \ref{fig1}]).

We note qualitatively that the failure of the approximation is well more 
evident for the $b$ set of ICVs ([Fig. \ref{fig2}C] and [Fig. \ref{fig2}D]), 
as it could be expected, since $\varepsilon^b > \varepsilon^a$. In both cases, 
in these figures the logarithmic scale highlights that the complex very 
short time behaviour is not captured at all, a well known failure of the sQSSA
(that is shared by the tQSSA, too). From this point of view, it is to be  
stressed that, as we already pointed out, whereas one can solve 
Eqs.~(\ref{QSSA}) by choosing $\tilde{s}^{out}_{0}(0)=1$, the original ICV for 
the complex concentration cannot {\em a priori} be satisfied within the 
present context. Therefore, the approximation is unable to predict the initial 
increase from zero of this quantity.

Noticeably, moreover, with the chosen kinetic constants and ICVs, 
the presence of more than one inflection point in logarithmic scale in
the curve for the substrate (hence the short time qualitative behaviour
of this quantity) is not reproduced at all. In fact, also the maximum 
reached by the complex during its evolution is lower than the ICV
within the sQSSA in both of the considered cases. 
In particular, this last qualitative feature of a complex maximum lower 
than the sQSSA ICV is not at all observed in the previously recalled simpler 
considered cases, in which the (not shown) curves for the substrates display 
a single inflection point in logarithmic scale, supporting the hypothesis that 
the various effects are 
related. Reasonably, it is for this reason that, in case $b$, it is naked-eye 
evident that the sQSSA fails in reproducing the long time behaviour, too 
(see [Fig. \ref{fig2}C] and [Fig. \ref{fig2}D]). Indeed, this case is even 
more demanding because of the quite large $\varepsilon=\varepsilon^b=0.5$.
In detail, one observes here sQSSA long time behaviours that tend to
zero definitely more quickly than the numerical solutions of the original 
problem. Thus, as anticipated, this is a particularly demanding situation for 
correctly approximating MM dynamics and studying the standard PE that goes 
beyond the sQSSA.

\section{The  perturbation expansion beyond the sQSSA}
\label{review2}
\noindent
The standard PE method, also in the case of MM kinetics 
\cite{BeBeDAPe,MaMo,LiSe,Mu,EdKe}, is based on mathematical results 
for systems of ODEs with both singular and regular perturbations. 
Precisely within this framework, system (\ref{outer}) is a well known 
example of singular perturbation, that is to be obeyed by the outer solutions of
the original system. In the same framework, system (\ref{outer}) needs to be 
considered together with the system that one obtains by taking 
$t \rightarrow \tau=t/\varepsilon$, {\em i.e.}, in the opposite limit of 
short times, that is: 
\begin{eqnarray}
\left\{
\begin{array}{lcl}
\dot{\tilde{s}}^{in}(\tau)&=& \varepsilon [\tilde{s}^{in}(\tau)+m] 
\left [\tilde{c}^{in}(\tau) - 
\frac{\textstyle  \tilde{s}^{in}(\tau)}
{\textstyle \tilde{s}^{in}(\tau)+m} \right  ]\\
\dot{\tilde{c}}^{in}(\tau)&=&-[\tilde{s}^{in}(\tau)+M] 
\left [\tilde{c}^{in}(\tau) - \frac{\textstyle \tilde{s}^{in}(\tau)}
{\textstyle \tilde{s}^{in}(\tau)+M} 
\right ],
\end{array}
\right.
\label{inner}
\end{eqnarray}
 to be instead obeyed by the {\em inner} solutions.

Correspondingly \cite{BeBeDAPe,MaMo,LiSe,Mu,EdKe}, one searches solutions
in the form:
\begin{eqnarray}
\left\{
\begin{array}{lclclcl}
{\tilde{s}}^{in}(\tau)&=&{ \sum\limits_{i=0}^{\infty}}
{\tilde{s}}^{in}_i(\tau)\varepsilon^i;
&\hspace{.2in}&
{\tilde{c}}^{in}(\tau)&=&{ \sum\limits_{i=0}^{\infty}}
{\tilde{c}}^{in}_i(\tau)\varepsilon^i;
\\
{\tilde{s}}^{out}(t)&=&{ \sum\limits_{i=0}^{\infty}}
{\tilde{s}}^{out}_i(t)\varepsilon^i;
&\hspace{.2in}&
{\tilde{c}}^{out}(t)&=&{ \sum\limits_{i=0}^{\infty}}
{\tilde{c}}^{out}_i(t)\varepsilon^i;
\end{array}
\right.
\end{eqnarray}
by requiring that 
$\left\{ {\tilde{s}}^{in}_i(\tau), {\tilde{c}}^{in}_i(\tau) \right \}$ 
satisfy the system (\ref{inner}) in the case of the inner solutions and that 
$\left\{ {\tilde{s}}^{out}_i(t), {\tilde{c}}^{out}_i(t) \right \}$ 
satisfy the system (\ref{outer}) in the case of the outer solutions, at each 
order in $\varepsilon$. One then imposes the appropriate MCs and takes, as UAs 
to the correct solutions at a given order in $\varepsilon$, the sum of the 
inner and of the outer solutions at that order minus the common terms 
\cite{LiSe}.  

The whole procedure appears therefore quite a standard PE method.  
For the sake of consistency, we also remind that one needs to impose 
appropriate MCs since, whereas one clearly takes inner solutions that also 
satisfy the ICVs ({\em i.e.}, ${\tilde{s}}^{in}(0)=1$ and 
${\tilde{c}}^{in}(0)=0$), the ICVs of the outer solutions  are to be determined 
from the behaviours of the inner ones in the large $\tau$ limit, that have to 
correspond to their behaviours in the small $t$ limit \cite{BeBeDAPe,MaMo,LiSe}.
As we are going to discuss more in detail in the following, a particularly 
delicate point in the case of MM kinetics is represented just by the choice of 
the MCs, that actually, already at the 1st order, implies the need for 
considering MCs involving the 1st derivative of the outer solution, too 
\cite{LiSe}.

In detail, in the present work, in the considered case of this kind of PE 
beyond the sQSSA with $\varepsilon=e_0/s_0$, we reproduce the calculations as 
also in \cite{BeBeDAPe,MaMo} at the 0th order, whereas we follow both the 
original paper by Heineken, Tsushiya and Aris \cite{HeTsAr} and 
the discussion in \cite{LiSe} for the complete 1st order contribution.

Let us first of all recall the 0th order inner solutions, obtained by setting 
$\varepsilon=0$ in Eqs.~(\ref{inner}), that thereby solve:
\begin{eqnarray}
\left\{
\begin{array}{lcl}
\dot{\tilde{s}}^{in}_0(\tau)&=& 0\\
\dot{\tilde{c}}^{in}_0(\tau)&=&-[\tilde{s}^{in}_0(\tau)+M] 
\left [\tilde{c}^{in}_0(\tau) - \frac{\textstyle \tilde{s}^{in}_0(\tau)}
{\textstyle \tilde{s}^{in}_0(\tau)+M} 
\right ],
\end{array}
\right.
\label{inner0}
\end{eqnarray}
with $\tilde{s}^{in}_0(0)=1$ and $\tilde{c}^{in}_0(0)=0$. The solutions are
\cite{BeBeDAPe,MaMo}:
\begin{eqnarray}
\left\{
\begin{array}{lcl}
\tilde{s}^{in}_0(\tau)&=& 1\\
\tilde{c}^{in}_0(\tau)&=& \frac{\textstyle 1}{\textstyle 1+M} \left [ 1-
e^{\textstyle -(1+M)\tau} \right ].
\end{array}
\right.
\label{solinner0}
\end{eqnarray}
At this point, we pass to consider the 0th order outer solution of the system 
(\ref{QSSA}) with ICV $\tilde{s}^{out}_0(0)=\tilde{s}^{out *}_0$, where 
$\tilde{s}^{out *}_0$ needs to be determined by the MC for the substrate. 
Nevertheless, it is clear that, in the relatively simple 0th order case, 
one expects that the substrate solution verifies $\tilde{s}^{out *}_0=1$. 
In fact, with this choice, one has: 
\begin{eqnarray}
\left\{
\begin{array}{lcccl}
\lim_{\tau \rightarrow \infty}\tilde{s}^{in}_0(\tau)&=& 1&=& \lim_{t \rightarrow 0}
\tilde{s}^{out}_0(t)\\
\lim_{\tau \rightarrow \infty}\tilde{c}^{in}_0(\tau)&=& 
\frac{\textstyle 1}{\textstyle 1+M}&=&\lim_{t \rightarrow 0}
\tilde{c}^{out}_0(t).
\end{array}
\right.
\label{matchsol0}
\end{eqnarray}
Hence, consistently \cite{LiSe}, the needed MC for the substrate satisfies 
the one for the complex, too.

In conclusion, the 0th order PE UAs (that we label $u$), are given by 
\cite{BeBeDAPe,MaMo}:
\begin{eqnarray}
\left\{
\begin{array}{lcl}
\tilde{s}^{u}_0(t)=\tilde{s}^{in}_0(t/\varepsilon)+\tilde{s}^{out}_0(t)-1\\
\tilde{c}^{u}_0(t)=\tilde{c}^{in}_0(t/\varepsilon)+\tilde{c}^{out}_0(t)- 
\frac{\textstyle 1}{\textstyle 1+M}.
\end{array}
\right.
\label{solun0}
\end{eqnarray}

\begin{figure}[t]
\begin{center}
\leavevmode
\epsfig{figure=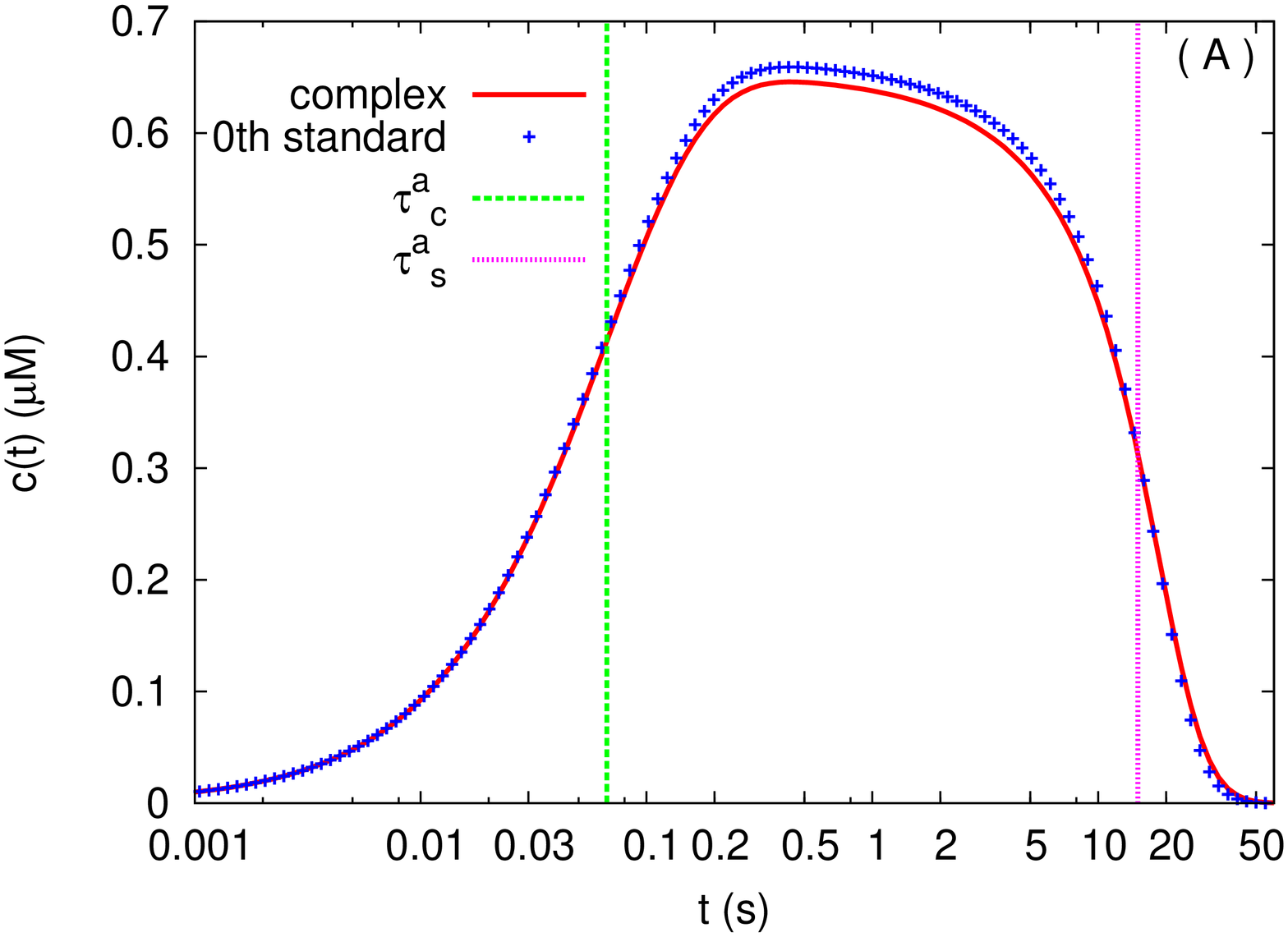,angle=270,width=8cm}
\epsfig{figure=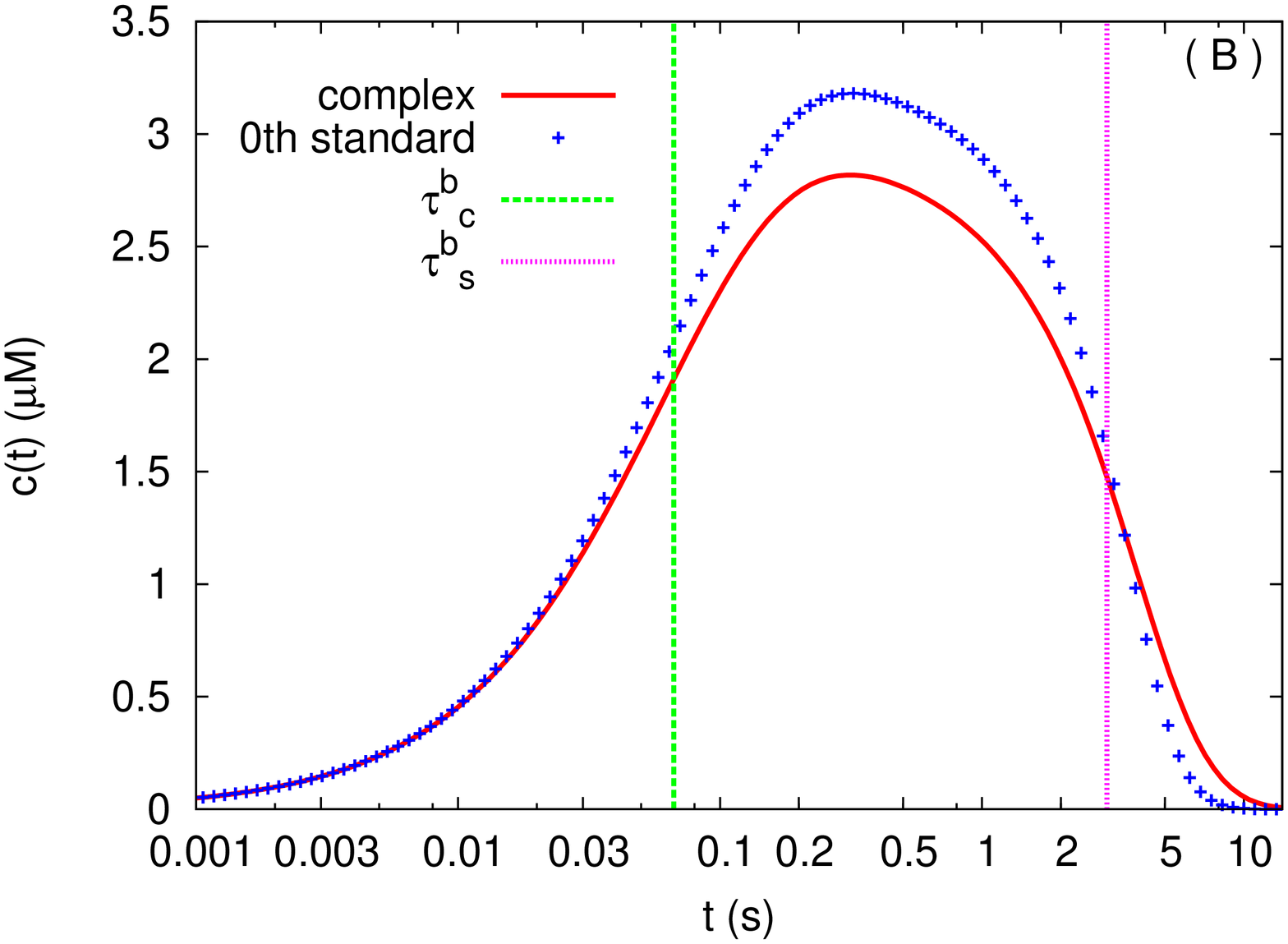,angle=270,width=8cm}
\caption{{\footnotesize In $A)$ we present the behaviour of the concentration 
of the complex $c(t)$ for the $a$ set of ICVs given in (\ref{initialvalues}), 
whereas in $B)$ we present the one for the $b$ set. Hence, it is  
$\varepsilon=\varepsilon^a=0.1$ in the first case and 
$\varepsilon=\varepsilon^b=0.5$ in the second one.
We plot both the numerical solutions of  Eqs.~(\ref{mmineq}) already shown in 
the previous figures, and the analytical solutions computed from the 
adimensional 0th order PE UAs (by using a standard numerical 
approximation for the Lambert function) as given in (\ref{solun0}).
We finally plot our corresponding rough evaluations of the two different 
time scales involved, too, with $\tau_s$ describing the substrate decay time 
and $\tau_c$ the complex saturation time. Notice that the time is in 
logarithmic scale.}}
\label{fig3}
\end{center}
\end{figure}

We plot in [Fig. \ref{fig3}] our results on the complex concentrations (the
UA for the substrate being coincident with the sQSSA for this quantity) for 
the two considered sets of ICVs, in comparison with the 
numerical solutions of the original problem (\ref{mmineq}) (the same 
curves as in [Fig. \ref{fig1}]), to show that this approximation already 
captures the most characteristic features of the whole system's dynamics, 
and in particular the rapid initial increase of the complex, too. 

Nevertheless, the approximation over-evaluates the numerically obtained 
maximum values for the complex, as it is definitely more evident in case $b$,
corresponding to the higher considered value of the expansion 
parameter, $\varepsilon=\varepsilon^b=0.5$ ([Fig. \ref{fig3}B]). On the other 
hand, as we already outlined, one could notice in the curves for the substrate
([Fig. \ref{fig2}A] and [Fig. \ref{fig2}C]) the sQSSA failure
in capturing the presence of the first two inflection points
(in logarithmic scale). In fact, it is now clear that this failure in 
correctly reproducing the qualitative short time substrate behaviour
is shared by the 0th order PE UA, too. We underline once more that
these observations appear explainable because the present choice of 
the kinetic constants (and of the initial values in case $b$) makes the 
dynamical behaviour particularly complex. 

Let us continue to sketch the procedure that is usually carried on, by 
recalling the calculation of the 1st order contribution \cite{LiSe,HeTsAr}. 
The 1st order inner solutions solve the system: 
\begin{eqnarray}
\left\{
\begin{array}{lcl}
\dot{\tilde{s}}^{in}_1(\tau)&=& (1+m) 
\left [ \tilde{c}^{in}_0(\tau) - 
\frac{\textstyle 1}{\textstyle 1+m} \right  ],\\
\dot{\tilde{c}}^{in}_1(\tau)&=&- (1+M)\tilde{c}^{in}_1(\tau)-
\left [ \tilde{c}^{in}_0(\tau)-1 \right ] \tilde{s}^{in}_1(\tau).
\end{array}
\right.
\label{inner1}
\end{eqnarray}
Here, we already used $\tilde{s}^{in}_0(\tau)=1$, thus the ICVs  
are $\tilde{s}^{in}_1(0)=\tilde{c}^{in}_1(0)=0$. In fact, one has
in particular to solve an ODE for ${\tilde{c}}^{in}_1$ in the form
$\dot{y}(\tau)=-(1+M)y(\tau)+f(\tau)$, whose solution is 
$y(\tau)=a(\tau)e^{-(1+M)\tau}$, with $a(\tau)=\int_0^\tau f(z) e^{(1+M)z} dz$. 
One obtains \cite{LiSe,HeTsAr}:
\begin{eqnarray}
\left\{
\begin{array}{lcl}
\tilde{s}^{in}_1(\tau)&=& -\frac{\textstyle M-m}{\textstyle 1+M} \tau
- \frac{\textstyle 1+m}{\textstyle (1+M)^2}  \left [ 1-
e^{\textstyle -(1+M)\tau} \right ]\\
\tilde{c}^{in}_1(\tau)&=& -\frac{\textstyle M(M-m)}{\textstyle (1+M)^3} \tau
- \frac{\textstyle M(1+2m-M)}{\textstyle (1+M)^4}   
 \left [ 1-e^{\textstyle -(1+M)\tau} \right ]+\\
&-& \left [ \frac{\textstyle (1-M)(1+m)}
{\textstyle (1+M)^3} \tau +\frac{\textstyle M-m}{\textstyle (1+M)^2} 
\frac{\textstyle \tau^2}{\textstyle 2} \right ]e^{\textstyle -(1+M)\tau}
+ \\
&+&\frac{\textstyle (1+m)}
{\textstyle (1+M)^4}e^{\textstyle -(1+M)\tau} \left [
1-e^{\textstyle -(1+M)\tau} \right ].
\end{array}
\right.
\label{solinner1}
\end{eqnarray}

On the other hand, the 1st order outer solutions are more 
involved, since they solve the system (that consists
once again of one ODE and of one algebraic relation):
\begin{eqnarray}
\left\{
\begin{array}{lcl}
\dot{\tilde{s}}^{out}_1(t)&=& \frac{\textstyle M(M-m)}
{\textstyle \left [\tilde{s}^{out}_0(t)+M \right ]^4}\tilde{s}^{out}_0(t) 
\left [\tilde{s}^{out}_0(t)+m \right ]-\frac{\textstyle M(M-m)}
{\textstyle \left [\tilde{s}^{out}_0(t)+M \right ]^2}\tilde{s}^{out}_1(t)\\
\tilde{c}^{out}_1(t)&=&\frac{\textstyle M(M-m)}
{\textstyle \left [ \tilde{s}^{out}_0(t)+M \right ]^4}+\frac{\textstyle M}
{\textstyle \left [\tilde{s}^{out}_0(t)+M  \right]^2}\tilde{s}^{out}_1(t),\\
\end{array}
\right.
\label{outer1}
\end{eqnarray}
in which $\tilde{s}^{out}_0(t)$ is reported, in terms of the Lambert function, 
in (\ref{sout0}). One can verify that the solution for $\tilde{s}^{out}_1(t)$ 
is given by \cite{LiSe}:
\begin{equation}
\tilde{s}^{out}_1(t)= \frac{\textstyle {s}^{out}_0(t)}
{\textstyle \tilde{s}^{out}_0(t)+M } \left \{ \frac{\textstyle m}{\textstyle M }
\log \left [ \frac{\textstyle \tilde{s}^{out}_0(t)+M}{\textstyle (1+ M) 
\tilde{s}^{out}_0(t)} \right ] - \frac{\textstyle \tilde{s}^{out}_0(t)+m}
{\tilde{s}^{out}_0(t)+M} \right \}.
\label{solsouter1}
\end{equation}
Correspondingly, the 1st order outer solution for the complex is:
\begin{equation}
\tilde{c}^{out}_1(t)= \frac{\textstyle \tilde{s}^{out}_0(t)}
{\textstyle \left [ \tilde{s}^{out}_0(t)+M \right ]^3} \left \{ m 
\log \left [ \frac{\textstyle \tilde{s}^{out}_0(t)+M}{\textstyle (1+ M) 
\tilde{s}^{out}_0(t)} \right ] + \frac{\textstyle 2M (M-m)}
{\tilde{s}^{out}_0(t)+M} -M \right \}.
\label{solcouter1}
\end{equation}

Indeed, if we  were to neglect the {\em secular} terms ({\em i.e.}, the
terms proportional to $\tau$ in (\ref{solinner1})), these solutions would also 
correctly satisfy the MCs, since one has:
\begin{equation}
\tilde{s}^{out}_1(0)= -\frac{\textstyle 1+m}{\textstyle (1+M)^2} 
\hspace{.3in} \tilde{c}^{out}_1(0)=-\frac{\textstyle M(1+2m-M)}
{\textstyle (1+M)^4},
\label{match1a}
\end{equation}
and these are also the constant terms in the 1st order inner solutions.

From the point of view of the present work, that aims to underline similarities
and differences between the PE and the SPDERG approach, it appears important 
to stress that, within the standard method framework, one reasonably justifies 
the disappearance of the 1st order secular terms with the imposition of two 
term MCs \cite{LiSe}. In fact, these MCs involve the 1st order derivatives of 
the 0th order outer solutions, too. Moreover \cite{LiSe}, one iteratively 
expects that the higher order divergences (hence in particular the presence of 
secular terms proportional to higher powers of $\tau$ in the higher order
inner solutions) could be absorbed by a possibly increasing number of terms 
in the corresponding MCs.

In detail, from Eqs.~(\ref{QSSA}), one has:
\begin{equation}
\dot{\tilde{s}}^{out}_0(0)= -\frac{\textstyle M-m}{\textstyle (1+M)}; 
\hspace{.3in} \dot{\tilde{c}}^{out}_0(0)=-\frac{\textstyle M(M-m)}
{\textstyle (1+M)^3};
\label{match1b}
\end{equation}
Correspondingly, it turns out to be verified a two term MCs \cite{LiSe}, 
roughly summarizable as:
\begin{eqnarray}
\left\{
\begin{array}{lcl}
\lim_{\tau \rightarrow \infty} \left \{ \left [ \tilde{s}^{in}_0(\tau)
+\varepsilon \tilde{s}^{in}_1(\tau) \right ] - \left [ \tilde{s}^{out}_0(0)
+t \dot{\tilde{s}}^{out}_0(0) + \varepsilon \tilde{s}^{out}_1(0)
 \right ] \right \}&=&0, \\
\lim_{\tau \rightarrow \infty} \left \{ \left [ \tilde{c}^{in}_0(\tau)
+\varepsilon \tilde{c}^{in}_1(\tau) \right ] - \left [ \tilde{c}^{out}_0(0)
+t \dot{\tilde{c}}^{out}_0(0) + \varepsilon \tilde{c}^{out}_1(0)
 \right ] \right \}&=&0. \\
\end{array}
\right.
\label{match1}
\end{eqnarray}
Noticeably, it can be also proved \cite{LiSe} that the 1st order MCs, that we 
reported here, apply on a time interval (the matching region) that 
ranges from $\varepsilon$ to $\sqrt{\varepsilon}$, and that the whole PE, 
with analogous MCs, approaches uniformly the correct solutions for 
$t \in [0, \infty )$.

\begin{figure}[t]
\begin{center}
\leavevmode
\epsfig{figure=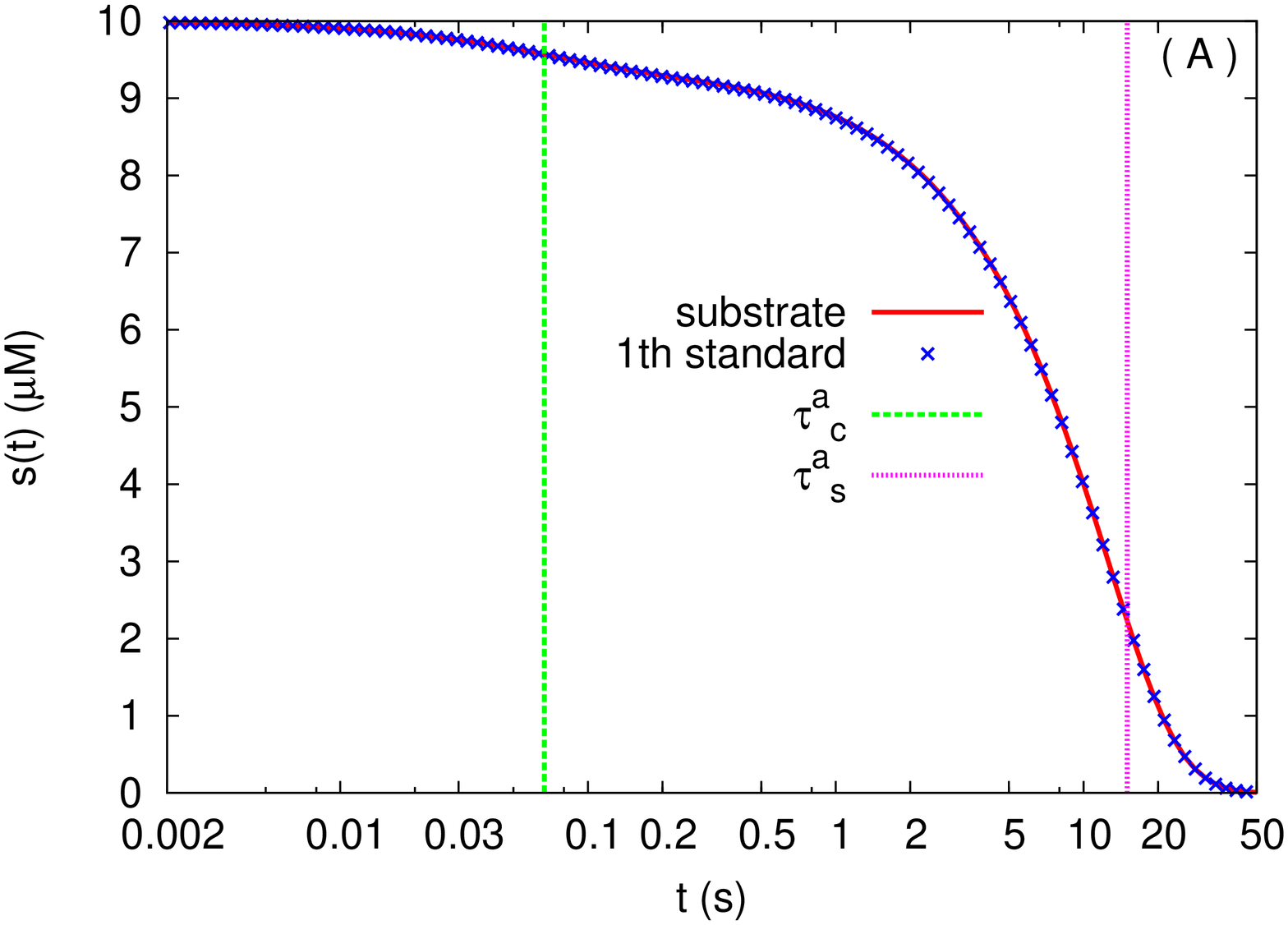,angle=270,width=8cm}
\epsfig{figure=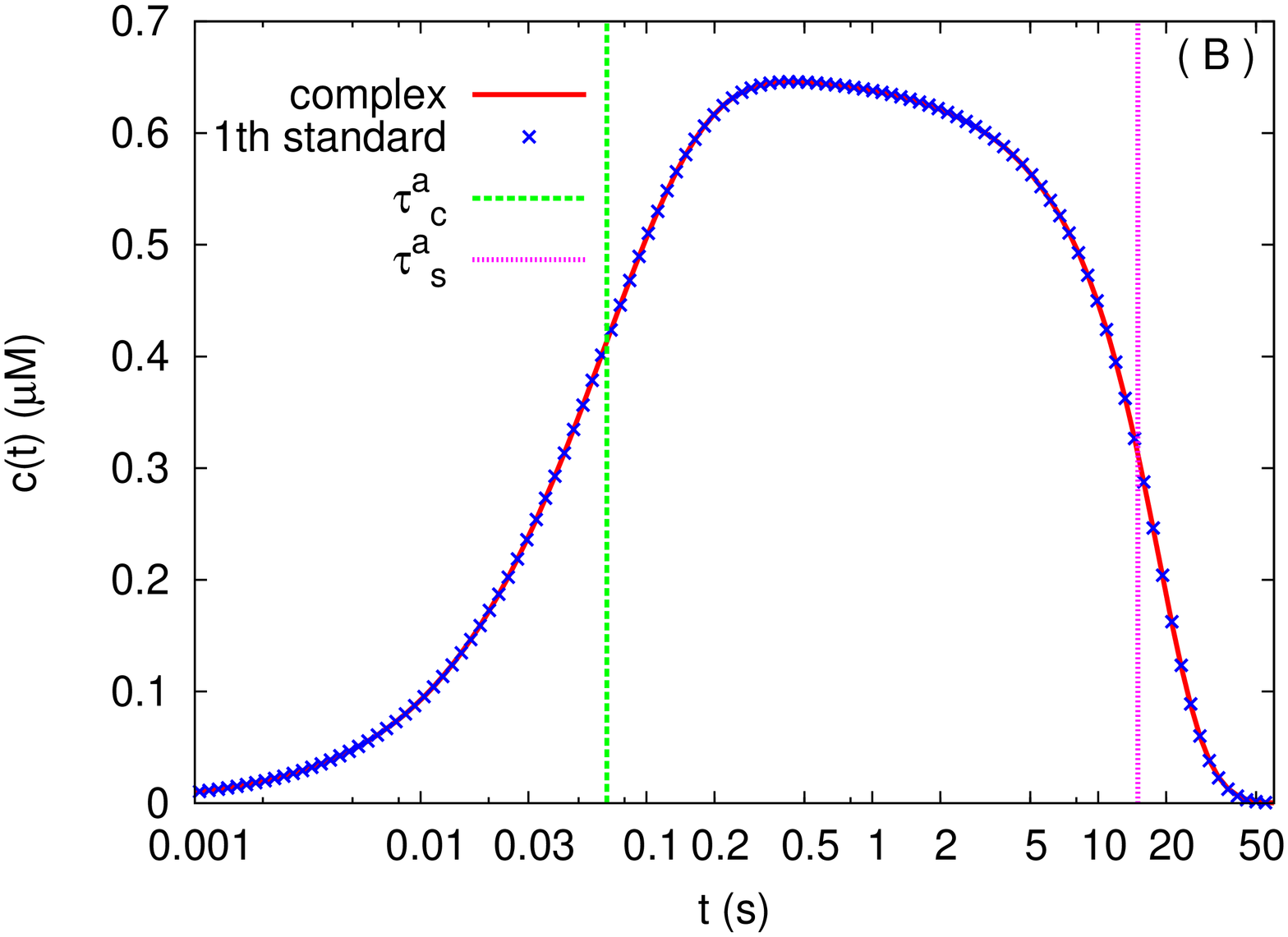,angle=270,width=8cm}
\epsfig{figure=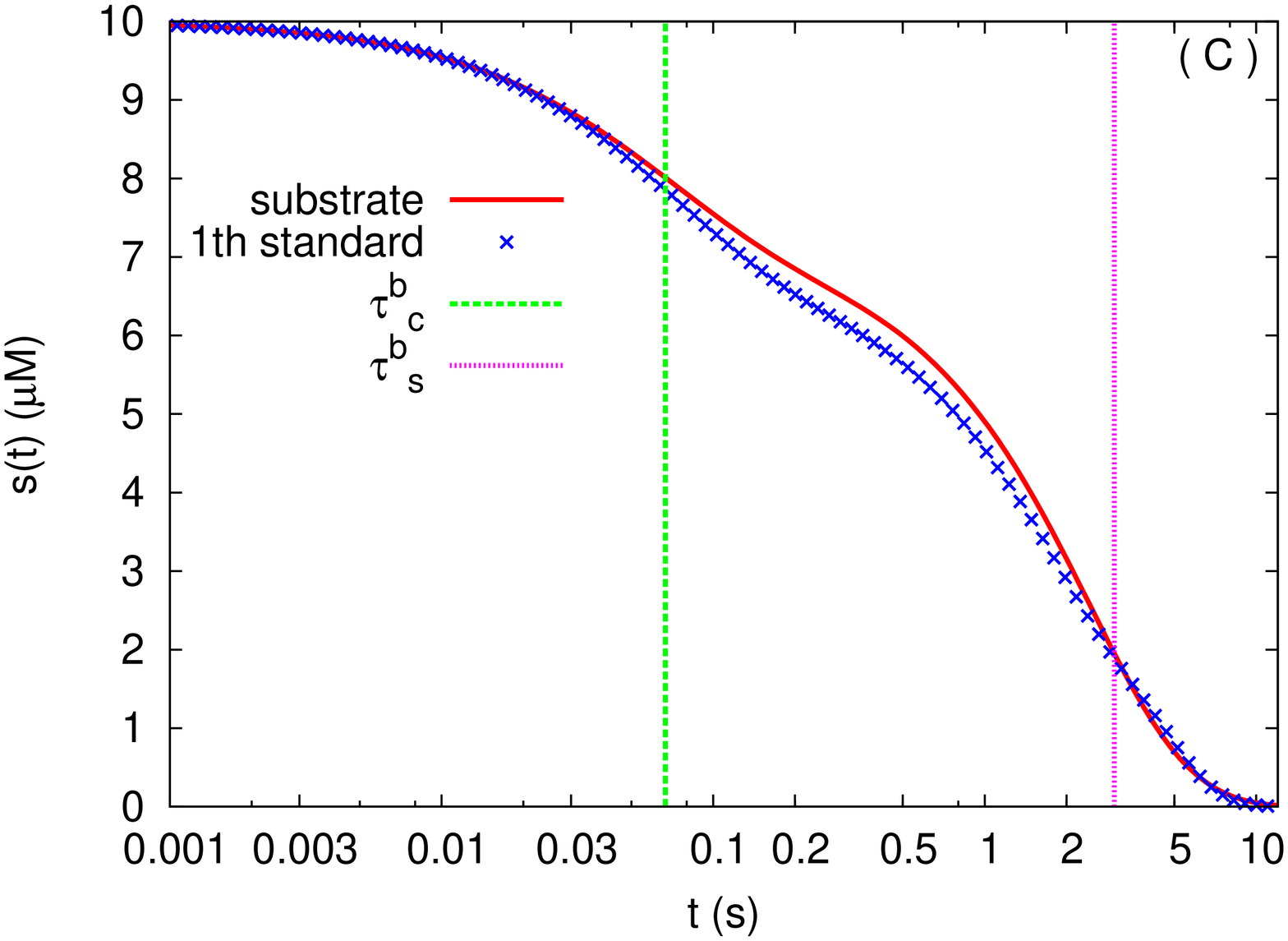,angle=270,width=8cm}
\epsfig{figure=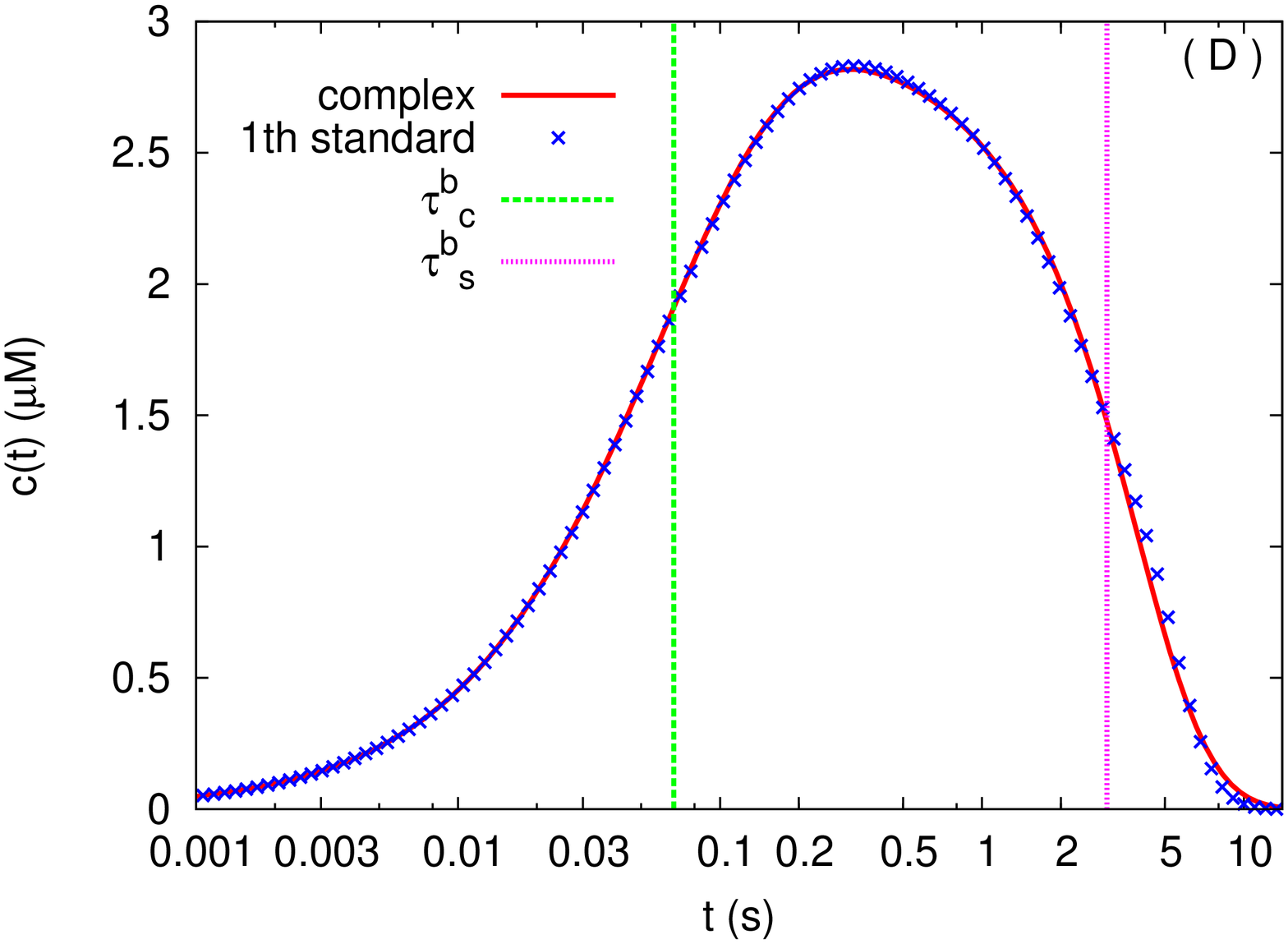,angle=270,width=8cm}
\caption{{\footnotesize In $A)$ and $C)$ we present the behaviour of the 
concentrations of the substrate $s(t)$, whereas in $B)$ and $D)$ we present 
the ones of the complex $c(t)$ for the $a$ and $b$ sets of ICVs given in  
(\ref{initialvalues}), respectively. Hence, in $A)$ and $B)$ we are 
in the case with $\varepsilon=\varepsilon^a=0.1$, whereas in $C)$ and $D)$ 
we are in the one with $\varepsilon=\varepsilon^b=0.5$.
We plot both the numerical solutions of Eqs.~(\ref{mmineq}) 
already shown in the previous figures, and the analytical solutions 
computed from the 1st order PE UAs (by using a standard numerical 
approximation for the Lambert function) as given in (\ref{solun1}).
We finally plot our corresponding rough evaluations of the two different 
time scales involved, too, with $\tau_s$ describing the substrate decay time 
and $\tau_c$ the complex saturation time. Notice that the time is in 
logarithmic scale.}}
\label{fig4}
\end{center}
\end{figure}

Therefore, one takes as PE UAs to the solutions at the 1st order in 
$\varepsilon$:
\begin{eqnarray}
\left\{
\begin{array}{lcl}
\tilde{s}^{u}_1(t)&=&\left [\tilde{s}^{in}_0(t/\varepsilon)+\tilde{s}^{out}_0(t)
\right ]+ \varepsilon \left [ \tilde{s}^{in}_1(t/\varepsilon)+
\tilde{s}^{out}_1(t) 
\right ]-\left [ 1 - \varepsilon \frac{\textstyle 1+m}{\textstyle (1+M)^2}
-\frac{\textstyle M-m}{\textstyle (1+M)}t \right ] \\
\tilde{c}^{u}_1(t)&=&\left [\tilde{c}^{in}_0(t/\varepsilon)+\tilde{c}^{out}_0(t)
\right ]+ \varepsilon \left [ \tilde{c}^{in}_1(t/\varepsilon)+
\tilde{c}^{out}_1(t) 
\right ]+\\
&-&\left [\frac{\textstyle 1}{\textstyle 1+M}-
\varepsilon \frac{\textstyle M(1+2m-M)}{\textstyle (1+M)^4}  
-\frac{\textstyle M(M-m)}{\textstyle (1+M)^3}t \right ].
\end{array}
\right.
\label{solun1}
\end{eqnarray}

We plot in [Fig. \ref{fig4}] our results on the substrate and complex 
concentrations for the two considered sets of ICVs, in comparison with the 
numerical solutions of the original problem (\ref{mmineq}) (the same curves 
as in [Fig. \ref{fig1}]). The figures make evident that in the case $a$, 
([Fig. \ref{fig4}A] and [Fig. \ref{fig4}B]) corresponding to a small
$\varepsilon=\varepsilon^a=0.1$, the 1st order PE UAs are 
indistinguishable, within the plot precision (in fact, they are nearly 
indistinguishable within our numerical precision, too), from the correct 
solution on the whole relevant time interval, despite of the particularly 
demanding situation, that is characterized by the presence of 
three inflection points in the curve for the substrate and of a sQSSA ICV for 
the complex that is higher than its correct maximum. 

On the other hand, in case $b$, that corresponds
to the higher $\varepsilon=\varepsilon^b=0.5$, there is a naked-eye 
detectable difference for the substrate ([Fig. \ref{fig4}C]), 
for $t \sim 0.1 \div 2s$. This region actually encompasses the matching 
one, which is expected to extend here from 
$t=\varepsilon^b/\delta^b=0.1 s$ to $t=\sqrt{\varepsilon^b}/\delta^b \simeq 
0.14 s$. Moreover, again in case $b$, when looking 
carefully at [Fig. \ref{fig4}D], one can notice that 
the complex tends to zero still slightly too 
rapidly. In fact, also the maximum reached 
by the complex during its evolution is still slightly over-evaluated.
Both of these last observations, that 
concern the 1st order UA for the complex within
the PE framework, will appear more evident
in the following [Fig. \ref{fig8}C] and [Fig. \ref{fig8}D].

\section{The SPDERG approach to boundary layer problems}
\label{alter}
\noindent
The SPDERG approach and its connections with the renormalization group, 
with attention to the context of boundary layer problems, have been 
extensively reviewed in \cite{Ra}, and
a preliminary attempt to apply it to MM kinetics within the tQSSA 
framework is presented in that work, too. Moreover, a recent review 
of the general method, in a context  that is instead different from the 
boundary layer one, is presented in \cite{Ki}. Here we limit ourselves to 
recall in some detail the original discussion in \cite{ChGoOo1,ChGoOo2}, 
where the approach was generally proposed for multiple-scale problems, in the 
case of the boundary layer ones. Indeed, in those works, the method was already 
shown successful in some of these last cases. 

As MM kinetics \cite{BeBeDAPe,MaMo,LiSe}, these  
problems \cite{BeOr} are generally characterized 
by a boundary layer of a given small 
thickness (that is here $O(\varepsilon)$, but that could also be, for instance, 
$O(\sqrt{\varepsilon})$), in which the solution is rapidly varying.
Correspondingly, in order to predict the system's dynamics, one needs to solve 
singularly perturbed ODEs and one usually resorts to a standard PE method. 

To start to sketch the general SPDERG procedure in these cases, one starts 
from the singular 2nd order ODE for $y(t)$ and makes the transformation 
$t \rightarrow \tau=t/\varepsilon$, thereby studying the ODE to be obeyed by 
the inner solution $Y(\tau)$. The approach basically consists in 
focusing on the large $\tau$ behaviour of this solution,  
by outlining the secular terms, with the aim of proposing first of all 
the correct renormalization of the integration constants, that allows 
to eliminate them. 

At this point, we anticipate that in the end one usually gets 
\cite{ChGoOo1,ChGoOo2} a physically meaningful solution, that also contains 
the leading terms of the outer expansion, and that is therefore to be 
interpreted as the UA within this approach, to be compared with the one 
of the standard PE methods. As we are going to verify in detail in the present 
application, this approximation is expected \cite{ChGoOo1,ChGoOo2}
to work even better than the corresponding UA solution of the PE at the previous
order in the matching region. Moreover, as we are going to discuss, it appears 
possible to make an analogy between the basic mechanism of the
SPDERG approach to boundary layer problems and the one that allows
to impose the correct MCs in the standard methods.  

More precisely \cite{ChGoOo2}, one studies the 
solution $Y(\tau)$ with given boundary conditions at a given time $\tau_0$
and {\em renormalizes} the $Y_{div}(\tau)$ 
part of this solutions that contains the Cauchy data, the other
leading order terms and the terms that do not tend to zero or to
a constant value at large times. This is obtained by renormalizing the 
corresponding {\em bare} integration constants ({\em i.e.}, the Cauchy data of 
the problem, though these integration constants can also more generally 
{\em contain} them in some given form), let us say 
$A_0(\tau_0)$ and $B_0(\tau_0)$, as   
$A_0(\tau_0) \rightarrow Z_1 A(\lambda)$, 
$B_0(\tau_0) \rightarrow Z_2 B(\lambda)$.  
In fact, the renormalization constants $Z_1$ and $Z_2$ depend both on $\tau_0$ 
and $\lambda$, their basic role being to change, in the part to be 
renormalized of the function, the initial dependence of the  Cauchy  
data on $\tau_0$ in a dependence on the arbitrary time $\lambda$. 

In detail, they are assumed to have the expansion (with $a_0=b_0=1$):
\begin{equation}
Z_1=\sum_{n=0}^\infty a_n(\tau_0,\lambda) \varepsilon^n;
\hspace{.2in}
Z_2=\sum_{n=0}^\infty b_n(\tau_0,\lambda) \varepsilon^n.
\label{constant_expansion}
\end{equation}
Since one can always take $(\tau-\tau_0)=(\tau-\lambda)+(\lambda-\tau_0)$, the 
secular terms in $(\lambda -\tau_0)$ are correspondingly absorbable in 
appropriate redefinitions of $A_0$ and $B_0$, that have to be made by correctly 
choosing the coefficients $\{a_n\},\{b_n\}$. As it will become clear in the
application that we are going to present, one can usually safely 
expect to be able to absorb in the coefficients of the renormalization 
constants also other possible secular terms, for instance those corresponding 
to higher powers of $(\tau -\tau_0)$. In particular, we will see in the present
2nd order calculations that, when writing $(\tau-\tau_0)^2=
(\tau-\lambda)^2+(\lambda-\tau_0)^2+2(\tau-\lambda)(\lambda-\tau_0)$,
only the second term is to be absorbed, whereas the last one is cancelled by 
the previously chosen 1st order renormalization coefficients, in the same way 
as in a case that is considered in \cite{ChGoOo2} (see Section B
in that work).

Summarizing \cite{ChGoOo1,ChGoOo2}, the basic hypothesis, that is 
analogous to the scaling one in the renormalization group theory, is that the 
bare quantities need to be renormalized in such a way that the solution turns 
out to be independent of the arbitrary time $\lambda$, as it has 
reasonably to be. Hence, $\lambda$ plays a key role in the 
whole approach. Intuitively, one can therefore think to $\lambda$ as the 
equivalent of the {\em unknown time} at which the matching needs to be 
obeyed, and from this point of view there is an analogy with the imposition of 
the MCs in the standard PE methods. Nevertheless, though this analogy can be 
useful, as we will show, it is not to be taken too rigorously, particularly in 
the present peculiar case in which the 1st order matching requires two term 
conditions, and one expects that the correct matching could
require even more terms at higher orders \cite{LiSe}.

Going back to the recalling of the main details of the approach in the case 
of boundary layer problems \cite{ChGoOo2}, the appropriate choice of the 
coefficients of the renormalization constants, and the appropriate 
redefinitions of $A_0$ and $B_0$, turns out in the replacement of these bare 
quantities with $A(\lambda)$ and $B(\lambda)$, and in the corresponding change 
of variable $\tau_0 \rightarrow \lambda$ in $Y_{div}(\tau)$. Then, on this 
$\lambda$-depending $Y_{div}$, one imposes the {\em scaling} condition 
$dY_{div}/d\lambda=0$, that allows to obtain the 1st order ODEs to be obeyed by 
$A(\lambda)$ and $B(\lambda)$. Finally, the change of variable
$\lambda \rightarrow \tau$ in the renormalized $Y_{div}$, in which the 
integration constants are replaced by the solutions of these equations, and 
the imposition of the original boundary conditions of the problem, allow to 
get the physically meaningful result. 

\section{Results and discussion: i) First order contribution}
\label{rdI}
Here we use, for the first time to our knowledge, the SPDERG approach  
to study MM kinetics beyond the sQSSA, whose known PE UAs we 
already recalled in Sections \ref{review1} and \ref{review2}. 
In order to make the presentation as clear as possible, 
we stress from the beginning that we introduce a procedure that is slightly 
different from the one considered in \cite{ChGoOo1,ChGoOo2}. 
In fact, we make explicit the dependence from the ICVs in the solutions,
thus outlining that, though in principle one can renormalize even general
integration constants that contain them, the bare quantities that are 
basically to be renormalized are just the ICVs of the problem.

We notice first of all that one can rewrite the system (\ref{inner}) for the
inner solutions in the form of a 2nd order ODE 
(see in particular \cite{Sw} for the first study of MM 
kinetics in terms of a 2nd order ODE) to be 
obeyed by the adimensional substrate concentration:
\begin{equation}
\ddot{\tilde{s}}^{in}(\tau)-\frac{\textstyle \left 
[ \dot{\tilde{s}}^{in}(\tau) \right ]^2}
{\textstyle {\tilde{s}}^{in}(\tau)+m}+ 
\left[ \tilde{s}^{in}(\tau)+M \right ]\dot{\tilde{s}}^{in}(\tau)
+\varepsilon \left  [\frac{\textstyle m}
{\textstyle \tilde{s}^{in}(\tau)+m}\dot{\tilde{s}}^{in}(\tau) + 
( M-m ) \tilde{s}^{in}(\tau)\right ] =0, 
\end{equation}
with ICVs (Cauchy data) 
${\tilde{s}}^{in}(0)=1$ and
$\dot{\tilde{s}}^{in}(0)=-\varepsilon$. 
Analogously, one can write the same system (\ref{inner}) in the form
of a 2nd order ODE to be obeyed by the adimensional complex concentration:
\begin{equation}
\left [ 1- {\tilde{c}}^{in}(\tau) \right ] \ddot{\tilde{c}}^{in}(\tau)
+\left[ \dot{\tilde{c}}^{in}(\tau) \right ]^2 +M {\tilde{c}}^{in}(\tau)
\dot{\tilde{c}}^{in}(\tau)+
\varepsilon \left  [ 1- {\tilde{c}}^{in}(\tau) \right ]^2 \left  [ 
\dot{\tilde{c}}^{in}(\tau)+(M-m) {\tilde{c}}^{in}(\tau) \right ]=0,
\end{equation}
with ICVs (Cauchy data) ${\tilde{c}}^{in}(0)=0$ and $\dot{\tilde{c}}^{in}(0)=1$.

Actually, the physically meaningful solutions of these 2nd order ODEs are
identical to the solutions of the original system (\ref{inner}). In fact,
at the 1st order in $\varepsilon$, the equations are satisfied by 
${\tilde{s}}^{in}(\tau)={\tilde{s}}^{in}_0(\tau)+\varepsilon 
{\tilde{s}}^{in}_1(\tau)$ and ${\tilde{c}}^{in}(\tau)={\tilde{c}}^{in}_0(\tau)+
\varepsilon {\tilde{c}}^{in}_1(\tau)$, with ${\tilde{s}}^{in}_0(\tau)$,
${\tilde{c}}^{in}_0(\tau)$ given by (\ref{solinner0}), and 
${\tilde{s}}^{in}_1(\tau)$, ${\tilde{c}}^{in}_1(\tau)$ given by 
(\ref{solinner1}), respectively. 

Therefore, instead of attempting to apply the approach to 
these 2nd order ODEs, we study the SPDERG ({\em i.e., renormalization group}, 
that we label $rg$) adimensional substrate and complex concentrations, 
${\tilde{s}}^{rg}(\tau)$ and ${\tilde{c}}^{rg}(\tau)$, that are solutions of 
the system (\ref{inner}). The difference, with respect to the
${\tilde{s}}^{in}(\tau)$ and  ${\tilde{c}}^{in}(\tau)$ that we previously 
considered within the PE method, is that here the ICVs are given at a time 
$\tau_0$, to be considered in principle different from zero, as 
${\tilde{s}}^{rg}(\tau_0)={\tilde{s}}^*$ and 
${\tilde{c}}^{rg}(\tau_0)={\tilde{c}}^*$, respectively. Indeed, these 
${\tilde{s}}^*$ and ${\tilde{c}}^*$ values are the bare 
quantities to be renormalized, whereas the original ICVs of the problem, 
{\em i.e.}, ${\tilde{s}}(0)=1$ and ${\tilde{c}}(0)=0$, will
be taken into account after the renormalization procedure.

For the sake of clarity, at the cost of being somehow repetitive, 
at the 1st order in $\varepsilon$, we search once again solutions in the form:
\begin{eqnarray}
\left\{
\begin{array}{lcl}
\tilde{s}^{rg}(\tau)=\tilde{s}^{rg}_0(\tau)+\varepsilon\tilde{s}^{rg}_1(\tau)\\
\tilde{c}^{rg}(\tau)=\tilde{c}^{rg}_0(\tau)+\varepsilon 
\tilde{c}^{rg}_1(\tau).
\end{array}
\right.
\label{formsolrg}
\end{eqnarray}
In these formulas, ${\tilde{s}}^{rg}_0(\tau)$ and ${\tilde{c}}^{rg}_0(\tau)$ 
solve as usual the 0th order system (\ref{inner0}), but the ICVs 
are given at $\tau=\tau_0$, and their values are ${\tilde{s}}^{rg}_0(\tau_0)=
{\tilde{s}}^*_0$ and ${\tilde{c}}^{rg}_0(\tau_0)={\tilde{c}}^{*}_0$. 
These solutions are: 
\begin{eqnarray}
\left\{
\begin{array}{lcl}
\tilde{s}^{rg}_0(\tau)&=& \tilde{s}^*_0\\
\tilde{c}^{rg}_0(\tau)&=& \tilde{c}^*_0 
e^{\textstyle -(\tilde{s}^*_0+M)(\tau-\tau_0)}+
\frac{\textstyle \tilde{s}^*_0}{\textstyle \tilde{s}^*_0+M} \left [ 1-
e^{\textstyle -(\tilde{s}^*_0+M)(\tau-\tau_0)} \right ].
\end{array}
\right.
\label{solrg0}
\end{eqnarray}
On the other hand, ${\tilde{s}}^{rg}_1(\tau)$ and ${\tilde{c}}^{rg}_1(\tau)$ 
need to solve the 1st order system (that is slightly different from 
Eqs.~(\ref{inner1}) because of the presence of $\tilde{s}^*_0 \neq 1$):
\begin{eqnarray}
\left\{
\begin{array}{lcl}
\dot{\tilde{s}}^{rg}_1(\tau)&=& (\tilde{s}^*_0+m)\tilde{c}^{rg}_0(\tau) 
-\tilde{s}^*_0\\
\dot{\tilde{c}}^{rg}_1(\tau)&=&- (\tilde{s}^*_0+M)\tilde{c}^{rg}_1(\tau)-
\left [ \tilde{c}^{rg}_0(\tau)-1 \right ] \tilde{s}^{rg}_1(\tau),
\end{array}
\right.
\label{rg1}
\end{eqnarray}
with ICVs, at $\tau=\tau_0$, ${\tilde{s}}^{rg}_1(\tau_0)=
{\tilde{s}}^*_1$ and ${\tilde{c}}^{rg}_1(\tau_0)={\tilde{c}}^{*}_1$.
We obtain, in the case of the substrate:
\begin{eqnarray}
\tilde{s}^{rg}_1(\tau)&=& \tilde{s}^*_1-(M-m)
\frac{\textstyle \tilde{s}^*_0}{\textstyle  \tilde{s}^*_0+M }
(\tau-\tau_0)+ \nonumber \\
&-&\frac{\textstyle \tilde{s}^*_0+m}
{\textstyle  ( \tilde{s}^*_0+M )^2} 
\left [ \tilde{s}^*_0- \tilde{c}^*_0 ( \tilde{s}^*_0+M ) \right ]
\left [1 -e^{\textstyle -(\tilde{s}^*_0+M)(\tau-\tau_0)}  \right ],
\label{solsrg1}
\end{eqnarray}
whereas the solution for the complex, $\tilde{c}^{rg}_1(\tau)$, is given in 
Appendix A.

Let us start by studying $\tilde{s}^{rg}(\tau)$. Following \cite{ChGoOo2}, we
look at $\tilde{s}^{rg}_{div}(\tau)$, that contains just the ICVs ({\em i.e.}, 
the terms that give 
$\tilde{s}^{rg}(\tau_0)=\tilde{s}^*_0+ \varepsilon \tilde{s}^*_1$), 
the other possible leading terms at this order (absent in the present case),
and the secular terms (here the one proportional to $(\tau-\tau_0)$ in  
$\tilde{s}^{rg}_1(\tau)$). Therefore, we write:
\begin{equation}
\tilde{s}^{rg}(\tau) = \tilde{s}^{rg}_{div}(\tau)+ 
\varepsilon {\cal R}^s_1(\tau)+O(\varepsilon^2),
\label{srgin1}
\end{equation}
by grouping in ${\cal R}^s_1$ all the sub-leading terms that tend to zero or
to a constant value in the large time limit and that are not to be renormalized.
One has:
\begin{equation}
\tilde{s}^{rg}_{div}(\tau)=\tilde{s}^*_0+ \varepsilon \tilde{s}^*_1
-\varepsilon (M-m) \frac{\textstyle \tilde{s}^*_0}{\textstyle  \tilde{s}^*_0+M }
(\tau-\tau_0).
\end{equation}

As previously recalled, the SPDERG approach \cite{ChGoOo1,ChGoOo2}, at this 
point, is based on writing $(\tau-\tau_0)=(\tau-\lambda)+(\lambda-\tau_0)$, by 
correspondingly assuming, in the present case, that the secular term 
proportional to $(\lambda-\tau_0)$ can be absorbed in an appropriate 
redefinition of the bare constants. In detail, by labelling $\tilde{s}^{rg *}_0$ 
and $\tilde{s}^{rg *}_1$ the contributions to the renormalized ICV
at the 0th and at the 1st order, respectively, we put 
$\tilde{s}^{*}_0=Z_{s_0} \tilde{s}^{rg *}_0(\lambda)$ and 
$\tilde{s}^{*}_1=Z_{s_1} \tilde{s}^{rg *}_1(\lambda)$. The renormalization 
constants, $Z_{0}^{s}$ and $Z_{1}^{s}$, are assumed to have the same 
$\varepsilon$-expansions as the ones given in (\ref{constant_expansion}). 
This expansion, at the 1st order in $\varepsilon$, that we considered here, 
implies $\tilde{s}^{*}_0=
[1+\varepsilon z_{s_0,1}(\tau_0,\lambda)] \tilde{s}^{rg *}_0(\lambda)$
and $\tilde{s}^{*}_1= \tilde{s}^{rg *}_1(\lambda)$. 
Thus, the present secular term can be absorbed by choosing:
\begin{equation}
z_{s_0,1}(\tau_0,\lambda)=\frac{(M-m)}{\tilde{s}^{rg *}_0+M}(\lambda-\tau_0).
\label{zs_s0_1}
\end{equation}

Therefore, we end up to study:
\begin{equation}
\tilde{s}^{rg}_{div}(\tau,\lambda)=\tilde{s}^{rg *}_0(\lambda)+ 
\varepsilon \tilde{s}^{rg *}_1(\lambda)
-\varepsilon (M-m) \frac{\textstyle \tilde{s}^{rg *}_0(\lambda)}
{\textstyle  \tilde{s}^{rg *}_0(\lambda)+M }(\tau-\lambda).
\end{equation}
Hence, by imposing the scaling condition \cite{ChGoOo1,ChGoOo2}, 
$d\tilde{s}^{rg}_{div}(\tau,\lambda)/d\lambda=0$, we get the 
1st order ODEs to be obeyed by the renormalized quantities. 

In detail, at the 1st order in $\varepsilon$, we obtain:
\begin{equation}
\frac{\textstyle d{\tilde{s}}^{rg *}_{0}(\lambda)}{\textstyle d\lambda}
+\varepsilon \frac{\textstyle d{\tilde{s}}^{rg *}_{1}(\lambda)}
{\textstyle d\lambda}-\varepsilon (M-m)\frac{\textstyle \tilde{s}^{rg *}_0
(\lambda)}{\textstyle  \tilde{s}^{rg *}_0(\lambda)+M }=0,
\end{equation}
and, correspondingly:
\begin{eqnarray}
\left\{
\begin{array}{lcl}
\frac{\textstyle d\tilde{s}^{rg *}_{0}(\lambda)}{\textstyle d\lambda}&=&
-\varepsilon (M-m)\frac{\textstyle \tilde{s}^{rg *}_0(\lambda)}
{\textstyle  \tilde{s}^{rg *}_0(\lambda)+M }\\
\frac{\textstyle d\tilde{s}^{rg *}_{1}(\lambda)}{\textstyle d\lambda}&=&0.\\
\end{array}
\right.
\label{scalingcondition1}
\end{eqnarray}

Consistently, when making the further transformation 
\cite{ChGoOo2} $\lambda \rightarrow \tau= t/\varepsilon$, the first of these 
equations is just the ODE to be obeyed by the 0th order outer 
adimensional substrate concentration, that we encountered in Eqs.~(\ref{QSSA}). 
Its solution is therefore $\tilde{s}^{rg *}_{0}(t)=\tilde{s}^{out}_{0}(t)$, 
that is given in (\ref{sout0}) by means of the Lambert function, 
and that also satisfies the ICV  $\tilde{s}^{rg *}_0(0)=1$. 

Clearly, this result already suggests that, also in the present particular 
demanding case of MM kinetics, the SPDERG approach proposed in 
\cite{ChGoOo1,ChGoOo2} turns out to be able to reproduce the leading order 
terms of the PE UAs. 

On the other hand, the result $d\tilde{s}^{rg *}_{1}(\lambda)/ d\lambda=0$
is here to be interpreted as the verification that the $\tilde{s}^{rg *}_{1}$
term can be neglected at this order. In fact, we find 
$\tilde{s}^{rg *}_{1}=const=0$ by imposing the ICV (more correctly the ICV, 
that is fixed here to the value $\tilde{s}(0)=1$, should be imposed on the 
solution at the end, but this is not influential in the present case, since one 
generally has ${\cal R}(0)=0$ for the appropriately calculated contribution of 
the parts not to be renormalized of the inner solutions). 

Finally, we get the renormalized result:
\begin{equation}
\tilde{s}^{rg,u}_1(t) = \tilde{s}^{out}_{0}(t)+  
\varepsilon {\cal R}^s_1(t/\varepsilon)+O(\varepsilon).
\end{equation}
Notice that, when we were  considering $\tilde{s}^{rg}(\tau)$ as the
inner solution, in (\ref{srgin1}), before applying the 
SPDERG approach, we knew that it was correct up to order $O(\varepsilon^2)$.
Here, instead, the renormalized $\tilde{s}^{rg,u}_1(t)$ 
(that for this reason we label {\em rg,u}, by moreover making explicit
that it is a 1st order approximation) is to be interpreted as the SPDERG UA to 
the solution of the original problem, clearly bearing in mind that, 
from this point of view, it is correct only up to order $O(\varepsilon)$. 
Actually, it is only expected to contain the leading order terms
of the outer solution, as one can check that is indeed the case, 
by comparing it with (\ref{solun1}).

Notice moreover that the terms in ${\cal R}^s_1$, that did not need to be 
renormalized, are to be evaluated in $\tau_0=0$, by using the correct ICVs 
for the bare constants ({\em i.e.}, $\tilde{s}^*_0=1, \: \tilde{s}^*_1=0, \: 
\tilde{c}^*_0=0$). Thus, the SPDERG 1st order UA to the solution for the 
adimensional substrate concentration does instead contain the 1st order terms 
of the PE UA to the correct solution that comes from 
$\tilde{s}^{in}_{1}(t/\varepsilon)$. On the other hand, correspondingly, the 
usual asymptotically vanishing solution behaviour that one finds when applying 
the recalled PE method to MM kinetics ({\em i.e.}, 
$\lim_{t \rightarrow \infty}\tilde{s}(t)=0$), is not verified here (one has 
$\lim_{t \rightarrow \infty}\tilde{s}^{rg, u}_1(t)=O(\varepsilon)$). We will
better discuss this failure of the present application of the SPDERG approach 
in the following, by proposing a way to bypass it, too.

Let us now study the complex. First of all, despite 
of the presence of a large number of terms in (\ref{solcrg1}) (the formula 
given in Appendix A), that describes the behaviour of the 1st order inner 
adimensional complex concentration within the SPDERG approach, 
$\tilde{c}^{rg}_1(\tau)$, the part of the function to be 
renormalized up to the 1st order in $\varepsilon$ is 
as manageable as in the case of the substrate. 

In detail, we write the same kind of expression as in (\ref{srgin1}):
\begin{equation}
\tilde{c}^{rg}(\tau) = \tilde{c}^{rg}_{div}(\tau)+ \varepsilon {\cal R}^c_1(\tau)+
O(\varepsilon^2),
\label{crgin1}
\end{equation}
by collecting in ${\cal R}^c_1$ all the sub-leading terms that  
tend to zero or to a constant value in the large time limit. 
Correspondingly, we find:
\begin{equation}
\tilde{c}^{rg}_{div}(\tau)=(\tilde{c}^*_0+\varepsilon \tilde{c}^*_1) 
e^{\textstyle -(\tilde{s}^*_0+M)(\tau-\tau_0)}+
\frac{\textstyle \tilde{s}^*_0}{\textstyle \tilde{s}^*_0+M} 
-\varepsilon \frac{\textstyle M(M-m)}
{\textstyle \left (\tilde{s}^*_0 +M \right )^3}  
\tilde{s}^*_0 (\tau-\tau_0).
\end{equation}

In fact, here we considered explicitly the terms that give the original 
ICV, ({\em i.e.}, 
$\tilde{c}^{rg}_{div}(0)=\tilde{c}^*_0 +\varepsilon \tilde{c}^*_1$). 
Nevertheless, both the 0th order contribution and the 1st order one to it are  
exponentially suppressed in the large time limit. Therefore, one can 
hypothesize (see an analogous case in \cite{ChGoOo2}, Section C) that also 
$\tilde{c}^{rg *}_0(\lambda)$, when renormalizing, should obey an ODE of the 
kind $d\tilde{c}^{rg *}_{0}(\lambda)/ d\lambda=0$. Hence, one expects to find
$\tilde{c}^{rg *}_{0}(\lambda)=const=0$ when imposing the ICV (fixed here
to $\tilde{c}(0)=0$) at the end of the renormalization procedure.  
Moreover, one also expects $\tilde{c}^{rg *}_{1}(\lambda)=const=0$, 
a result that is even more predictable at this order, since we 
already verified that $\tilde{s}^{rg *}_{1}$ gives no contribution in the case 
of the substrate. Correspondingly, we assume that 
these terms can be neglected in the present application
of the SPDERG approach to MM kinetics.

Therefore, the function to be studied is further reduced to:
\begin{equation}
\tilde{c}^{rg}_{div}(\tau) \left
\lvert_{\!\!\shortmid_{\!\shortmid_{\tilde{c}^*_1=0}^{\tilde{c}^{*}_0=0}}} \right. =
\frac{\textstyle \tilde{s}^*_0}
{\textstyle \tilde{s}^*_0+M}-\varepsilon (M-m) \frac{\textstyle M \tilde{s}^*_0}
{\textstyle \left (\tilde{s}^*_0 +M \right )^3}(\tau-\tau_0).
\end{equation}
As already recalled, in the context of the standard method, the MC that is 
verified by the adimensional substrate concentration does consistently 
also satisfy the matching of the inner and outer adimensional complex 
concentrations. For this reason, on the basis of the previously
sketched analogy, we find not particularly surprising that, when making the
transformations $\left \{ \tau_0 \rightarrow \lambda, \tilde{s}^*_0
\rightarrow \tilde{s}^{rg *}_0(\lambda) \right \}$, 
the scaling condition $d\tilde{c}^{rg}_{div}(\tau,\lambda)/d\lambda=0$
gives again the same ODE to be obeyed by $\tilde{s}^{rg *}_0(\lambda)$ as the 
one that we just found in the study of $\tilde{s}^{rg}_{div}(\tau,\lambda)$ 
({\em i.e.}, Eqs.~(\ref{scalingcondition1})). Hence (with the transformation 
$\lambda \rightarrow \tau=t/\varepsilon$), we get again  
$\tilde{s}^{rg *}_0(t)=\tilde{s}^{out}_{0}(t)$. One can check in particular that, 
since $d [\tilde{s}^*_0/(\tilde{s}^*_0+M)]/d\tilde{s}^*_0=M/(\tilde{s}^*_0+M)^2$,
when renormalizing the bare constant $\tilde{s}^*_0$ by using the value
of the coefficient $z_{s_0,1}$ that is already fixed by (\ref{zs_s0_1}), the 
contribution of the first term to the order $\varepsilon$ is exactly
equal to 
$\varepsilon(\lambda-\tau_0) M(M-m)\tilde{s}^{rg *}_0/(\tilde{s}^{rg *}_0+M)^2$, 
and therefore it suitably absorbs the secular term that is present in this case.

By recalling that 
$\tilde{c}^{out}_{0}(t)=\tilde{s}^{out}_{0}(t)/(\tilde{s}^{out}_{0}(t)+M)$ from 
Eqs.~(\ref{QSSA}), the obtained SPDERG 1st order UA to the correct solution is:
\begin{equation}
\tilde{c}^{rg,u}_1(t) = \tilde{c}^{out}_{0}(t)+ \varepsilon
{\cal R}^c_1(t/\varepsilon)+O(\varepsilon),
\end{equation}
with, as for the substrate, the terms in ${\cal R}^c_1$ to be evaluated in 
$\tau_0=0$, by using the correct ICVs for the bare constants 
({\em i.e.}, $\tilde{s}^*_0=1, \: \tilde{s}^*_1= \tilde{c}^*_0= 
\tilde{c}^*_1=0$).

\begin{figure}[t]
\begin{center}
\leavevmode
\epsfig{figure=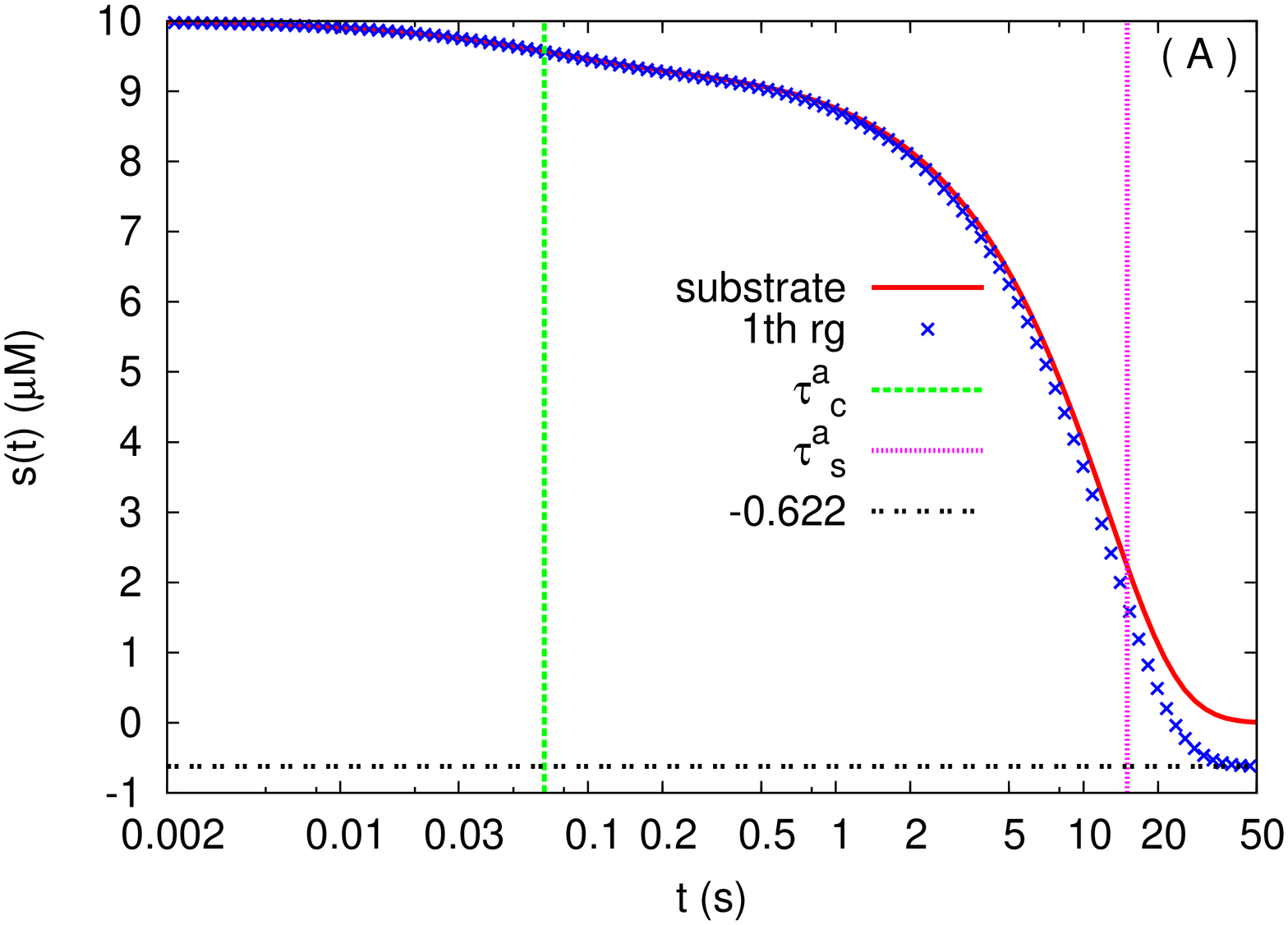,angle=270,width=8cm}
\epsfig{figure=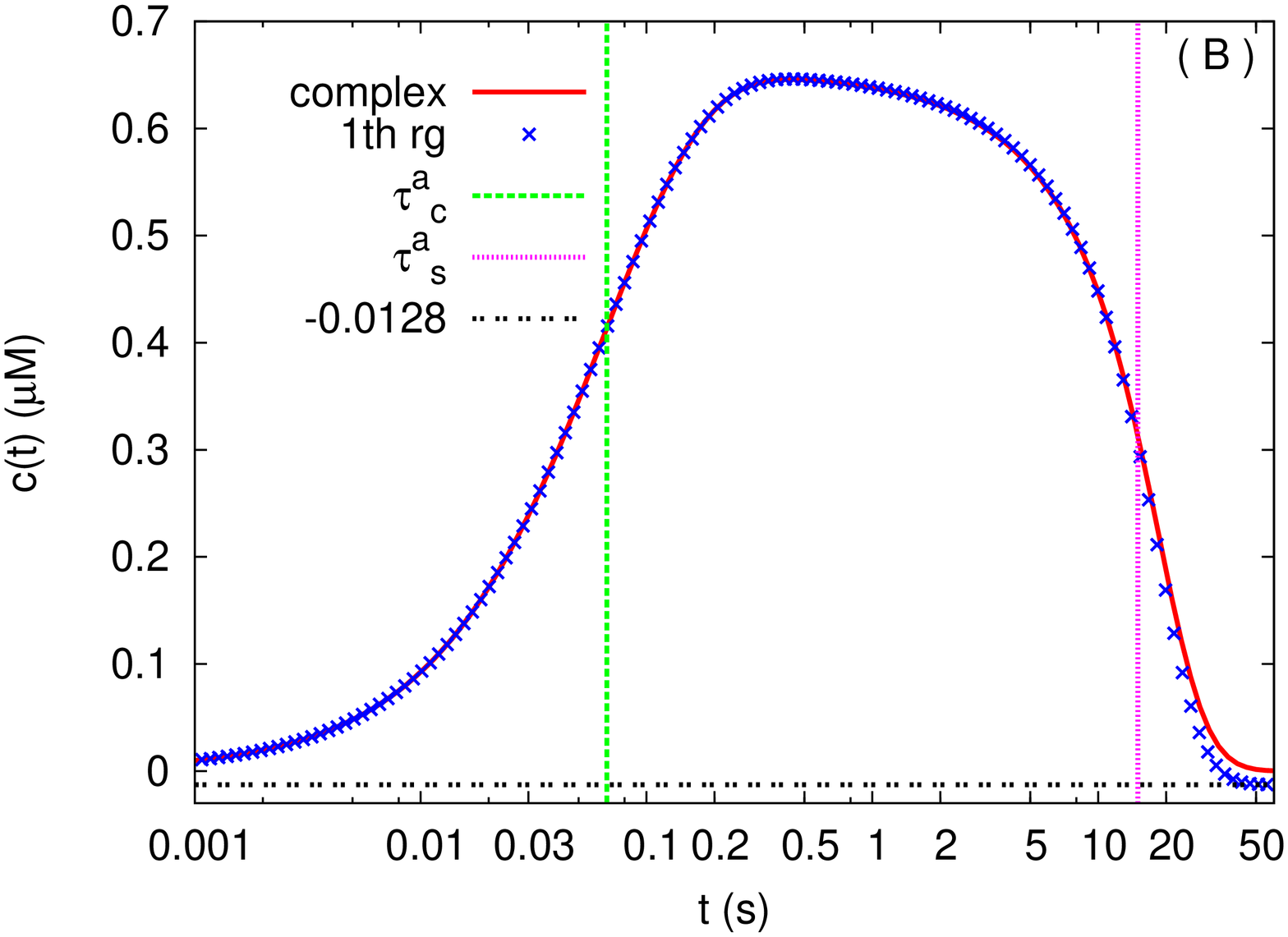,angle=270,width=8cm}
\epsfig{figure=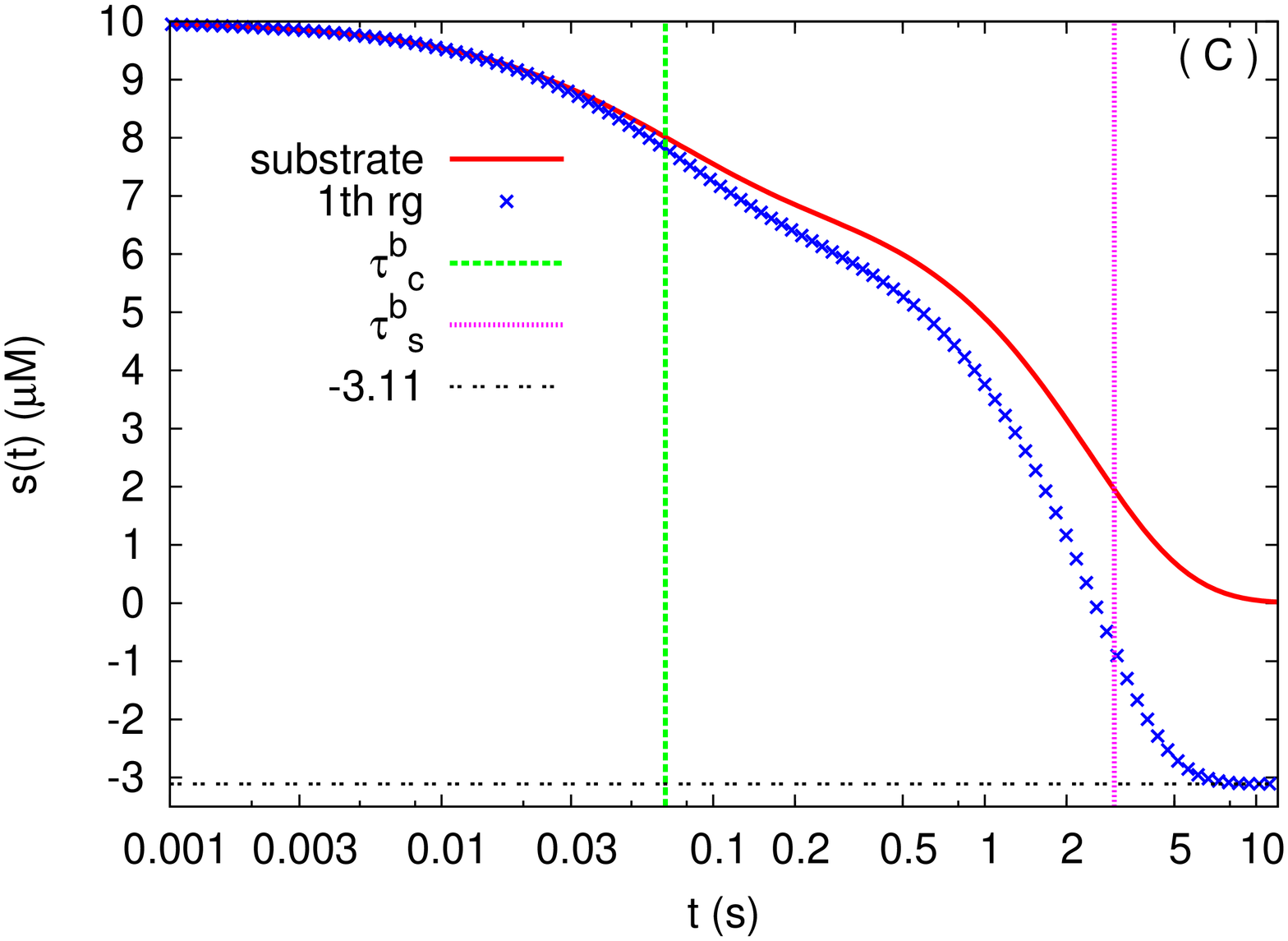,angle=270,width=8cm}
\epsfig{figure=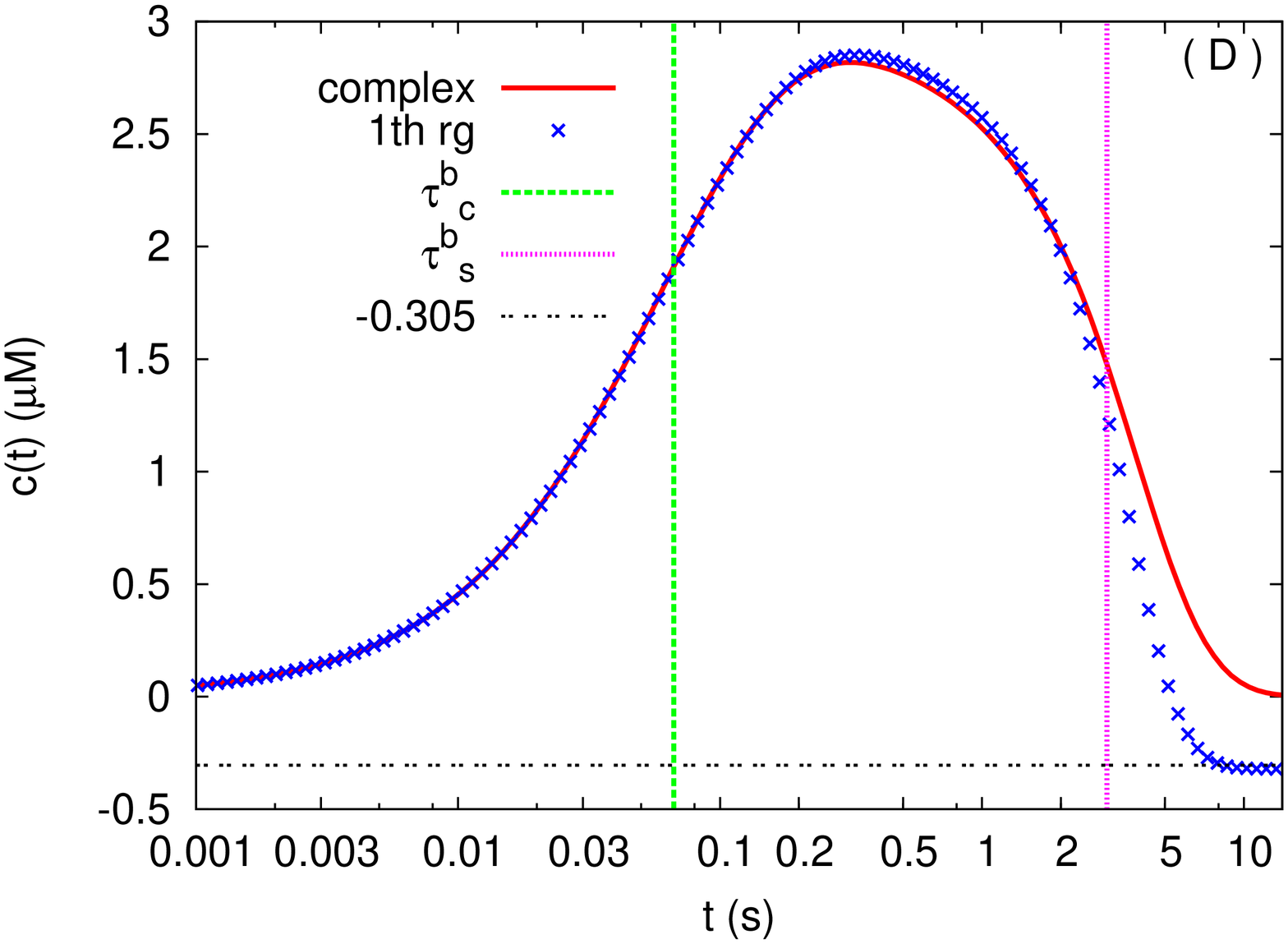,angle=270,width=8cm}
\caption{{\footnotesize In $A)$ and $C)$ we present the behaviour of the 
concentrations of the substrate $s(t)$, whereas in $B)$ and $D)$ we present 
the ones of the complex $c(t)$ for the $a$ and $b$ sets of ICVs given in 
(\ref{initialvalues}), respectively. Hence, in $A)$ and $B)$ we are in the case 
with $\varepsilon=\varepsilon^a=0.1$, whereas in $C)$ and $D)$ we are in the 
one with  $\varepsilon=\varepsilon^b=0.5$. We plot both the numerical 
solutions of Eqs.~(\ref{mmineq}) already shown in the previous figures, and 
the analytical solutions computed from the SPDERG 1st order UAs
(with a standard numerical approximation for the Lambert function), as given 
in (\ref{solrgu1}). We plot moreover the (physically meaningless)  
limits for $t\rightarrow \infty$ of the analytical 
solutions: $s_0^a \tilde{s}^{rg,u}_{1,\infty} \simeq -0.622$,  
$e_0^a \tilde{c}^{rg,u}_{1,\infty} \simeq -0.0122$, and 
$s_0^b \tilde{s}^{rg,u}_{1,\infty} \simeq -3.11$, 
$e_0^b \tilde{c}^{rg,u}_{1,\infty} \simeq -0.305$, respectively.
We  finally plot our corresponding rough evaluations of the two different 
time scales involved, too, with $\tau_s$ describing the substrate decay time 
and $\tau_c$ the complex saturation time. Notice that the time is in 
logarithmic scale.}}
\label{fig5}
\end{center}
\end{figure}

Writing explicitly the contribution of the different terms, we
obtain the following SPDERG 1st order UAs to the correct solutions
for the time behaviour of the adimensional substrate 
concentration and of the adimensional complex concentration in 
MM kinetics beyond the sQSSA, respectively 
(with $\omega$ the Lambert function):
\begin{eqnarray}
\left\{
\begin{array}{lcl}
\tilde{s}^{rg,u}_1(t)&=& M \omega(e^{\textstyle -(M-m)t/M+1/M}/M)+\\
&-&\varepsilon \frac{\textstyle 1+m}{\textstyle  (1+M )^2} 
\left [1 -e^{\textstyle -(1+M)t/\varepsilon}  \right ]+O(\varepsilon)\\
\tilde{c}^{rg,u}_1(t)&=& \frac{\textstyle  \omega(e^{\textstyle -(M-m)t/M+1/M}/M)}
{\textstyle   \omega(e^{\textstyle -(M-m)t/M+1/M}/M)+1}-
\frac{\textstyle 1}{\textstyle 1+M} e^{\textstyle -(1+M)t/\varepsilon}+\\
&-& \varepsilon \frac{\textstyle M(1+2m-M)}{\textstyle (1+M)^4}   
 \left [ 1-e^{\textstyle -(1+M)t/\varepsilon} \right ]+\\
&-& \varepsilon \left [ \frac{\textstyle (1-M)(1+m)}
{\textstyle (1+M)^3} (t/\varepsilon) +\frac{\textstyle M-m}{\textstyle (1+M)^2} 
\frac{\textstyle (t/\varepsilon)^2}{\textstyle 2} \right ]
e^{\textstyle -(1+M)t/\varepsilon}+ \\
&+&\varepsilon \frac{\textstyle (1+m)}
{\textstyle (1+M)^4}e^{\textstyle -(1+M)t/\varepsilon} \left [
1-e^{\textstyle -(1+M)t/\varepsilon} \right ]+O(\varepsilon). 
\end{array}
\right.
\label{solrgu1}
\end{eqnarray}

We plot in [Fig. \ref{fig5}] our corresponding results
for the two considered sets of ICVs,  
as usual in comparison with the numerical solutions of the original 
problem (\ref{mmineq}) (the same curves as in [Fig. \ref{fig1}]). 

The figure shows that, as expected, these results 
are more refined than the PE 0th order UA ([Fig. \ref{fig2}A], 
[Fig. \ref{fig3}A] and [Fig. \ref{fig2}C], [Fig. \ref{fig3}B] for the results 
for the $a$ and $b$ sets of ICVs, respectively), but not as much refined as 
the PE 1st order ones ([Fig. \ref{fig4}]). Moreover, the figure also makes
evident that, differently from the PE results, here the approximation fails in 
particular in the large time region. Actually, as expected, both the substrate 
and the complex approach (physically meaningless) $O(\varepsilon)$ values 
for $t \rightarrow \infty$. In fact, from (\ref{solrgu1}), one has 
$\lim_{t \rightarrow \infty} \tilde{s}^{rg,u}_1(t)=\tilde{s}^{rg,u}_{1,\infty}=
-\varepsilon (1+m)/(1+M)^2$ and 
$\lim_{t \rightarrow \infty} \tilde{c}^{rg,u}_1(t)=\tilde{c}^{rg,u}_{1,\infty}=
-\varepsilon M(1+2m-M)/(1+M)^4$ (the corresponding asymptotic constants for the 
two considered sets of ICVs are also plotted in the figures). Both this 
unphysical outcome and the lack of the 1st order contribution to the outer 
solution are particularly evident in the result for the substrate in the case 
of the $b$ set of ICVs, corresponding to the larger 
$\varepsilon=\varepsilon^b=0.5$ ([Fig. \ref{fig5}C]).

On the other hand, noticeably, the figure definitely
shows that the 1st order SPDERG approach works better than the 0th
order PE method in capturing the qualitative features of
the correct solutions. In detail, one can observe the presence of the
three inflection points in the curves for the substrate both in 
[Fig. \ref{fig5}A] and in [Fig. \ref{fig5}C]. From the quantitative
point of view, also the correct complex maximum values turn out to be 
better predicted than in the corresponding [Fig. \ref{fig3}],
further confirming that the approach is particularly successful
in the proximity of the matching region. 

Therefore, on the basis of these initial results, that appear
as a whole to support the SPDERG usefulness, both for 
further testing its correctness and for possibly obtaining
better approximations to the correct solutions 
than the ones given by the PE 1st order UAs, 
we also consider the 2nd order in $\varepsilon$.

\section{Results and discussion: ii) Second order contribution}
\label{rdII}
In order to simplify the calculations, here we only look at the case in 
which the complex ICVs, in $\tau=\tau_0$, are 
fixed at $\tilde{c}^*_0=\tilde{c}^*_1=\tilde{c}^*_2=0$ from the 
beginning, $\forall \tau_0$. This could appear unrealistic, but 
it is not expected to influence the result, since we already checked that 
there is no need for renormalizing these bare constants at the 1st order, and
the same thing should reasonably apply to the 2nd order, too.
 
From a different point of view, this choice is not expected to influence
the results since  at the end one is interested in taking 
as initial time $\tau_0=0$, and  at the 2nd order, as at the 1st one, the 
renormalized part of $\tilde{c}^{rg}$ should be obtainable by simply using the 
renormalized ICV for the substrate (after appropriately 
redefining the corresponding bare quantities for removing the 
secular terms). We will anyway verify in the following that indeed
one gets the same ODE to be obeyed by $\tilde{s}_1^{rg *}$ both from
the study of the substrate and from the study of the complex, as it can be 
once again better understood within the analogy with the matching in the 
PE, where there is freedom for fixing only one of the two conditions,
whereas the other turns out to be consistently also satisfied.

We take moreover $\tilde{s}^*_2=0$ from the beginning, since we 
expect that it can be anyway neglected within this approach at the 2nd 
order, in the same way as $\tilde{s}^*_1$ turned out to be negligible 
at the 1st order (notice that we should instead allow for $\tilde{s}^*_2\neq 0$ 
if we were to use these solutions to calculate the 3rd order contribution, too).
 
Clearly, one looks for 2nd order solutions in the form:
\begin{eqnarray}
\left\{
\begin{array}{lcl}
\tilde{s}^{rg}(\tau)=\tilde{s}^{rg}_0(\tau)+
\varepsilon\tilde{s}^{rg}_1(\tau)+\varepsilon^2\tilde{s}^{rg}_2(\tau)\\
\tilde{c}^{rg}(\tau)=\tilde{c}^{rg}_0(\tau)+\varepsilon 
\tilde{c}^{rg}_1(\tau)+\varepsilon^2\tilde{c}^{rg}_2(\tau).
\end{array}
\right.
\label{formsolrg2}
\end{eqnarray}
Here, $\tilde{s}^{rg}_0(\tau)$ and $\tilde{c}^{rg}_0(\tau)$ are 
still given by (\ref{solrg0}), with $\tilde{c}^*_0=0$ in the case of the 
complex. Moreover, $\tilde{s}^{rg}_1(\tau)$ and $\tilde{c}^{rg}_1(\tau)$ are 
still given by (\ref{solsrg1}) and (\ref{solcrg1}), respectively, with  
$\tilde{c}^*_1=0$ in the case of the complex, whereas the condition 
$\tilde{c}^*_0=0$ does also apply to the solution for the substrate.

Correspondingly, we end up to study the system:
\begin{eqnarray}
\left\{
\begin{array}{lcl}
\dot{\tilde{s}}^{rg}_2(\tau)&=& (\tilde{s}^*_0+m) 
\tilde{c}^{rg}_1(\tau) + \left [\tilde{c}^{rg}_0(\tau)-1 \right ]
\tilde{s}^{rg}_1(\tau)\\
\dot{\tilde{c}}^{rg}_2(\tau)&=&- (\tilde{s}^*_0+M)\tilde{c}^{rg}_2(\tau)-
\left [ \tilde{c}^{rg}_0(\tau)-1 \right ] \tilde{s}^{rg}_2(\tau)
-\tilde{s}^{rg}_1(\tau)\tilde{c}^{rg}_1(\tau),\\
\end{array}
\right.
\label{rg2}
\end{eqnarray}
whose solutions, $\tilde{s}^{rg}_2(\tau)$ and $\tilde{c}^{rg}_2(\tau)$, 
with ICVs, at $\tau=\tau_0$, $\tilde{s}^{rg}_2(\tau_0)=0$ and 
$\tilde{c}^{rg}_2(\tau_0)=0$, are given in Appendix B.

The obtained formulas are anyway cumbersome, as one could expect: in fact, 
they contain both (implicitly) the 1st order contributions to the outer 
solutions and (explicitly) the 2nd order contributions to the inner ones. 
As anticipated, these last 2nd order contributions are calculated in the 
present work for the first time to our knowledge.

By carrying on the study as in the previous section, we start from
the part of the substrate solution to be renormalized. 
Interestingly, beyond the 2nd order contribution that originates from the 
first three terms in the solution for $\tilde{s}^{rg}_2(\tau)$ given by 
(\ref{solsrg2}) (that are the secular ones, whose coefficients are reported in 
(\ref{coeffs2rgdiv})), the correct function needs also to contain the constant 
term that appears in $s^{rg}_1(\tau)$ (\ref{solsrg1}), since this is now a 
leading order term. In fact, it depends on one of the bare constants 
($\tilde{s}^*_0$), and it does not tend to zero for 
$(\tau-\tau_0) \rightarrow \infty$. 

Hence, one finds:
\begin{eqnarray}
\tilde{s}^{rg}_{div}(\tau)&=&\tilde{s}^{*}_0+\varepsilon \tilde{s}^{*}_1
-\varepsilon\frac{\tilde{s}^{*}_0+m}{(\tilde{s}^{*}_0+M)^2}\tilde{s}^{*}_0
-\varepsilon \frac{\textstyle  (M-m)\tilde{s}^{*}_0}
{\textstyle  {\tilde{s}^{*}_0}+M }(\tau-\tau_0) + \nonumber \\ 
&+& \varepsilon^2 \frac{\textstyle 2 M (M-m) 
{\tilde{s}^{*}_0}}{\textstyle \left ({\tilde{s}^{*}_0}+M \right)^4}
\left ({\tilde{s}^{*}_0}+m\right)(\tau-\tau_0)
- \varepsilon^2 \frac{\textstyle  M (M-m) {\tilde{s}^{*}_1}}
{\textstyle \left ({\tilde{s}^{*}_0}+M\right)^2}(\tau-\tau_0)+ 
\nonumber \\
&+& \varepsilon^2 \frac{\textstyle M (M-m)^2 {\tilde{s}^{*}_0}}
{\textstyle 2 \left ({\tilde{s}^{*}_0}+M\right)^3}(\tau-\tau_0)^2.
\label{sdivrgII}
\end{eqnarray}

We then replace the bare constants with the renormalized ones, {\em i.e.},
$\tilde{s}^{*}_0=Z_{s_0}^{s} (\tau_0,\lambda) \tilde{s}^{rg *}_0(\lambda)$ and 
$\tilde{s}^{*}_1=Z_{s_1}^{s} (\tau_0,\lambda) \tilde{s}^{rg *}_1(\lambda)$, with, 
at the 2nd order in $\varepsilon$: 
\begin{eqnarray}
\left \{
\begin{array}{lcl}
Z_{s_0}(\tau_0,\lambda)&=&
1+\varepsilon z_{s_0,1}(\tau_0,\lambda)+
\varepsilon^2 z_{s_0,2}(\tau_0,\lambda), \\
Z_{s_1}(\tau_0,\lambda)&=& 1+\varepsilon z_{s_1,1}(\tau_0,\lambda).
\end{array}
\right.
\end{eqnarray} 
Correspondingly, we write $(\tau-\tau_0)=(\tau-\lambda)+(\lambda-\tau_0)$
in (\ref{sdivrgII}) and we absorb the secular terms in $(\lambda-\tau_0)$ by 
appropriately choosing the renormalization constant coefficients. In detail, 
$z_{s_0,1}(\tau_0,\lambda)$ is already fixed by (\ref{zs_s0_1}). 
Nevertheless, the contributions proportional to this coefficient to the third 
term and to the fourth one in (\ref{sdivrgII}) are to be taken carefully into 
account, since they turn out to be $O(\varepsilon^2)$. 

In particular, in the case of the third term, when renormalizing 
$\tilde{s}^{*}_0$, one has: 
\begin{equation}
-\varepsilon\frac{(\tilde{s}^{*}_0+m)\tilde{s}^{*}_0}
{(\tilde{s}^{*}_0+M)^2}  = - \varepsilon \frac{(\tilde{s}^{rg *}_0+m)
\tilde{s}^{rg *}_0}{(\tilde{s}^{rg *}_0+M)^2} 
- \varepsilon^2 \left \{ \frac{d}{d\tilde{s}^{*}_0} 
\left [  \frac{(\tilde{s}^{*}_0+m)\tilde{s}^{*}_0}
{(\tilde{s}^{*}_0+M)^2}\right ] \left  
\lvert_{\!\!\shortmid_{\!\shortmid_{\tilde{s}^*_0=\tilde{s}^{rg *}_0}}} \right. \right \}  
z_{s_0,1}(\tau_0,\lambda) {s}^{rg *}_0,
\label{explication1a}
\end{equation}
with:
\begin{eqnarray}
&\hspace{.2in}& -\varepsilon^2 \left \{ \frac{d}{d\tilde{s}^{*}_0} 
\left [  \frac{(\tilde{s}^{*}_0+m)\tilde{s}^{*}_0}
{(\tilde{s}^{*}_0+M)^2}\right ] 
\left \lvert_{\!\!\shortmid_{\!\shortmid_{\tilde{s}^*_0=\tilde{s}^{rg *}_0}}} \right. \right \} 
z_{s_0,1}(\tau_0,\lambda) {s}^{rg *}_0= 
\nonumber \\ 
&=& 
 -\varepsilon^2  \left [ \frac{(M-m)\tilde{s}^{rg *}_0 }{(\tilde{s}^{rg *}_0+M)^3}
+  \frac{M (\tilde{s}^{rg*}_0+m)}{(\tilde{s}^{rg *}_0+M)^3} \right ]
z_{s_0,1}(\tau_0,\lambda) \tilde{s}^{rg *}_0=
\nonumber \\  
&=&
-\varepsilon^2  \left [\frac{(M-m)^2(\tilde{s}^{rg *}_0)^2 }
{(\tilde{s}^{rg *}_0+M)^4}+ \frac{M (M-m)\tilde{s}^{rg *}_0}
{(\tilde{s}^{rg *}_0+M)^4}(\tilde{s}^{rg*}_0+m)\right ](\lambda-\tau_0).
\label{explication1b}
\end{eqnarray}
Clearly, both of these two terms do in fact contribute, 
and they have to be taken into account in the definition of 
$z_{s_0,2}(\tau_0,\lambda)$. Let us note
moreover that $d [\tilde{s}^{*}_0/(\tilde{s}^{*}_0+M)]/d\tilde{s}^{*}_0=
M/(\tilde{s}^{*}_0+M)^2$. Thus, one can check that the 2nd order 
contribution proportional to the same coefficient, $z_{s_0,1}(\tau_0,\lambda)$ 
(that originates from the fourth term in $\tilde{s}^{rg}_{div}$, in the same way 
as we just explained in detail with the formulas (\ref{explication1a}) and 
(\ref{explication1b})), partially absorbs the term proportional to
$(\tau-\lambda)(\lambda-\tau_0)$ (that originates from the last term), 
and partially contributes to $z_{s_0,2}(\tau_0,\lambda)$. 

Thus, by taking: 
\begin{eqnarray}
\left \{
\begin{array}{lcl}
z_{s_0,2}(\tau_0,\lambda)&=&\left [ 
\frac{\textstyle (M-m)^2\tilde{s}^{rg *}_0 }{\textstyle (\tilde{s}^{rg *}_0+M)^4}-
\frac{\textstyle M(M-m)}
{\textstyle (\tilde{s}^{rg *}_0+M)^4}(\tilde{s}^{rg *}_0+m)
\right ] (\lambda-\tau_0) +  \\ 
&+&\frac{\textstyle M(M-m)^2}{\textstyle 
2(\tilde{s}^{rg *}_0+M)^3}(\lambda-\tau_0)^2\\
z_{s_1,1}(\tau_0,\lambda)&=&\frac{\textstyle M(M-m)}
{\textstyle (\tilde{s}^{rg *}_0+M)^2}(\lambda-\tau_0), 
\end{array}
\right.
\label{zs_II}
\end{eqnarray} 
we end up with an expression for $\tilde{s}^{rg}_{div}(\tau,\lambda)$ that is 
exactly the same as the one given in (\ref{sdivrgII}), with 
$\tilde{s}^{*}_0\rightarrow \tilde{s}^{rg *}_0(\lambda)$, 
$\tilde{s}^{*}_1\rightarrow \tilde{s}^{rg *}_1(\lambda)$ and 
$\tau_0 \rightarrow \lambda$. Noticeably, moreover, though we wrote explicitly 
the first two terms in $z_{s_0,2}$ to make evident the different origins of the 
contributions, their sum can be clearly simplified in such a way that the 
coefficient of the term proportional to $(\lambda-\tau_0)$ is simply 
equal to $-m(M-m)/(\tilde{s}^{rg *}_0+M)^3$.

Therefore, we continue our analysis by studying directly the derivative
with respect to $\lambda$ of  $\tilde{s}^{rg}_{div}(\tau,\lambda)$, 
and we obtain, at the 2nd order:
\begin{eqnarray}
\frac{\textstyle d\tilde{s}^{rg}_{div}(\tau,\lambda)}{\textstyle d\lambda}&=&
\frac{\textstyle d\tilde{s}^{rg *}_0(\lambda)}{\textstyle d\lambda}+ 
\varepsilon \frac{\textstyle d\tilde{s}^{rg *}_1(\lambda)}{\textstyle d\lambda}+ 
\varepsilon \frac{\textstyle  (M-m)\tilde{s}^{rg *}_0(\lambda)}
{\textstyle  {\tilde{s}^{rg *}_0(\lambda)}+M }+
\varepsilon^2 \frac{\textstyle  (M-m)^2 \left [ \tilde{s}^{rg *}_0(\lambda) 
\right ]^2 }{\textstyle  \left [\tilde{s}^{rg *}_0(\lambda)+M \right ]^4}+
\nonumber \\ 
&+&\varepsilon^2 \frac{\textstyle M (M-m)\tilde{s}^{rg *}_0(\lambda)}
{\textstyle [\tilde{s}^{rg *}_0(\lambda)+M]^4}[\tilde{s}^{rg*}_0(\lambda)+m]+
\varepsilon^2 \frac{\textstyle M(M-m)^2\tilde{s}^{rg *}_0(\lambda)}{\textstyle 
[\tilde{s}^{rg *}_0(\lambda)+M]^3}(\tau-\lambda)+ \nonumber \\
&-& \varepsilon^2 \frac{\textstyle 2 M (M-m) 
{\tilde{s}^{rg *}_0(\lambda)}}{\textstyle 
\left [{\tilde{s}^{rg *}_0(\lambda)}+M \right]^4}
\left [{\tilde{s}^{rg *}_0(\lambda)}+m\right]
+ \varepsilon^2 \frac{\textstyle  M (M-m) {\tilde{s}^{rg *}_1}(\lambda)}
{\textstyle \left [{\tilde{s}^{rg *}_0(\lambda)}+M\right]^2}+ 
\nonumber \\
&-& \varepsilon^2 \frac{\textstyle M (M-m)^2 {\tilde{s}^{rg *}_0(\lambda)}}
{\textstyle  \left [{\tilde{s}^{rg *}_0(\lambda)}+M\right]^3}(\tau-\lambda).
\label{dsdivrgII}
\end{eqnarray}
In fact, one already makes use of the known 1st order result on   
$d\tilde{s}^{rg *}_0/d\lambda$, given in Eqs.~(\ref{scalingcondition1}), 
in the derivation of this equation.

Here, for the sake of clarity, we also wrote explicitly both of the 
2nd order terms proportional to $(\tau-\lambda)$, that obviously cancel 
each other. In fact, this appears a quite consistent result of the present 
approach, since they have completely different origins. Indeed,
the first of these terms originates from part of the derivative with respect to 
$\lambda$ of the 1st order term proportional to $(\tau-\lambda)$ in 
$\tilde{s}^{rg}_{div}(\tau,\lambda)$ ({\em i.e.}, from the term that is equal to 
$-\varepsilon (M-m) \{ d [ \tilde{s}^{rg *}_0/(\tilde{s}^{rg *}_0+M)] / 
d\lambda \} (\tau-\lambda)$). The second of these terms originates 
instead from the derivative with respect to $\lambda$ of the last term in 
$\tilde{s}^{rg}_{div}(\tau,\lambda)$, that is proportional to $(\tau-\lambda)^2$.

Moreover, interestingly, the second term proportional to $\varepsilon^2$ 
(part of the contribution that originates from the constant term in 
$\tilde{s}^{rg}_1(\tau)$), letting aside a factor 2, is the same as the one
that is found when deriving the first of the terms proportional to
$\varepsilon^2 (\tau-\lambda)$ in (\ref{sdivrgII}) with respect to $\lambda$ 
({\em i.e.}, the term in $\tilde{s}^{rg}_{div}(\tau,\lambda)$ that corresponds 
to the fourth term proportional to $\varepsilon^2$ here), but that 
has opposite sign. 

In fact, both of these results are obtained in a similar manner to the one
that we described in detail previously, in the context of the derivation of 
the appropriate renormalization constants (from Eqs.~(\ref{scalingcondition1}), 
one has $(\lambda-\tau_0)d\tilde{s}^{rg *}_0/d\lambda=-
\varepsilon z_{s_0,1}^{s}(\tau_0,\lambda) \tilde{s}^{rg *}_0$).

Finally, by imposing the scaling condition 
$d\tilde{s}^{rg}_{div}(\tau,\lambda)/d\lambda=0$, and by grouping the terms
in $\varepsilon$ and in $\varepsilon^2$, we derive the two ODEs
to be obeyed by $\tilde{s}^{rg *}_0(\lambda)$ and $\tilde{s}^{rg *}_1(\lambda)$,
respectively:
\begin{eqnarray}
\left\{
\begin{array}{lcl}
\frac{\textstyle d\tilde{s}^{rg *}_{0}(\lambda)}
{\textstyle d\lambda}&=&
- \varepsilon\frac{\textstyle  (M-m)\tilde{s}^{rg *}_0(\lambda)}
{\textstyle  \tilde{s}^{rg *}_0(\lambda)+M }\\
\frac{\textstyle d\tilde{s}^{rg *}_{1}(\lambda)}
{\textstyle d\lambda}&=&\varepsilon \left \{
{-\frac{\textstyle  (M-m)^2 \left [ \tilde{s}^{rg *}_0(\lambda) \right ]^2 }
{\textstyle \left [\tilde{s}^{rg *}_0(\lambda)+M \right ]^4}}+
\frac{\textstyle M (M-m) {\tilde{s}^{rg *}_0(\lambda)}}{\textstyle 
\left [{\tilde{s}^{rg *}_0(\lambda)}+M \right]^4}
\left [{\tilde{s}^{rg *}_0(\lambda)}+m\right] + \right. \\ 
&-& \left. \frac{\textstyle  M (M-m) {\tilde{s}^{rg *}_1}(\lambda)}
{\textstyle \left [{\tilde{s}^{rg *}_0(\lambda)}+M\right]^2} \right \}
= \varepsilon \left \{ \frac{\textstyle  m(M-m) \tilde{s}^{rg *}_0(\lambda) }
{\textstyle \left [\tilde{s}^{rg *}_0(\lambda)+M \right ]^3}-
\frac{\textstyle  M (M-m) {\tilde{s}^{rg *}_1}(\lambda)}
{\textstyle \left [{\tilde{s}^{rg *}_0(\lambda)}+M\right]^2} \right \}.
\\
\end{array}
\right.
\label{scalingcondition2}
\end{eqnarray}
The first of these ODEs is just the already known 
1st order result for $d\tilde{s}^{rg *}_0(\lambda)/d\lambda$ reported in 
Eqs.~(\ref{scalingcondition1}). Actually, at the 2nd order, 
the interesting ODE is the one to be obeyed by $\tilde{s}^{rg *}_1(\lambda)$.  

Let us remark that, when making the final transformation 
$\lambda \rightarrow \tau=t/\varepsilon$, and when recalling that 
$\tilde{s}^{rg *}_0(t)=\tilde{s}^{out}_0(t)$ (given in (\ref{sout0})), the ODE 
to be obeyed by $\tilde{s}^{rg *}_1(t)$ turns out to be different from
the one for the 1st order outer substrate within the PE method reported 
in Eqs.~(\ref{outer1}). Here, we wrote the equation both with and 
without the simplification due to the sum of the first two terms 
just in order to make evident the difference, since it is the first term in the 
not simplified expression that was absent there. Indeed, to get in particular 
the same coefficient as in the ODE in Eqs.~(\ref{outer1}) for the second term 
in the present not simplified expression, it is essential to 
correctly take into account the constant in $\tilde{s}^{rg}_1(\tau)$ in the 
part to be renormalized of the function (at the 2nd order). In fact, 
as outlined, the contribution of the only term proportional to $\varepsilon^2$ 
would give an incorrect (twice larger) coefficient.
 
Interestingly, the present ODE turns out to be simpler 
than the one encountered in the PE method. One 
can check that the solution is given by:
\begin{equation}
\tilde{s}^{rg *}_1(t)= \frac{\textstyle m {s}^{out}_0(t)}
{\textstyle M \left [\tilde{s}^{out}_0(t)+M\right ] } 
\log \left [ \frac{\textstyle \tilde{s}^{out}_0(t)+M}{\textstyle (1+ M) 
\tilde{s}^{out}_0(t)} \right ],
\label{solsrgstar1}
\end{equation}
with the choice $\tilde{s}^{rg *}_1(0)=0$, that is reasonable in this context, 
since it allows to get correctly $\tilde{s}^{rg, u}_2(0)=1$. It is also 
important to stress that the solution also satisfies the asymptotic condition 
$\lim_{t \rightarrow \infty}\tilde{s}^{rg *}_1(t)=0$, though this in fact
implies $\lim_{t \rightarrow \infty}\tilde{s}^{rg, u}_2(t)=O(\varepsilon^2)$.

On the other hand, consistently, when adding the 1st order term that originates 
from the replacement $\tilde{s}^{*}_0 \rightarrow \tilde{s}^{rg, *}_0(\lambda)$ 
in the 1st order constant term in the part to be renormalized of the function, 
the total 1st order {\em outer} contribution to the SPDERG 2nd order UA to the 
correct solution for the substrate is just: 
\begin{equation}
\tilde{s}^{rg, out}_1(t) = -\frac{[\tilde{s}^{out}_0(t)+m]\tilde{s}^{out}_0(t)}
{[\tilde{s}^{out}_0(t)+M]^2}+\tilde{s}^{rg *}_1(t)=\tilde{s}^{out}_1(t).
\label{srgout2}
\end{equation}
Thus, one recovers the 1st order outer contribution to the PE UA reported in 
(\ref{solsouter1}). As we will discuss more in detail in the following, this 
was not a result to be taken for granted \cite{ChGoOo2}. Obviously, the result 
implies that the total 1st order outer contribution satisfies the MC, too 
({\em i.e.}, in particular that $\tilde{s}^{rg, out}_1(0)=-(1+m)/(1+M)^2$).

Hence, we recall that ${\cal R}^s_2(\tau)$ is to be evaluated
from the remaining part of the solution in $\tilde{s}^{*}_0=1$,
$\tilde{s}^{*}_1=0$ and $\tau_0=0$. It is given by
(see (\ref{solsrg2}) and (\ref{coeffsolsrg2}), in Appendix B):
\begin{eqnarray}
{\cal R}^s_2(\tau)&=& D_{{s}^{rg}_2}(1)+
\left [ F_{{s}^{rg}_2}(1)+ H_{{s}^{rg}_2}(1) \tau+J_{{s}^{rg}_2}(1)\tau^2
\right ]e^{\textstyle- (1+M){\tau}} + K_{{s}^{rg}_2}(1)
e^{\textstyle-2 (1+M) \tau} = \nonumber \\
&=&-\frac{\textstyle 1}{\textstyle 2 (1+M)^5} [2M^2(2m+1)-M(6m^2+5m+3)-m^2+m]+
\nonumber \\  
&+&\frac{\textstyle 1}{\textstyle (1+M)^5} 
[M^2(2m+1)-M(3m^2+3m+2)-m^2-m-1]e^{\textstyle- (1+M){\tau}}+ \nonumber \\
&-& \left \{\frac{\textstyle 1}{\textstyle (1+M)^4} [M^2+M(m^2+1)-2m-1]\tau
-\frac{\textstyle (M-m)}{\textstyle 2 (1+M)^3} (m+1) \tau^2 
\right \} e^{\textstyle- (1+M){\tau}}+ \nonumber \\
&+&\frac{\textstyle (m+1)}{\textstyle 2 (1+M)^5} (M+m+2)
e^{\textstyle- 2(1+M){\tau}},
\label{rs2last}
\end{eqnarray}
by using in particular the coefficient values reported in (\ref{coeffsolsrg2}). 

Correspondingly, we get the complete result for the SPDERG
2nd order UA to the adimensional substrate solution:
\begin{equation}
\tilde{s}^{rg, u}_2(t)=\tilde{s}^{rg, u}_1(t)+\varepsilon 
\left  \{ -\frac{\tilde{s}^{out}_0(t)+m}
{[\tilde{s}^{out}_0(t)+M]^2}\tilde{s}^{out}_0(t) + \frac{1+m}
{(1+M)^2} + \tilde{s}^{rg *}_1(t)  \right \}+ 
\varepsilon^2 {\cal R}^s_2(t/\varepsilon)+O(\varepsilon^2),
\label{solsrgu2}
\end{equation}
with $\tilde{s}^{rg, u}_1(t)$ given in (\ref{solrgu1}), 
 $\tilde{s}^{out}_0(t)= \tilde{s}^{rg *}_0(t)$ given in (\ref{sout0}), and 
$\tilde{s}^{rg *}_1(t)$, ${\cal R}^s_2(t/\varepsilon)$, given in 
(\ref{solsrgstar1}), (\ref{rs2last}), respectively. 
We wrote the solution in the present form to make more evident
the analogies and the differences with the 1st order 
PE result. In particular, the constant term 
that is equal to $-(1+m)/(1+M)^2$ originates here from
the constant term depending on  $\tilde{s}^{*}_0$ that appeared 
(calculated in $\tilde{s}^{*}_0=1$) in $\tilde{s}^{rg, u}_1(t)$, the one that 
has been now included in $\tilde{s}^{rg}_{div}$ at the 2nd order. 
Moreover, one can notice that the result implies, as expected,
$\lim_{t\rightarrow \infty}\tilde{s}^{rg, u}_2(t)=\tilde{s}^{rg,u}_{2,\infty}= 
\varepsilon^2 D_{{s}^{rg}_2}(1)=O(\varepsilon^2)$, where $D_{{s}^{rg}_2}(1)$ 
is the first of the terms in (\ref{rs2last}), reported in Appendix B, too.

For the complex, the 2nd order terms to be added to the part of the solution 
to be renormalized are the three ones that appear in 
$\tilde{c}^{rg}_{2,div}(\tau)$, that is reported in (\ref{solcrg2div}), 
in Appendix A. Moreover, we have again to take into account
also the constant term in $\tilde{c}^{rg}_{1}(\tau)$, {\em i.e.}, the one 
corresponding to the coefficient 
$B_{{c}^{rg}_1}(\tilde{s}^*_0,\tilde{s}^*_1,\tilde{c}^*_0)$ 
that is reported in (\ref{coeffsolcrg1}), in Appendix A, and that needs to be
calculated in $\tilde{c}^*_0=0$. We obtain:
\begin{eqnarray}
\tilde{c}^{rg}_{div}(\tau)&=&
\frac{\textstyle \tilde{s}^*_0}
{\textstyle \tilde{s}^*_0+M}-
\varepsilon \frac{\textstyle M(\tilde{s}^*_0-M+2m)}
{\textstyle \left (\tilde{s}^*_0 +M \right )^4}
\tilde{s}^*_0+ \varepsilon
\frac{\textstyle M\tilde{s}^*_1}{\textstyle \left (\tilde{s}^*_0 +M \right )^2}
-\varepsilon \frac{\textstyle M  (M-m) \tilde{s}^*_0}
{\textstyle \left (\tilde{s}^*_0 +M \right )^3}(\tau-\tau_0)
+ \nonumber \\ 
&+&\varepsilon^2 A_{{c}^{rg}_{2,div}}({\tilde{s}^*_0}) (\tau-\tau_0)
+\varepsilon^2 B_{{c}^{rg}_{2,div}}({\tilde{s}^*_0},{\tilde{s}^*_1}) (\tau-\tau_0)
+\varepsilon^2 C_{{c}^{rg}_{2,div}}({\tilde{s}^*_0}) (\tau-\tau_0)^2,
\label{cdivrgII}
\end{eqnarray}
where the coefficients $A_{{c}^{rg}_{2,div}}$, $B_{{c}^{rg}_{2,div}}$ and
$C_{{c}^{rg}_{2,div}}$ of the three 2nd order secular terms  
are given in (\ref{coeffsolcrg2div}), in Appendix B.

Though the proof involves more demanding calculations in the present 
case, one can check that here, in the same way as we showed in detail in the 
case of the substrate, with the same renormalization coefficients of the ICVs 
($z_{s_0,1}$ already fixed in (\ref{zs_s0_1}), 
and $z_{s_0,2}$, $z_{s_1,1}$ already fixed in (\ref{zs_II}), respectively),
one obtains a function $\tilde{c}^{rg}_{div}(\tau,\lambda)$ 
exactly corresponding to (\ref{cdivrgII}), with $\tilde{s}^{*}_0\rightarrow 
\tilde{s}^{rg *}_0(\lambda)$, $\tilde{s}^{*}_1\rightarrow 
\tilde{s}^{rg *}_1(\lambda)$ and $\tau_0 \rightarrow \lambda$. 
This was in fact the expected result from the point of view of the analogy 
with the matching in the PE.  

Moreover, as furthermore expected within the same context, we verify that, 
when imposing the scaling condition 
$d\tilde{c}^{rg}_{div}(\tau,\lambda)/d\lambda=0$,
one  recovers once again the two ODEs in (\ref{scalingcondition2}),
to be obeyed by $\tilde{s}^{rg *}_0(\lambda)$ and $\tilde{s}^{rg *}_1(\lambda)$.
Thus, one obtains in particular the same result on 
$\tilde{s}^{rg *}_1(\lambda)$. 

In detail, also in the case of the complex, one needs to use the known 1st 
order result on $d\tilde{s}^{rg *}_0(\lambda)/d\lambda$ in the derivation of 
the 2nd order one on $d\tilde{s}^{rg *}_1(\lambda)/d\lambda$. Finally, also
in this case the result on the derivative with respect to $\lambda$ is obtained 
in a very similar way to the proof of the correspondence between 
$\tilde{c}^{rg}_{div}(\tau,\lambda)$ and the original 
$\tilde{c}^{rg}_{div}(\tau)$ given by (\ref{cdivrgII}). 

Clearly, in the present case, there is a definitely larger number of 
relevant terms that anyway either cancel each other or contribute in the 
correct way to the final result. Therefore, the verification of the 
expectations we made on the basis of the recalled analogy, that is reported in 
Appendix C, appears to give further consistency to the whole approach. 

Correspondingly, we obtain a 1st order {\em outer} contribution to the SPDERG 
2nd order UA to the correct solutions for the adimensional complex 
concentration that turns out to be exactly equal to the 1st order 
PE outer contribution given in (\ref{solcouter1}). It is
obtained here from an algebraic relation that could appear different from
the one in Eqs.~(\ref{outer1}), but that is in fact equivalent. This 
becomes evident when writing $\tilde{s}^{rg *}_1(t)$ in terms of 
$\tilde{s}^{out}_0(t)$ and $\tilde{s}^{out}_1(t)$ by means of (\ref{srgout2}):
\begin{equation}
\tilde{c}^{rg, out}_1(t)=
-\frac{\textstyle M(\tilde{s}^{rg *}_0(t)-M+2m)}
{\textstyle \left [\tilde{s}^{rg *}_0(t) +M \right ]^4}\tilde{s}^{rg *}_0(t)+ 
\frac{\textstyle M\tilde{s}^{rg *}_1(t)}
{\textstyle \left [\tilde{s}^{rg *}_0(t) +M \right ]^2}=\tilde{c}^{out}_1(t).
\label{soloutcrgu2}
\end{equation}
The equality can also be easily checked by reminding that 
$\tilde{s}^{rg *}_0(t)=\tilde{s}^{out}_0(t)$, and by using the known result for 
$\tilde{s}^{rg *}_1(t)$ reported in (\ref{solsrgstar1}).

The SPDERG 2nd order UA to the correct solution for the complex is then 
obtainable by taking into account also the remaining part of the 2nd order 
inner solution, and is given by:
\begin{eqnarray}
\tilde{c}^{rg, u}_2(t)&=&\tilde{c}^{rg, u}_1(t) + \varepsilon \left  \{
-\frac{\textstyle M(\tilde{s}^{rg *}_0(t)-M+2m)}
{\textstyle \left [\tilde{s}^{rg *}_0(t) +M \right ]^4}
\tilde{s}^{rg *}_0(t)+ \frac{\textstyle M\tilde{s}^{rg *}_1(t)}
{\textstyle \left [\tilde{s}^{rg *}_0(t) +M \right ]^2}+
\frac{\textstyle M(1-M+2m)}
{\textstyle \left (1 +M \right )^4}\right \}+
\nonumber \\ 
&+& \varepsilon^2 {\cal R}^c_2(t/\varepsilon)+O(\varepsilon^2).
\label{solcrgu2}
\end{eqnarray}
Here, we outline once again similarities and differences with the 
PE result. In fact, because of (\ref{soloutcrgu2}), one can equivalently
write the terms in curly brackets as $\tilde{c}^{out}_1(t)$ minus the constant
terms that in the standard method appear twice. 

In detail, the various terms that appear
in (\ref{solcrgu2}) are reported: $\tilde{c}^{rg, u}_1(t)$  
in (\ref{solrgu1}); $\tilde{s}^{rg *}_0(t)=\tilde{s}^{out}_0(t)$ in
(\ref{sout0}); $\tilde{s}^{rg *}_1(t)$ in (\ref{solsrgstar1}); and
${\cal R}^c_2(\tau)$, correctly evaluated in $\tau_0=0$, $\tilde{s}^{*}_0=1$
and $\tilde{s}^{*}_1=0$ in (\ref{solrcrg2}), with the coefficients
given in (\ref{coeffsolrcrg2}), in Appendix B.

This UA to the correct solution verifies 
the ICV $\tilde{c}^{rg, u}_2(0)=0$, whereas one finds 
$\lim_{t \rightarrow \infty}\tilde{c}^{rg, u}_2(t)
=\tilde{c}^{rg,u}_{2,\infty}=\varepsilon^2 
A_{{\cal R}^c_2}(1,0)=O(\varepsilon^2)$, with $A_{{\cal R}^c_2}(1,0)$ 
the constant term in ${\cal R}^c_2(\tau)$, whose
detailed dependence on $\tilde{s}^{*}_0$ and $\tilde{s}^{*}_1$ is reported in
(\ref{arcrg2}), and that is calculated in  $\tilde{s}^{*}_0=1$,  
$\tilde{s}^{*}_1=0$ in (\ref{coeffsolrcrg2}), in Appendix B.
 
\begin{figure}[t]
\begin{center}
\leavevmode
\epsfig{figure=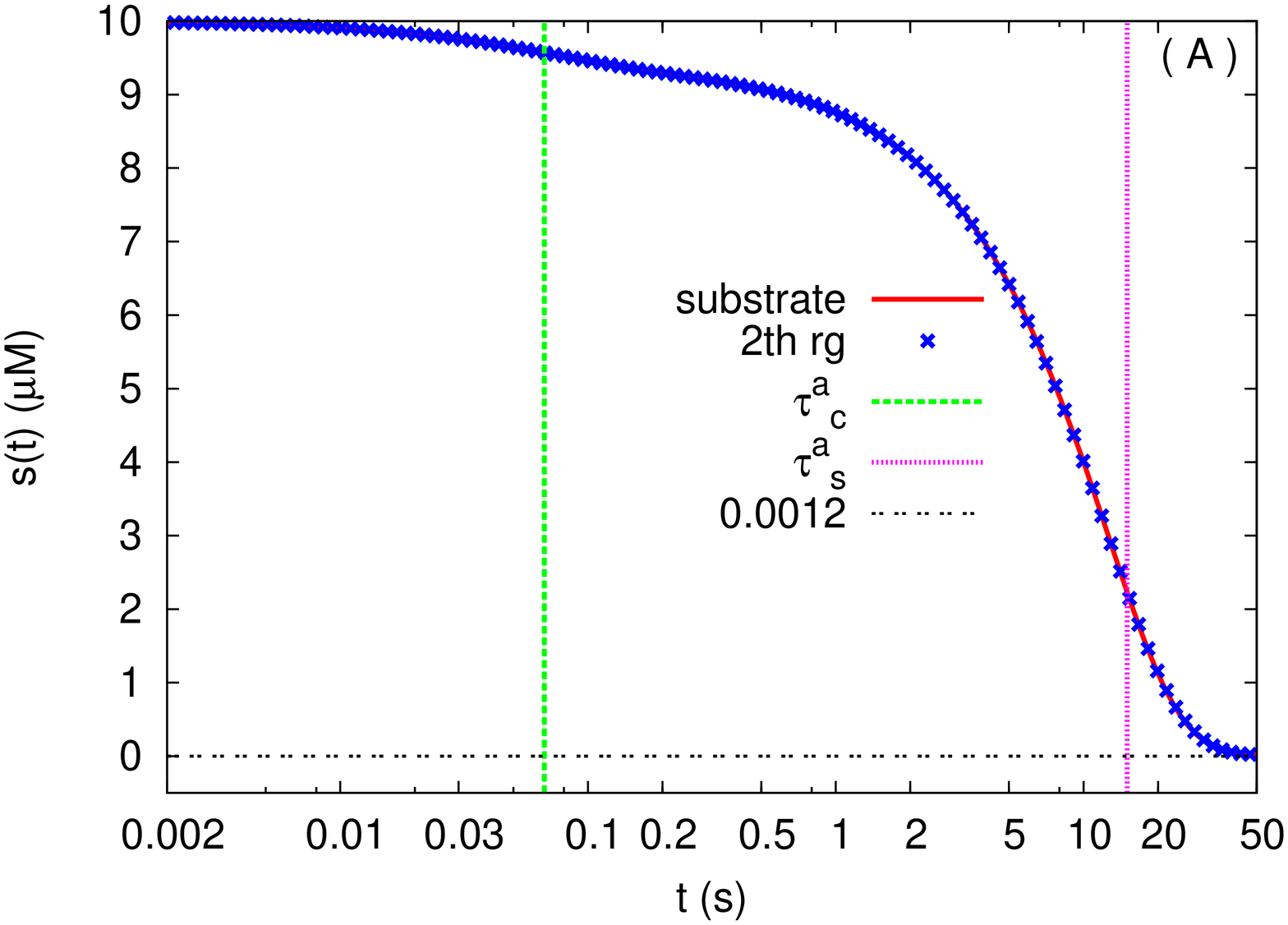,angle=270,width=8cm}
\epsfig{figure=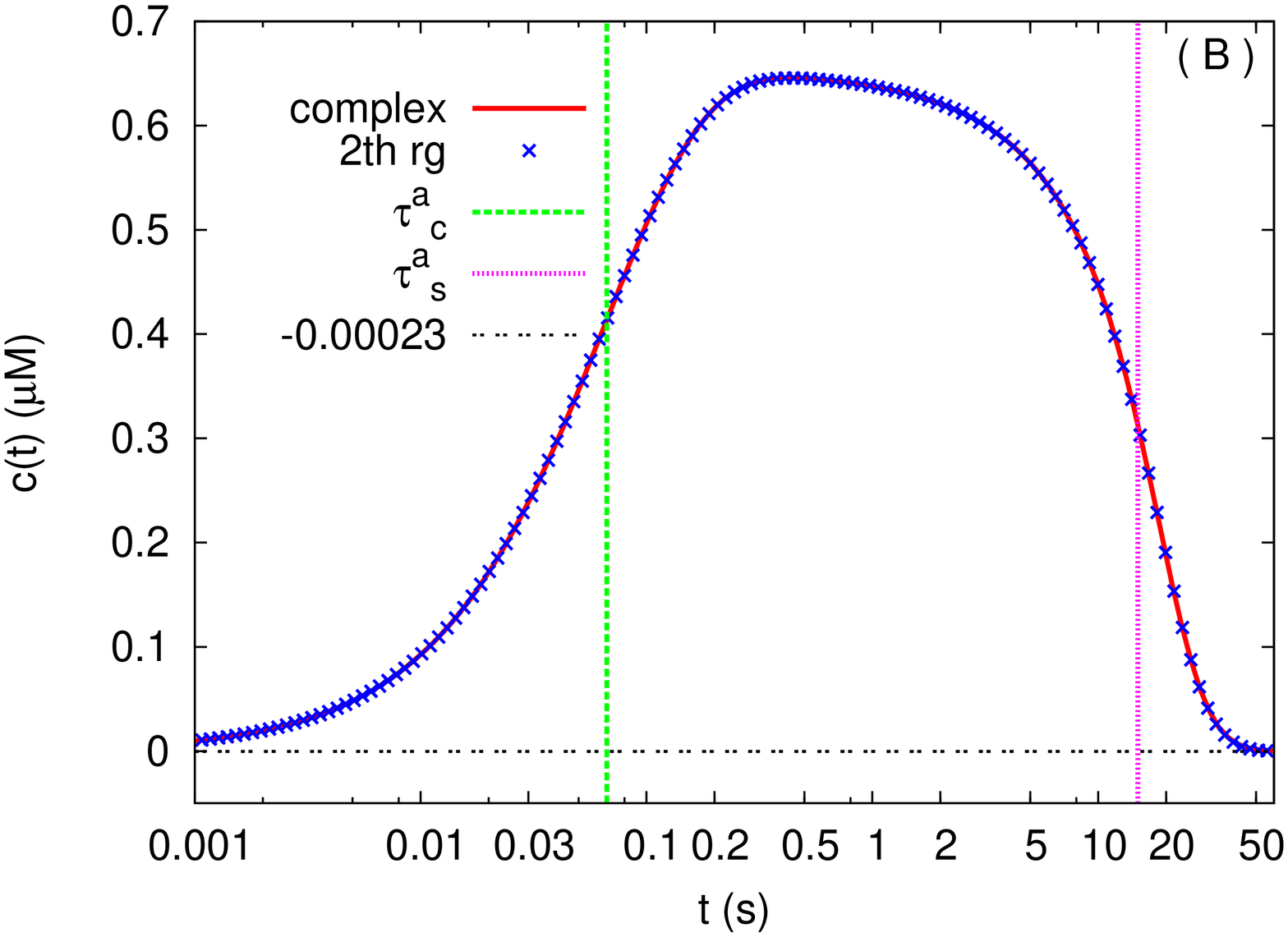,angle=270,width=8cm}
\epsfig{figure=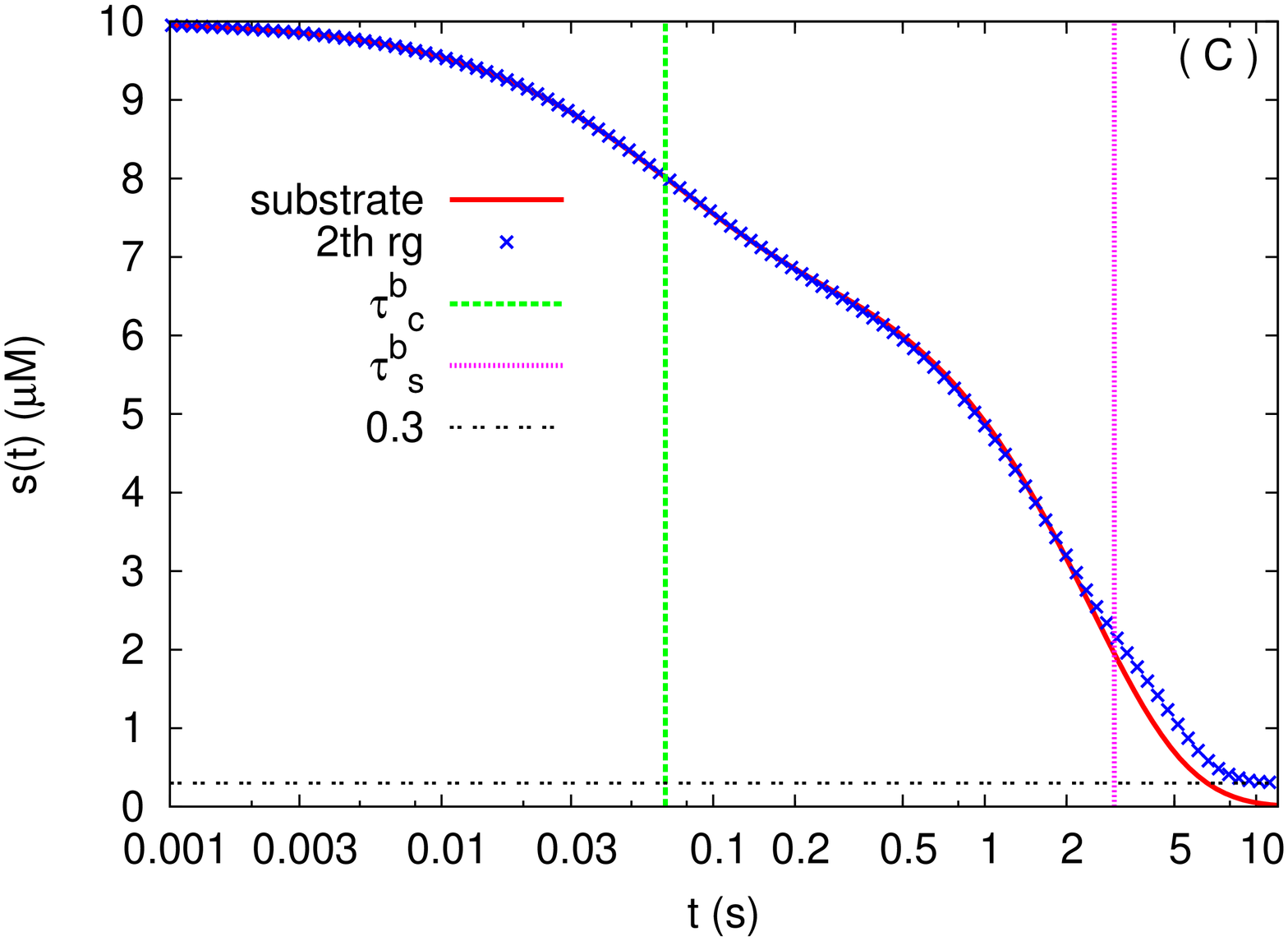,angle=270,width=8cm}
\epsfig{figure=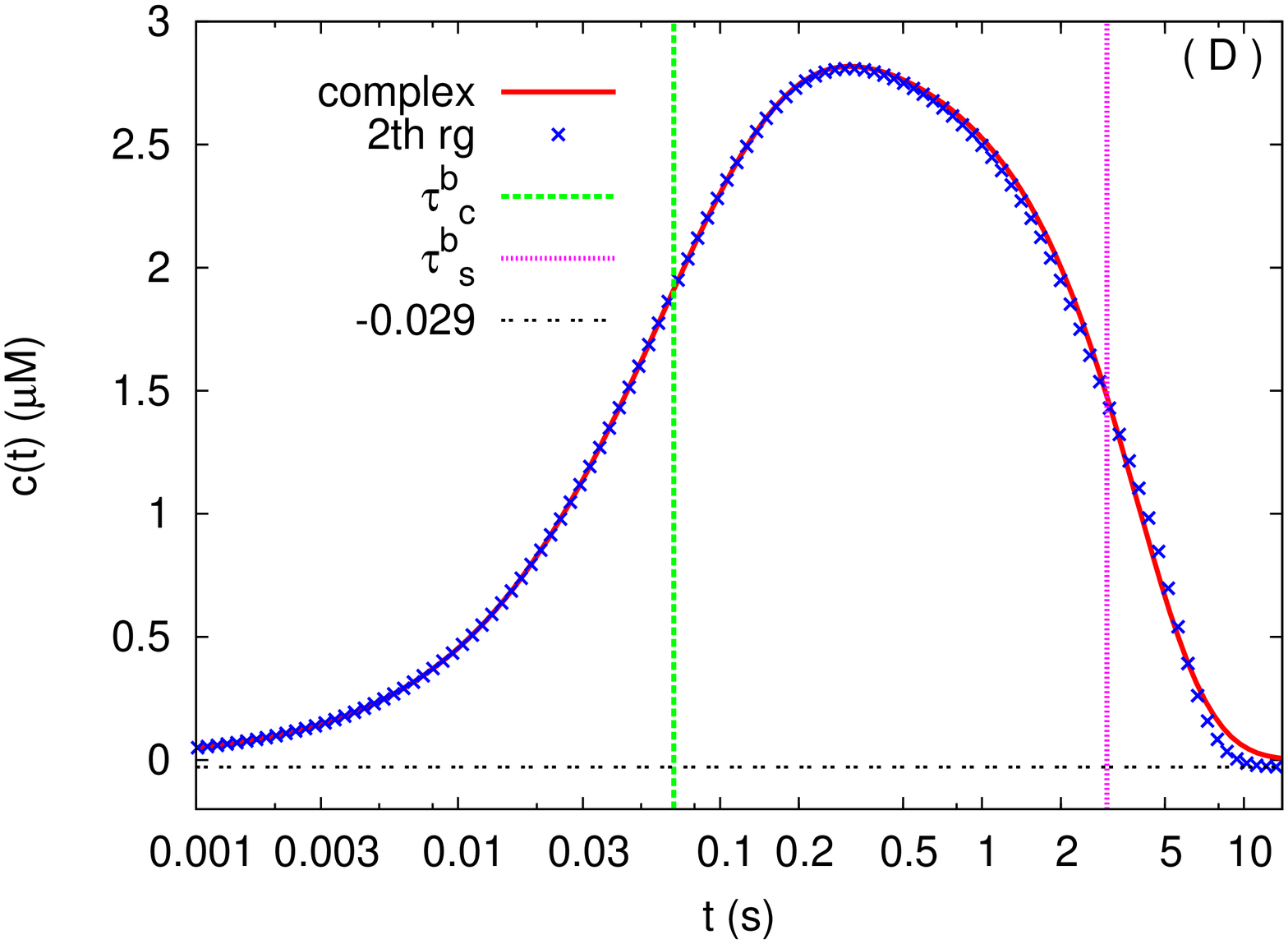,angle=270,width=8cm}
\caption{{\footnotesize In $A)$ and $C)$ we present the behaviour of the 
concentrations of the substrate $s(t)$, whereas in $B)$ and $D)$ we present 
the ones of the complex $c(t)$ for the $a$ and $b$ sets of ICVs given in 
(\ref{initialvalues}), respectively. Hence, in $A)$ and $B)$ we are in the case 
with $\varepsilon=\varepsilon^a=0.1$, whereas in $C)$ and $D)$ we are in the 
one with $\varepsilon=\varepsilon^b=0.5$. We plot both the numerical solutions 
of Eqs.~(\ref{mmineq}) already shown in the previous figures, and the 
analytical solutions computed from the SPDERG 2nd order UAs (with a standard 
numerical approximation for the Lambert function), as given in (\ref{solsrgu2})
and (\ref{solcrgu2}), respectively. We plot moreover the (physically 
meaningless) asymptotic limits of these analytical solutions: 
$s_0^a \tilde{s}^{rg,u}_{2,\infty} \simeq 0.0012$, 
$e_0^a \tilde{c}^{rg,u}_{2,\infty} \simeq -0.00023$, and 
$s_0^b \tilde{s}^{rg,u}_{2,\infty} \simeq 0.3$, 
$e_0^b \tilde{c}^{rg,u}_{2,\infty} \simeq -0.029$,
for the $a$ and $b$ ICV sets, respectively. We finally plot our corresponding 
rough evaluations of the two different time scales involved, too, with 
$\tau_s$ describing the substrate decay time and $\tau_c$ the complex 
saturation time. Notice that the time is in logarithmic scale.}}
\label{fig6}
\end{center}
\end{figure}

We plot in [Fig. \ref{fig6}] our numerical results on the
SPDERG 2nd order UAs to the solutions  for the substrate and
the complex, respectively, for the two considered sets of ICVs. The plots are 
as usual in comparison with the numerical solutions of the original 
problem (\ref{mmineq}) (the same curves as in [Fig. \ref{fig1}]). 

As it is in fact explicable on the basis that they contain the 2nd order terms 
of the inner solutions, the approximations work better than the 1st order  
PE ones in a region that encompasses the matching one. Indeed, the results here 
are nearly indistinguishable, within our numerical precision, from the correct 
ones on a definitely larger time window. This is true also in the particularly 
demanding case of the substrate in  [Fig. \ref{fig6}C], and the outcome is 
clearly different from the one observed in the same case at the 1st order, 
when applying the standard method, that is reported in [Fig. \ref{fig4}C].

Nevertheless, one can still note a minor discrepancy at large times, that 
is at least partially to be related to the 
failure of the approximations in reproducing the physically correct 
asymptotically vanishing solutions. Actually, on the basis 
of the results that we already obtained in the present study, this failure  
appears easily correctable in a reasonable way. In the following, we 
just consider the SPDERG 2nd order UAs that can be proposed that 
satisfy the asymptotic conditions $\lim_{t \rightarrow \infty} 
s(t)=\lim_{t \rightarrow \infty} c(t)=0$, too.

\section{Results and discussion: iii) Second order with proposed refinement }
\label{rdIIref}
Within the framework of the already obtained results, by 
assuming that the found solution behaviours could be iterated, one
can hypothesize that, in the SPDERG approach (at least in its present kind
of application to a boundary layer problem, in which we directly renormalized 
the bare initial constant values), the constant term at a given order will 
contribute to the outer component of the solution at the following order. 
Actually, this appears to us the SPDERG approach ingredient that is equivalent 
to take into account both the MCs and the removing 
of one of the constants that appear twice in the standard PE
method, though the observation needs to be better
formalized for investigating its possible generalizations. 

In fact, in the present case, when passing from the 2nd to the 3rd order, 
these iteratively expected solution behaviours should be obtainable
by means of the substitutions $\tilde{s}^{*}_0 \rightarrow \tilde{s}^{rg *}_0(t)$
and $\tilde{s}^{*}_1 \rightarrow \tilde{s}^{rg *}_1(t)$ in the constant terms 
$D_{{s}^{rg}_2}(\tilde{s}^{*}_0)+E_{{s}^{rg}_2}(\tilde{s}^{*}_0,\tilde{s}^{*}_1)$
and $A_{{\cal R}^c_2}(\tilde{s}^{*}_0,\tilde{s}^{*}_1)$. These terms appear in the 
2nd order inner solution for the substrate and the complex, respectively, that 
are given in (\ref{coeffsolsrg2}) and (\ref{arcrg2}), in Appendix B.  
We notice that, despite of the substitutions, they will remain terms of 2nd 
order, in agreement with the general consideration that, at the $n$-th order, 
within the SPDERG approach, one in fact obtains the $(n-1)$-th order outer 
components of the corresponding PE UAs, to be interpreted as their leading 
order terms \cite{ChGoOo1,ChGoOo2}.

\begin{figure}[t]
\begin{center}
\leavevmode
\epsfig{figure=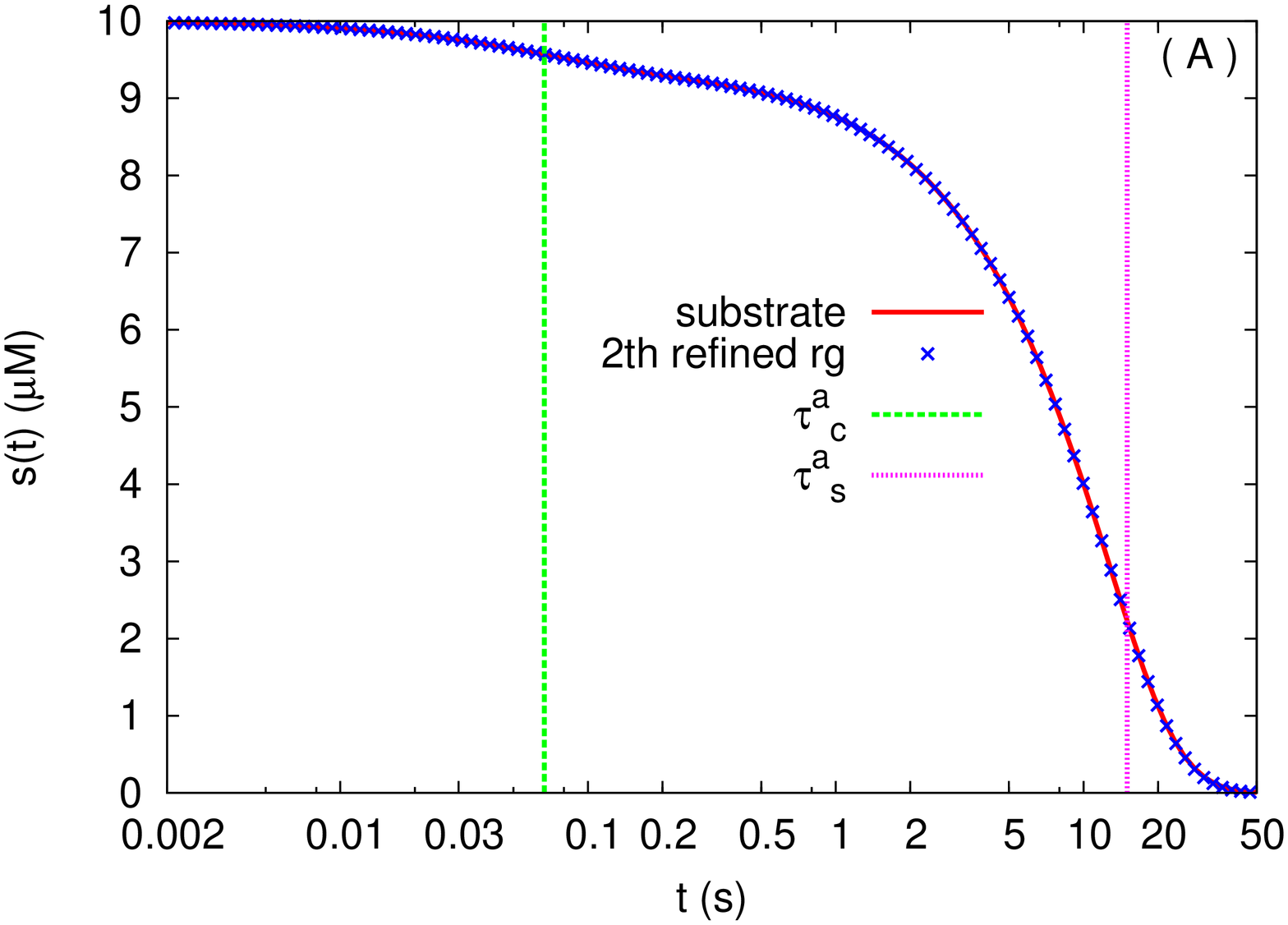,angle=270,width=8cm}
\epsfig{figure=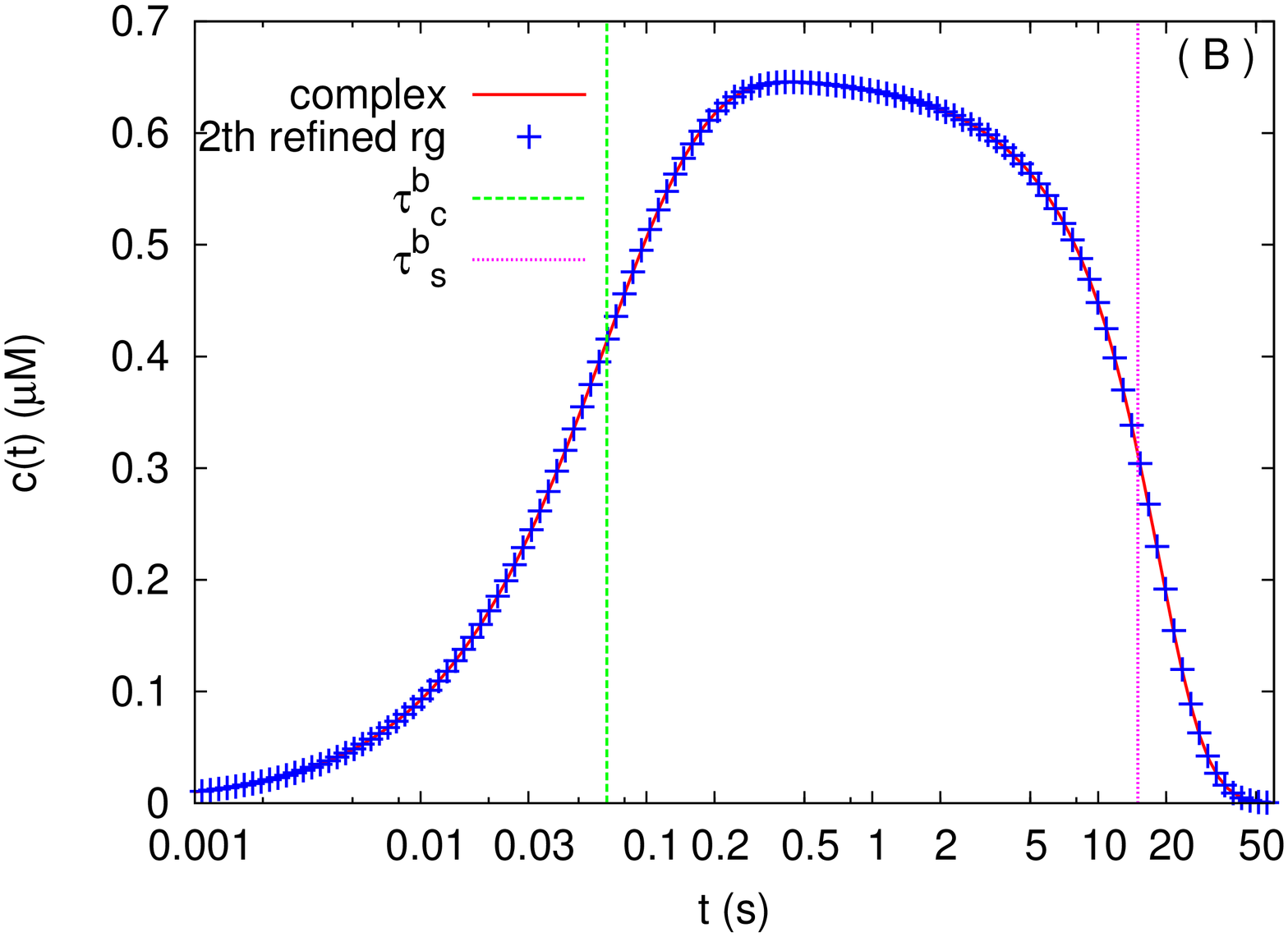,angle=270,width=8cm}
\epsfig{figure=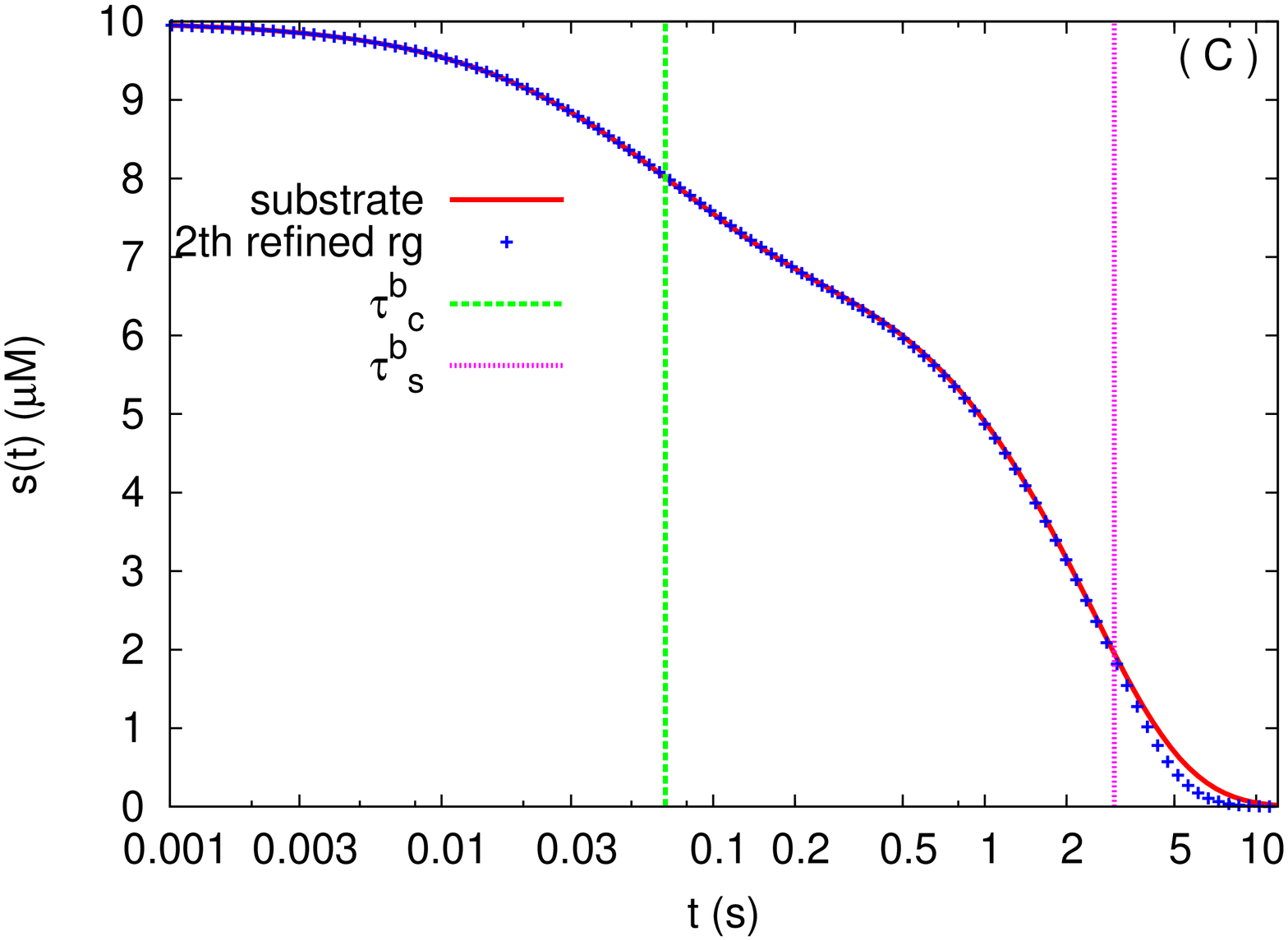,angle=270,width=8cm}
\epsfig{figure=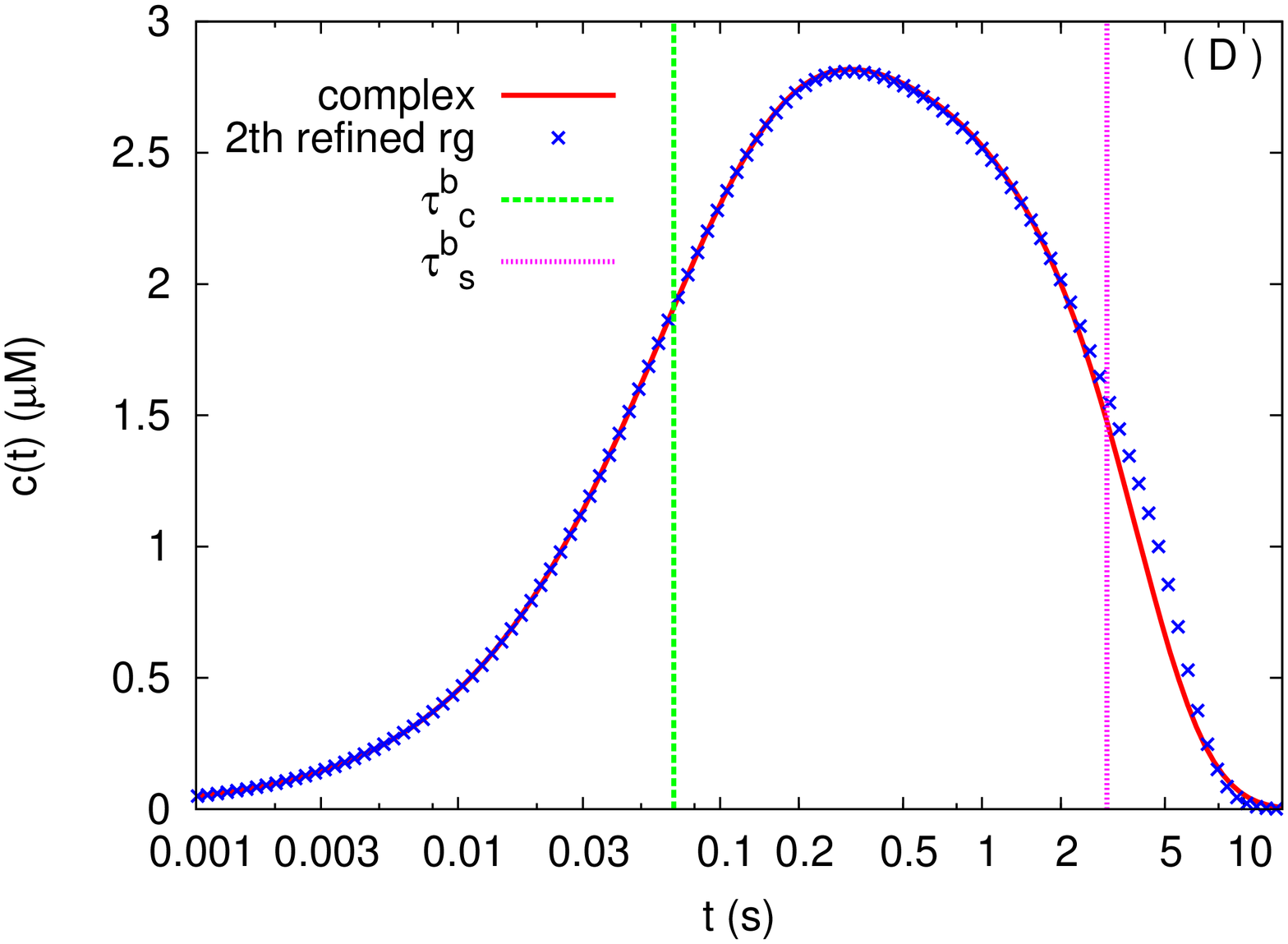,angle=270,width=8cm}
\caption{{\footnotesize In $A)$ and $C)$ we present the behaviour of the 
concentrations of the substrate $s(t)$, whereas in $B)$ and $D)$ we present 
the ones of the complex $c(t)$ for the $a$ and $b$ sets of ICVs given in 
(\ref{initialvalues}), respectively. Hence, in $A)$ and $B)$ we are in the case 
with $\varepsilon=\varepsilon^a=0.1$, whereas in $C)$ and $D)$ we are in the 
one with $\varepsilon=\varepsilon^b=0.5$. We plot both the numerical solutions 
of Eqs.~(\ref{mmineq}) already shown in the previous figures, and the  
analytical solutions computed from the refined SPDERG 2nd order UAs (with a 
standard numerical approximation for the Lambert function), as given in 
(\ref{solrgu2r}). We  finally plot our corresponding rough evaluations of the 
two different time scales involved, too, with $\tau_s$ describing the 
substrate decay time and $\tau_c$ the complex saturation time. Notice that the 
time is in logarithmic scale.}}
\label{fig7}
\end{center}
\end{figure}

Therefore, we can finally consider as refined SPDERG 2nd order UAs to the 
correct solutions the functions $\tilde{s}^{rg, u}_{2,r}(t)$ and 
$\tilde{c}^{rg, u}_{2,r}(t)$, that satisfy by construction the 
physically meaningful asymptotic conditions  
$\lim_{t \rightarrow \infty} \tilde{s}^{rg, u}_{2,r}(t)=
\lim_{t \rightarrow \infty}\tilde{c}^{rg, u}_{2,r}(t) =0$, given by:
\begin{eqnarray}
\left \{
\begin{array}{lcl}
\tilde{s}^{rg, u}_{2,r}(t)&=&\tilde{s}^{rg, u}_{2}(t)+ 
\varepsilon^2 \left \{  D_{{s}^{rg}_2}[\tilde{s}^{rg *}_0(t)] +  
E_{{s}^{rg}_2}[\tilde{s}^{rg *}_0(t),\tilde{s}^{rg *}_1(t)] -
 D_{{s}^{rg}_2}(1)\right \}+O(\varepsilon^2)  \\ 
\tilde{c}^{rg, u}_{2,r}(t)&=&\tilde{c}^{rg, u}_{2}(t)+ 
\varepsilon^2 \left \{  A_{{\cal R}^c_2}[\tilde{s}^{rg *}_0(t),
\tilde{s}^{rg *}_1(t)] -
 A_{{\cal R}^c_2}(1,0) \right \}+O(\varepsilon^2), \\
\end{array}
\right.
\label{solrgu2r}
\end{eqnarray}
We remind that $E_{{s}^{rg}_2}(1,0)=0$, whereas $\tilde{s}^{rg,u}_2(t)$
and $\tilde{c}^{rg,u}_2(t)$ are reported in (\ref{solsrgu2}) and 
(\ref{solcrgu2}), respectively, $\tilde{s}^{rg *}_0(t)=\tilde{s}^{out}_0(t)$ 
in (\ref{solsouter1}), and $\tilde{s}^{rg *}_1(t)$ in (\ref{solsrgstar1}). 

It is not to be taken for granted that these approximations could work better 
than the previously considered ones, since they anyway lack of a part of the 
2nd order outer contribution. On the other hand, these appear to us the most 
refined SPDERG 2nd order UAs that one can propose by exploiting as much as 
possible the obtained results. 

We present the corresponding numerical solutions for the two different 
considered ICV sets in [Fig. \ref{fig7}]. The plots are 
as usual in comparison with the numerical solutions of the original 
problem (\ref{mmineq}) (the same curves as in [Fig. \ref{fig1}]).
 
In fact, one could already observe in the previous [Fig. \ref{fig6}A], 
[Fig. \ref{fig6}B] that the SPDERG 2nd order UAs appeared 
indistinguishable from the correct solutions in the case of the $a$ set of 
ICVs. Indeed, this set corresponds to the relatively
small $\varepsilon=\varepsilon^a=0.1$, and  the PE 1st order UAs 
appeared indistinguishable from the correct solutions for
this ICV set ([Fig. \ref{fig4}A], [Fig. \ref{fig4}B]), 
too. In this case, we limit ourselves to underline that the  
present proposed 2nd order approximations, that are presented in 
[Fig. \ref{fig7}A] for the substrate concentration and in [Fig. \ref{fig7}B] 
for the complex one, are moreover correctly asymptotically vanishing.
In fact, as previously discussed, in the not refined 2nd order 
approximations the limits for $t \rightarrow \infty$ 
were not zero, though they gave a practically negligible 
contribution to the plotted curves.

On the other hand, when looking at [Fig. \ref{fig7}C] and [Fig. \ref{fig7}D],
that present the behaviour of the substrate and 
complex concentration, respectively, for the $b$ ICV case,
{\em i.e.}, for the larger value of the expansion parameter
$\varepsilon=\varepsilon^b=0.5$, the plots appear not enough detailed 
for understanding up to which point the present approximated solutions 
are better than the ones without the refinement. Both for this reason and for 
roughly quantifying the differences between the PE 1st order UAs and the 
present SPDERG 2nd order results, we are led to a more careful study.

\section{Results and discussion: iv) A conclusive comparison}
\label{rdcomp}
\begin{figure}[t]
\begin{center}
\leavevmode
\epsfig{figure=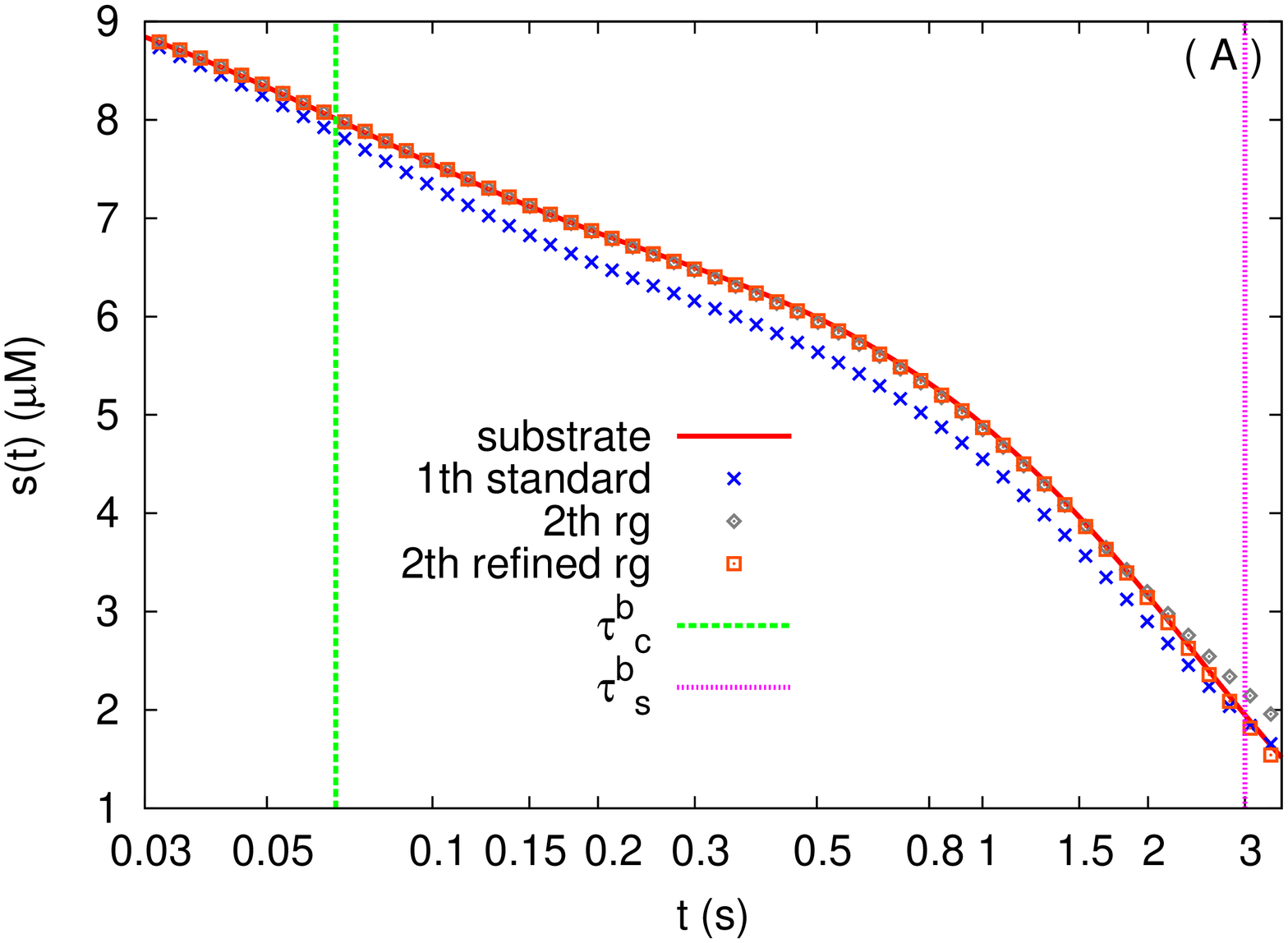,angle=270,width=8cm}
\epsfig{figure=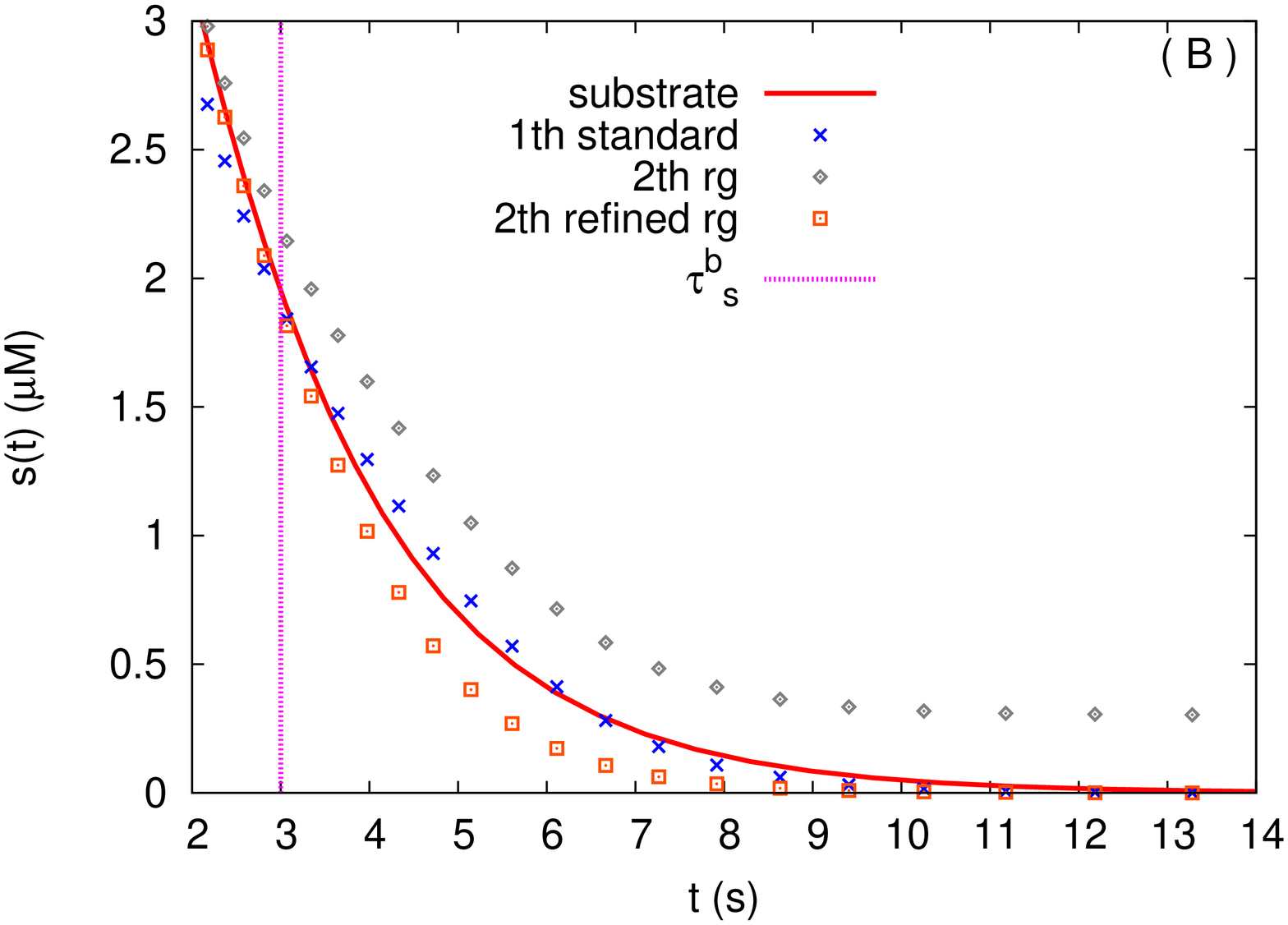,angle=270,width=8cm}
\epsfig{figure=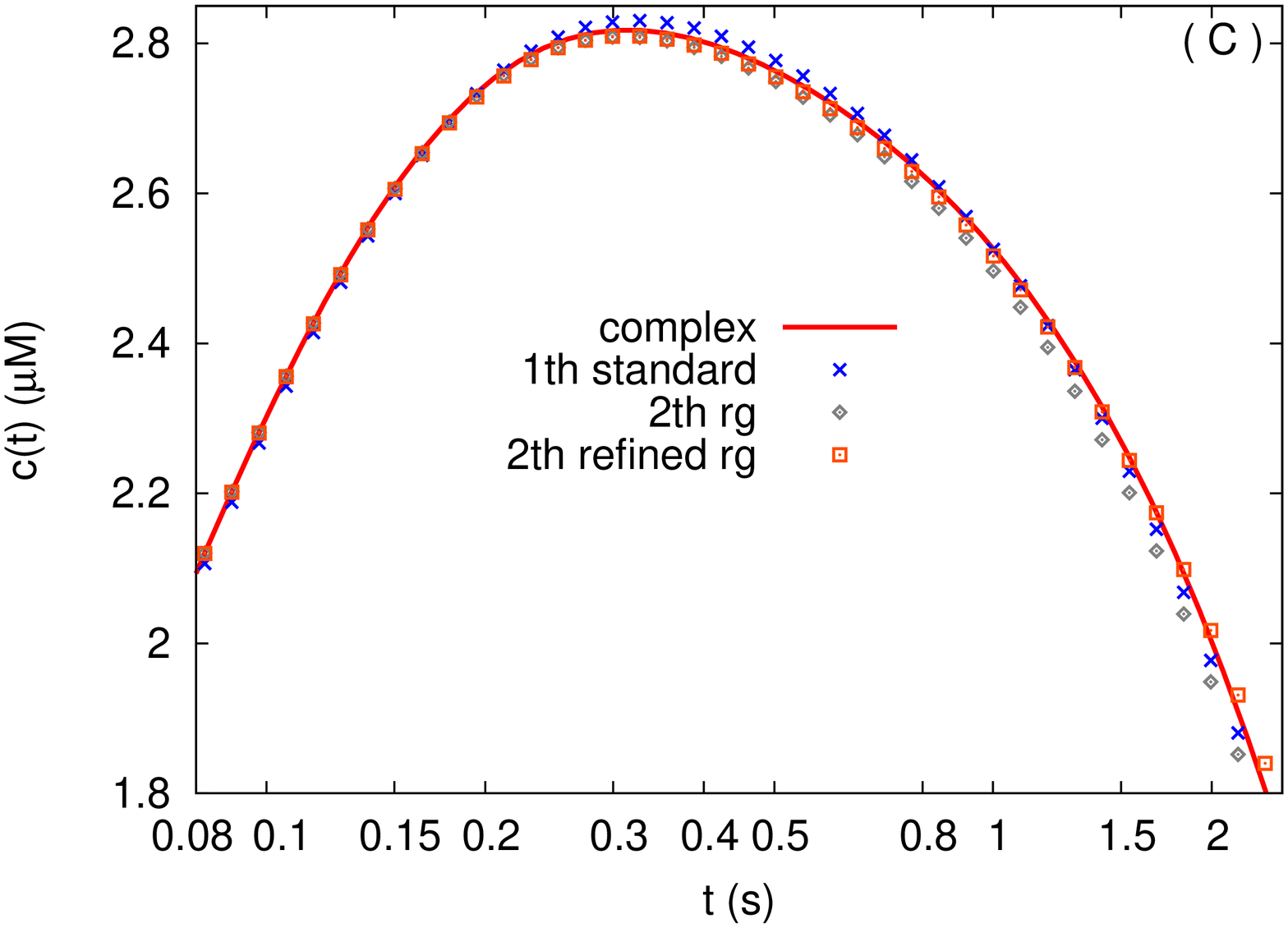,angle=270,width=8cm}
\epsfig{figure=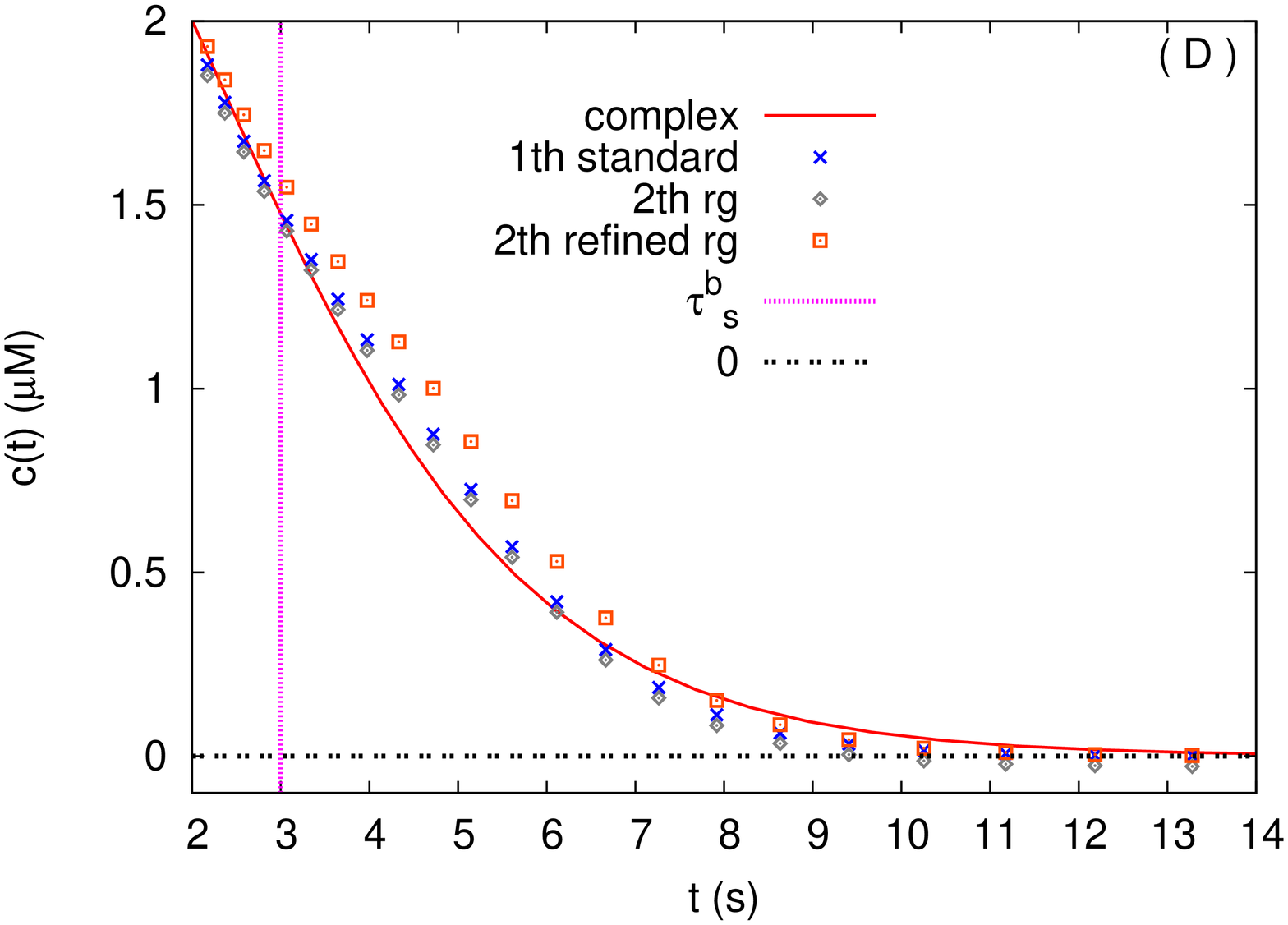,angle=270,width=8cm}
\caption{{\footnotesize In $A)$ (respectively $C)$) we present the behaviour 
of the concentration of the substrate $s(t)$ (respectively the complex $c(t)$) 
in the central part of the time window, in logarithmic time scale, whereas in 
$B)$ (respectively $D)$) we present the large time behaviour of these two 
quantities. The results are for the $b$ set of ICVs given in 
(\ref{initialvalues}), {\em i.e.}, with $\varepsilon=\varepsilon^b=0.5$.
We plot the numerical solutions of Eqs.~(\ref{mmineq}), 
already shown in the previous figures, and we compare with
these correct behaviours: i) the PE 1st order UAs (as given in (\ref{solun1}), 
already presented in [Fig. \ref{fig4}C], [Fig. \ref{fig4}D]); ii) the SPDERG 
2nd order UAs (as given in (\ref{solsrgu2}), (\ref{solcrgu2}), already 
presented in [Fig. \ref{fig6}C], [Fig. \ref{fig6}D]); iii) 
the refined SPDERG 2nd order UAs  (as given in (\ref{solrgu2r}), already 
presented in [Fig. \ref{fig7}C], [Fig. \ref{fig7}D]). All these UAs to the 
correct solutions are computed by means of the same standard numerical 
approximation for the Lambert function. When they belong to the relevant time 
window, we finally plot our corresponding rough evaluations of the two 
different time scales involved, too, with $\tau^b_s$ describing the substrate 
decay time and $\tau^b_c$ the complex saturation time.}}
\label{fig8}
\end{center}
\end{figure}

\noindent
We present in [Fig. \ref{fig8}] the detailed time depending behaviours of the
substrate concentration, $s(t)$, and of the complex one, $c(t)$, for the
more demanding ICV case with $\varepsilon=\varepsilon^b=0.5$, in the
two relevant parts of the time window, by comparing the different
{\em best} approximations we considered:  i) the PE 1st order UAs
(as given in (\ref{solun1}), already presented in [Fig. \ref{fig4}C], 
[Fig. \ref{fig4}D]); ii) the SPDERG 2nd order UAs (as given in (\ref{solsrgu2}),
(\ref{solcrgu2}), already presented in [Fig. \ref{fig6}C], [Fig. \ref{fig6}D]); 
iii) the refined SPDERG 2nd order UAs (as given in (\ref{solrgu2r}), already 
presented in [Fig. \ref{fig7}C], [Fig. \ref{fig7}D]).

Here, we neglect first of all the initial time interval, up to $t= 0.03s$ for
the substrate and to $t= 0.08 s$ for the complex, respectively. 
Indeed, in this interval, the different results are indistinguishable, 
within our numerical precision, both each other 
and with the correct numerical solutions. 
Instead, we consider in detail, as usual in logarithmic time scale, the central 
intervals,  {\em i.e.}, the ones that encompass the matching region. These 
intervals are  $t \in [0.03,3.5]s$ for the substrate
 and  $t \in [0.08,2.5]s$ for the complex, respectively. Finally, we present 
in non logarithmic time scale the relevant large time window,
[Fig. \ref{fig8}B] and  [Fig. \ref{fig8}D], for the substrate and the 
complex, respectively. In fact, this window ranges up to $t \sim 12 s$ for the 
substrate and up to $t \sim 14 s$ for the complex, since at larger times 
(as it is clear from [Fig. \ref{fig8}B] in particular) the solutions have 
already reached, within our numerical precision, their asymptotic values 
(that are different from zero in the case of the non refined SPDERG).
 
Interestingly, the curves shown in [Fig. \ref{fig8}A] not only make once more 
evident that, in the case of the substrate, the SPDERG 
2nd order UAs we considered works definitely better than the PE 1st order ones,
both with and without the refinement, but it also allows to naked-eye evaluate 
the range in which this happens. In fact, this range corresponds to
$t \sim 0.05 \div 2 s$, hence it covers about the  $15 \%$ of the whole
relevant time window in the case without refinement, whereas 
it corresponds to $t \sim 0.05 \div 3 s$ (with the larger time being of the
same order of $\tau_s^b$), and hence it covers about the 
$25 \%$ of the whole relevant time window, in the case with the refinement. 
On the other hand, for $t \simg 3s$ ([Fig. \ref{fig8}B]), the refined SPDERG 
2nd order UA appears to tend to zero slightly too rapidly, with respect both 
to the correct solution and to the PE 1st order result, though it makes  
a smaller error than the same SPDERG approximation without the refinement.

In the case of the complex, as already anticipated, the curves shown in 
[Fig. \ref{fig8}C] make clear that the PE 1st order UA slightly over-evaluates 
the value of the maximum of this quantity. In fact, it appears to 
over-evaluate the dynamical behaviour of this quantity in the whole range 
$t \sim 0.25 \div 0.7 s$, that is a time interval roughly centered
around the maximum abscissa. On the other hand, both of the  
SPDERG 2nd order UAs that we considered are successful in correctly 
capturing the maximum value of the complex numerical solution.  
In particular, the refined SPDERG approximation turns out
to be the one more in agreement with the correct result
up to the larger time $t \sim 2 s$, {\em i.e.}, 
in about the $15 \%$ of the whole relevant time window. 
It is  moreover both as much correct as the 
1st order result of the standard method (though
going to zero slightly more rapidly than the correct solution) and definitely 
better than the same approximation without refinement at large times. In 
particular, this is true for $t \sim 9 \div 14 s$, that
covers about a remaining $35 \%$ of the whole relevant time window. 
Nevertheless, the refined SPDERG approximation turns out to be the one that
makes the largest error, by over-evaluating the correct complex
concentration behaviour, in the remaining part of the
relevant time window  (in detail, for $t \sim 2.5 \div 6.5 s$).

Thus, the refined SPDERG 2nd order UAs we proposed successfully 
capture the correct dynamical behaviour in a large part of the relevant time 
window, despite of the considered case being very demanding. First of all, 
this further confirms both the correctness and the utility of the 
SPDERG approach in general. Moreover, these findings support the correctness 
of the present proposed way for obtaining asymptotically vanishing solutions, 
that exploits as much as possible the results, too. At least within our kind of 
SPDERG application procedure, the proposed refined SPDERG UAs appear easily
generalizable to other similar cases. At the same time, the present analysis 
makes clear that the remaining part, to be analytically calculated, of the 
2nd order outer contributions would be important for 
an approximation to MM kinetics, beyond the sQSSA, 
that could work in the whole relevant time window, whatever the kinetic 
constants are, for values of the expansion parameter as large as 
$\varepsilon=e_0/s_0 \sim 0.5$.

\section{Conclusions}
\label{conc}
In the present work, we start by recalling the standard PE method in the case 
of MM kinetics, beyond the sQSSA \cite{BeBeDAPe,MaMo,LiSe,HeTsAr}, {\em i.e.}, 
the method that is generally used to deal with the problem. Against this 
background, we are able to successfully apply to this case 
the alternative SPDERG approach to boundary layer problems 
\cite{ChGoOo1,ChGoOo2}, by clarifying similarities and differences.

The procedure that we choose for applying the approach makes use of a basic 
observation in \cite{ChGoOo2}, that the bare quantities to be effectively 
renormalized are the ICVs of the problem (in fact, here, the substrate 
one). By starting from the 1st order ODEs to be obeyed by the 
inner solutions, with ICVs given at a generic time, and by performing the 
calculations up to the 2nd order, besides generally outlining the 
contribution of the single terms in the functions to the behaviours 
captured by the obtained approximations, we are able to 
show that one gets exactly the same outer component, as in the standard 
PE 1st order UA, for the substrate (thus, as expected, also for the complex). 

First of all, this result was not to be taken for
granted. In fact, in the context of the boundary layer problems studied in the 
original works \cite{ChGoOo2}, the approach turned out to be able to possibly 
give not exactly the same terms as in the corresponding PE UA,
but anyway a better approximation in the matching region.
Therefore, we interpret the outcome of the present study
also as a further confirmation, from a very different 
point of view, of the correctness of the known PE results up to 1st order. 

At the same time, within our application, the SPDERG approach turns out  
to make possible to correctly manage both the secular terms and the constant 
ones, without the imposition of MCs (that in the present case, 
already at the 1st order, need in fact to be two term conditions, that involve 
the first derivatives of the outer solutions, too \cite{LiSe}). 

Actually, one can see an analogy between the present approach and the MCs of 
the standard method, since the key ingredient here, 
{\em i.e.}, the fact that the renormalized part of the solution has to be
 independent of the arbitrary time $\lambda$, appears similar to the
imposition of the MCs at an unknown time. Then, one notices that, when imposing 
the MCs in the standard PE method, the one for the substrate needs to apply 
to the complex, too. Thus, within this context, it appears expectable 
that, as we verified, the ODE to be obeyed by the renormalized 
substrate ICV is the only one to be obtained. In detail, we 
verified that the one found from the study of the substrate is 
the same as the one found in the case of the complex.

On the one hand, within the SPDERG approach, it is necessary 
to perform the calculations up to the 2nd order for recovering the
1st order outer contribution of the standard PE method. On the other
hand, this allowed us to present the 2nd order contribution of the inner 
solutions for the first time to our knowledge.
Moreover, this also allowed us to hypothesize, by assuming
that the observed solution behaviours could be iterated, that 
the constant terms at a given order, not to be renormalized, play the role of a
part of the outer component at the following order. Indeed, this 
appears equivalent to both the imposition of the first term of the MC and the 
need for cancelling one of the constant terms that appear twice in the UA, 
within the PE. In fact, it is reasonable to expect that this is a general 
characteristic of the SPDERG approach to boundary layer problems, at least in 
the present kind of application procedure, in which one directly renormalizes 
the bare ICVs. In particular, this observation allowed us 
to propose refined SPDERG 2nd order UAs, that contain the parts of the
2nd order outer components that are predictable on the basis of our
hypothesis, and that are thus asymptotically vanishing, too. 

The conclusive comparison among the best different approximations
that we considered, in the more demanding case that we studied (the
one with the larger value of the expansion parameter, 
$\varepsilon=\varepsilon^b=0.5$), shows that the time region in which the 
SPDERG approach at the 2nd order works better than the PE method at the 1st 
order, definitely encompasses the matching region. In fact, the obtained SPDERG
approximations are nearly indistinguishable, within our numerical
precision, from the correct solutions of the problem in about the first
$15 \%$ of the relevant time window. In detail, the extension of this region is 
slightly different for the substrate and for the complex,
and it is in both of the cases larger when considering the refined 
UAs. Moreover, in the case of the complex, this last refined approximation 
works at least as well as the PE 1st order UAs in a part of the 
relevant time window that also includes the large times.

Actually, we studied particular demanding cases of MM kinetics. In fact, with 
the present kinetic constant choice, in logarithmic time scale, the curve for 
the substrate displays three inflection points instead of a single one 
already for values of the expansion parameter as small as $\varepsilon=0.1$. 
This observation appears related to an ICV for the sQSSA to the complex
behaviour that, apart from being as usual larger than zero, is
even larger than the maximum reached by the correct solution for this
quantity. Indeed, these qualitative observations would deserve a more 
careful study, that could allow a better understanding of the 
parameter's dependence of the relevant time scales in MM kinetics.

From a different point of view, an interesting advantage of the
present procedure for applying the SPDERG approach is that,  
at the 2nd order, one needs to solve a simpler ODE for 
the 1st order outer component than the corresponding one in the standard 
method (just because the other part of the outer contribution
is already known). This could turn out to be particularly useful  
when attempting to apply the same procedure within the different
framework of the tQSSA since, despite of this framework being more
largely applicable from the experimental point of view, 
the outer solution is not known explicitly at the 1st order
and at the 0th order (in any event, it is not known in terms of the Lambert 
function) \cite{TzEd,DABe}.
 
Finally, we notice that, despite of their being quite technical and  
cumbersome, there is no particular difficulty in the present calculations.
The performed detailed verifications, apart from being possibly
better formalized in the future, can be in any event taken for granted in 
other similar cases, too. Thus, the present results definitely support 
further applications of the SPDERG approach to boundary layer problems. 

\section*{Acknowledgments}
The analytical calculations have been partially made with 
the help of {\em Mathematica$^\circledR$}, whereas the numerical computations 
have been made by using {\em Matlab$^\circledR$}. We acknowledge 
interesting discussions with Emanuele Raccah, Gavriel Segre and 
Pierluigi Vellucci. 

\section*{Appendix A}
{\footnotesize
We report the solution $\tilde{c}^{rg}_1(\tau)$ of the system (\ref{rg1}),
{\em i.e.}, the 1st order inner solution for the adimensional complex
concentration in MM kinetics within the SPDERG approach, 
in which the initial values, to be renormalized, are fixed to 
$\tilde{s}^{rg}(\tau_0)=\tilde{s}^{*}_0+\varepsilon \tilde{s}^{*}_1$  and
$\tilde{c}^{rg}(\tau_0)=\tilde{c}^{*}_0+\varepsilon \tilde{c}^{*}_1$,
at $\tau=\tau_0$. In detail, we distinguish the terms 
corresponding to different functions of $(\tau-\tau_0)$. We find:
\begin{eqnarray}
\tilde{c}^{rg}_1(\tau)&=&
A_{{c}^{rg}_1}(\tilde{s}^*_0) (\tau-\tau_0)+B_{{c}^{rg}_1}(\tilde{s}^*_0,
\tilde{s}^*_1,\tilde{c}^*_0) +C_{{c}^{rg}_1}(\tilde{s}^*_0,
\tilde{s}^*_1,\tilde{c}^*_0,\tilde{c}^*_1)
e^{\textstyle -(\tilde{s}^*_0+M)(\tau-\tau_0)}+
\nonumber  \\
&+&\left [ D_{{c}^{rg}_1}(\tilde{s}^*_0,\tilde{s}^*_1,\tilde{c}^*_0) 
(\tau-\tau_0) + E _{{c}^{rg}_1}(\tilde{s}^*_0,\tilde{c}^*_0) (\tau-\tau_0)^2
\right ]e^{\textstyle -(\tilde{s}^*_0+M)(\tau-\tau_0)}+
\nonumber  \\
&+&F_{{c}^{rg}_1}(\tilde{s}^*_0,\tilde{c}^*_0)
e^{\textstyle -2(\tilde{s}^*_0+M)(\tau-\tau_0)},
\label{solcrg1}
\end{eqnarray}
with:
\begin{eqnarray}
A_{{c}^{rg}_1}(\tilde{s}^*_0)&=&
-\frac{\textstyle M(M-m)}
{\textstyle \left (\tilde{s}^*_0 +M \right )^3} \tilde{s}^*_0;
\nonumber \\
B_{{c}^{rg}_1}(\tilde{s}^*_0,\tilde{s}^*_1,\tilde{c}^*_0)&=&
-\frac{\textstyle M(\tilde{s}^*_0-M+2m)}
{\textstyle \left (\tilde{s}^*_0 +M \right )^4}
\tilde{s}^*_0+
\frac{\textstyle M}{\textstyle \left (\tilde{s}^*_0 +M \right )^2}
\tilde{s}^*_1+
\frac{\textstyle M(\tilde{s}^*_0+m)}
{\textstyle \left (\tilde{s}^*_0 +M \right )^3}\tilde{c}^*_0;
\nonumber \\
C_{{c}^{rg}_1}(\tilde{s}^*_0,\tilde{s}^*_1,\tilde{c}^*_0,\tilde{c}^*_1)&=&
\tilde{c}^*_1+\frac{\textstyle M(\tilde{s}^*_0-M+2m)}
{\textstyle \left (\tilde{s}^*_0 +M \right )^4}\tilde{s}^*_0
+\frac{\textstyle \tilde{s}^*_0+m}
{\textstyle \left (\tilde{s}^*_0 +M \right )^4}(\tilde{s}^*_0)^2
-\frac{\textstyle M}
{\textstyle \left (\tilde{s}^*_0 +M \right )^2}\tilde{s}^*_1+ \nonumber \\
&-&\frac{\textstyle (2\tilde{s}^*_0+M)({s}^*_0+m)}
{\textstyle \left (\tilde{s}^*_0 +M \right )^3}\tilde{c}^*_0+
\frac{\textstyle \tilde{s}^*_0+m}
{\textstyle \left (\tilde{s}^*_0 +M \right )^2}(\tilde{c}^*_0)^2;
\nonumber \\
D_{{c}^{rg}_1}(\tilde{s}^*_0,\tilde{s}^*_1,\tilde{c}^*_0)&=& 
-\frac{\textstyle (\tilde{s}^*_0+m)(\tilde{s}^*_0 -M )}
{\textstyle \left (\tilde{s}^*_0 +M \right )^3}\tilde{s}^*_0+
\frac{\textstyle  \tilde{s}^*_0\tilde{s}^*_1}
{\textstyle \tilde{s}^*_0 +M }
-\frac{\textstyle \tilde{s}^*_1\tilde{c}^*_0}
{\textstyle \left (\tilde{s}^*_0 +M \right )^2}+
\frac{\textstyle (\tilde{s}^*_0+m)(2\tilde{s}^*_0-M)}
{\textstyle \left (\tilde{s}^*_0 +M \right )^2}\tilde{c}^*_0+\nonumber \\
&-&\frac{\textstyle \tilde{s}^*_0+m}
{\textstyle \left (\tilde{s}^*_0 +M \right )^2}(\tilde{c}^*_0)^2;
\nonumber \\
E _{{c}^{rg}_1}(\tilde{s}^*_0,\tilde{c}^*_0)&=&
-\frac{\textstyle (M-m)}
{\textstyle 2 \left (\tilde{s}^*_0 +M \right )^2}(\tilde{s}^*_0)^2+
\frac{\textstyle (M-m)}
{\textstyle 2 (\tilde{s}^*_0 +M)}\tilde{s}^*_0\tilde{c}^*_0;
\nonumber \\
F_{{c}^{rg}_1}(\tilde{s}^*_0,\tilde{c}^*_0)&=&
-\frac{\textstyle \tilde{s}^*_0+m}
{\textstyle \left (\tilde{s}^*_0 +M \right )^4}(\tilde{s}^*_0)^2+
2\frac{\textstyle \tilde{s}^*_0+m}
{\textstyle \left (\tilde{s}^*_0 +M \right )^3}\tilde{s}^*_0\tilde{c}^*_0
-\frac{\textstyle \tilde{s}^*_0+m}
{\textstyle \left (\tilde{s}^*_0 +M \right )^2}(\tilde{c}^*_0)^2.
\label{coeffsolcrg1}
\end{eqnarray}
In particular, one can check that one has 
$\tilde{c}^{rg}_1(\tau_0)=B_{{c}^{rg}_1}+C_{{c}^{rg}_1}+F_{{c}^{rg}_1}=\tilde{c}^*_1$.
Moreover, for $\tilde{s}^*_0=1$, $\tilde{s}^*_1=\tilde{c}^*_0=\tilde{c}^*_1=0$ 
and $\tau_0=0$, the result on $\tilde{c}^{in}_1$ reported in (\ref{solinner1})
is correctly reproduced.}

\section*{Appendix B}
{\footnotesize
We report the solutions $\tilde{s}^{rg}_2(\tau)$ and $\tilde{c}^{rg}_2(\tau)$ 
of the system (\ref{rg2}), {\em i.e.}, the 2nd order inner solution for the 
adimensional substrate and complex concentrations in MM kinetics within the  
SPDERG approach, in which the ICVs are anyway already fixed, for the sake of 
simplicity, to 
$\tilde{s}^{rg}(\tau_0)=\tilde{s}^{*}_0+\varepsilon \tilde{s}^{*}_1$  and
$\tilde{c}^{rg}(\tau_0)=0$, at $\tau=\tau_0$. As discussed in the text,
this choice is done on the basis of the expectation, to be verified,
that the only quantities that need to be renormalized at the present order
are $\tilde{s}^{*}_0$ (in agreement with the previously obtained
result) and $\tilde{s}^{*}_1$.

Let us start with $\tilde{s}^{rg}_2(\tau)$. Notice that here we separate
both the terms that are different functions of $(\tau-\tau_0)$ and 
the terms that depend only on $\tilde{s}^{*}_0$ or both on $\tilde{s}^{*}_0$
and on $\tilde{s}^{*}_1$:
\begin{eqnarray}
\tilde{s}^{rg}_2(\tau)&=&
\left  [ A_{{s}^{rg}_2}({\tilde{s}^*_0})
+B_{{s}^{rg}_2}({\tilde{s}^*_0},{\tilde{s}^*_1})\right  ] (\tau-\tau_0)+
C_{{s}^{rg}_2}({\tilde{s}^*_0}) (\tau-\tau_0)^2 +  
D_{{s}^{rg}_2}({\tilde{s}^*_0})+
E_{{s}^{rg}_2}({\tilde{s}^*_0},{\tilde{s}^*_1})+\nonumber \\
&+&
\left \{ \left [ F_{{s}^{rg}_2}({\tilde{s}^*_0})
+G_{{s}^{rg}_2}({\tilde{s}^*_0},{\tilde{s}^*_1})\right  ] + 
\left  [ H_{{s}^{rg}_2}({\tilde{s}^*_0}) 
+I_{{s}^{rg}_2}({\tilde{s}^*_0},{\tilde{s}^*_1})\right  ] (\tau-\tau_0)
\right \}
e^{\textstyle- ({\tilde{s}^*_0}+M) ({\tau}-{\tau_0})} + \nonumber \\
&+& J_{{s}^{rg}_2}({\tilde{s}^*_0}) (\tau-\tau_0)^2
e^{\textstyle- ({\tilde{s}^*_0}+M) ({\tau}-{\tau_0})} + 
K_{{s}^{rg}_2}({\tilde{s}^*_0})
e^{\textstyle-2 ({\tilde{s}^*_0}+M) ({\tau}-{\tau_0})}.
\label{solsrg2}
\end{eqnarray}
In fact, only the first three terms contribute to the part to be renormalized
of the whole functions, $\tilde{s}^{rg}_{div}(\tau)$, at the 2nd order. 
Their coefficients are:
\begin{eqnarray}
A_{{s}^{rg}_2}({\tilde{s}^*_0})&=&
\frac{\textstyle 2 M (M-m) {\tilde{s}^*_0}}{\textstyle ({\tilde{s}^*_0}+M)^4}
({\tilde{s}^*_0}+m); \hspace{.5in}
B_{{s}^{rg}_2}({\tilde{s}^*_0},{\tilde{s}^*_1})=
-\frac{\textstyle  M (M-m) {\tilde{s}^*_1}}{\textstyle ({\tilde{s}^*_0}+M)^2};
\nonumber \\ 
C_{{s}^{rg}_2}({\tilde{s}^*_0})&=&
\frac{\textstyle M (M-m)^2 {\tilde{s}^*_0}}{\textstyle 2({\tilde{s}^*_0}+M)^3}.
\label{coeffs2rgdiv}
\end{eqnarray}

For the sake of completeness, we give the explicit dependence on
${\tilde{s}^*_0}$ and ${\tilde{s}^*_1}$ also of the other terms. One finds:
\begin{eqnarray}
D_{{s}^{rg}_2}({\tilde{s}^*_0})&=&
\frac{\textstyle {\tilde{s}^*_0}}{\textstyle 2 ({\tilde{s}^*_0}+M)^5}
\left [({\tilde{s}^*_0})^2 (3M-m)-{\tilde{s}^*_0}(2M^2-5Mm-m^2)
-2Mm(2M-3m) \right  ]; \nonumber \\
E_{{s}^{rg}_2}({\tilde{s}^*_0},{\tilde{s}^*_1})&=&
-G_{{s}^{rg}_2}({\tilde{s}^*_0},{\tilde{s}^*_1})=
-\frac{\textstyle {\tilde{s}^*_1}}{\textstyle ({\tilde{s}^*_0}+M)^3}
\left  [  {\tilde{s}^*_0} (2M-m)+Mm \right  ]; \nonumber \\
F_{{s}^{rg}_2}({\tilde{s}^*_0})&=&
-\frac{\textstyle {\tilde{s}^*_0} }{\textstyle ({\tilde{s}^*_0}+M)^5}
\left  [ ({\tilde{s}^*_0})^3+({\tilde{s}^*_0})^2(2M+m)  -
{\tilde{s}^*_0}(M^2-3Mm-m^2)-Mm(2M-3m) \right]; 
\nonumber \\ 
H_{{s}^{rg}_2}({\tilde{s}^*_0})&=&
\frac{\textstyle {\tilde{s}^*_0} }{\textstyle ({\tilde{s}^*_0}+M)^4}
\left  [ ({\tilde{s}^*_0})^3-({\tilde{s}^*_0})^2(M-2m)  
-{\tilde{s}^*_0}M^2   - Mm^2 \right ]; 
\nonumber 
\\ 
I_{{s}^{rg}_2}({\tilde{s}^*_0},{\tilde{s}^*_1})&=&
-\frac{\textstyle {\tilde{s}^*_0}{\tilde{s}^*_1}}
{\textstyle ({\tilde{s}^*_0}+M)^2}({\tilde{s}^*_0}+m);
\hspace{.5in}
J_{{s}^{rg}_2}({\tilde{s}^*_0})=
\frac{\textstyle ({\tilde{s}^*_0})^2(M-m) }{\textstyle 2 ({\tilde{s}^*_0}+M)^3}
({\tilde{s}^*_0}+m); 
\nonumber \\ 
K_{{s}^{rg}_2}({\tilde{s}^*_0})&=&
\frac{\textstyle ({\tilde{s}^*_0})^2({\tilde{s}^*_0}+m)}
{\textstyle 2 ({\tilde{s}^*_0}+M)^5} (2 {\tilde{s}^*_0}+M+m). 
\label{coeffsolsrg2}
\end{eqnarray}
One can check that $\tilde{s}^{rg}_2(\tau_0)=D_{{s}^{rg}_2}+E_{{s}^{rg}_2}+
F_{{s}^{rg}_2}+G_{{s}^{rg}_2}+K_{{s}^{rg}_2}=0$. Notice moreover that, when 
calculating the ${\cal R}^s_2(\tau)$ contribution to the SPDERG 2nd order UA 
to the correct solution, with the renormalized divergent part, one is 
interested in evaluating this quantity in $\tau_0=0$, for $\tilde{s}^*_0=1$ and 
$\tilde{s}^*_1=0$. Hence, one immediately gets 
$E_{{s}^{rg}_2}=G_{{s}^{rg}_2}=I_{{s}^{rg}_2}=0$, since all of these coefficients are 
proportional to $\tilde{s}^*_1$. The other coefficients can be easily 
calculated for $\tilde{s}^*_0=1$, and the corresponding ${\cal R}^s_2$ is 
reported in (\ref{rs2last}).
 
In the case of $\tilde{c}^{rg}_2(\tau)$, since the complete formula is 
definitely uselessly cumbersome, we limit ourselves 
to report explicitly the dependence on ${\tilde{s}^*_0}$ and ${\tilde{s}^*_1}$
only in the part of the whole function that belongs to the one to be 
renormalized ({\em i.e.}, in the coefficients of the terms that are
proportional to $(\tau-\tau_0)$ and to $(\tau-\tau_0)^2$). 

Correspondingly, we write from the beginning:
\begin{equation}
\tilde{c}^{rg}_2(\tau)= 
\tilde{c}^{rg}_{2,div}(\tau)+{\cal R}^c_2(\tau).
\label{crg2newform}
\end{equation}
Here, we are collecting in $\tilde{c}^{rg}_{2,div}(\tau)$ the terms that 
contribute to the part of the 2nd order complex function to be renormalized:
\begin{equation}
\tilde{c}^{rg}_{2,div}(\tau)=
\left [ A_{{c}^{rg}_{2,div}}({\tilde{s}^*_0})+
B_{{c}^{rg}_{2,div}}({\tilde{s}^*_0},{\tilde{s}^*_1}) \right ] (\tau-\tau_0)
+ C_{{c}^{rg}_{2,div}}({\tilde{s}^*_0})(\tau-\tau_0)^2,
\label{solcrg2div}
\end{equation}
with, in detail:
\begin{eqnarray}
A_{{c}^{rg}_{2,div}}({\tilde{s}^*_0})&=&
-\frac{\textstyle M (M-m){\tilde{s}^*_0}}{\textstyle ({\tilde{s}^*_0}+M)^6}
\left [  2 ({\tilde{s}^*_0})^2-5{\tilde{s}^*_0}(M-m)+M^2-3Mm \right]; 
\nonumber \\
B_{{c}^{rg}_{2,div}}({\tilde{s}^*_0},{\tilde{s}^*_1})&=&
\frac{\textstyle M (M-m){\tilde{s}^*_1}}{\textstyle ({\tilde{s}^*_0}+M)^4}
(2{\tilde{s}^*_0}-M); \hspace{.3in}
C_{{c}^{rg}_{2,div}}({\tilde{s}^*_0})=
-\frac{\textstyle M (M-m)^2{\tilde{s}^*_0}}{\textstyle 2({\tilde{s}^*_0}+M)^5}
(2{\tilde{s}^*_0}-M).
\label{coeffsolcrg2div}
\end{eqnarray}

On the other hand, ${\cal R}^c_2(\tau)$ contains all the terms
that remain constant or tend to zero in the
large $(\tau -\tau_0)$ limit, and it gives the 2nd order contribution
of the inner solution to the SPDERG UA to the correct
complex function (that is $O(\varepsilon^2)$). 
For $\tilde{s}^*_0=1$ and $\tilde{s}^*_1=0$, 
by also coherently evaluating it in $\tau_0=0$, one finds:
\begin{eqnarray}
{\cal R}^c_2(\tau) \left \lvert_{\!\!\shortmid_{\!\shortmid_{\tilde{s}^*_1=0}^{\tilde{s}^{*}_0=1}}}
\right. &=&
A_{{\cal R}^c_2}+ 
\left [ B_{{\cal R}^c_2}+C_{{\cal R}^c_2}\tau+
D_{{\cal R}^c_2}\tau^2+ E_{{\cal R}^c_2}\tau^3+F_{{\cal R}^c_2} \tau^4
\right ] e^{\textstyle -(1+M) {\tau}}+  \nonumber \\ 
&+& \left [G_{{\cal R}^c_2}+
H_{{\cal R}^c_2}\tau+I_{{\cal R}^c_2}\tau^2
\right ] e^{\textstyle-2 (1+M) {\tau}}+
J_{{\cal R}^c_2} e^{\textstyle-3 (1+M) {\tau}}.
\label{solrcrg2}
\end{eqnarray}

In fact, it is useful to report also the complete dependence on 
$\tilde{s}^*_0$, $\tilde{s}^*_1$, $M$ and $m$ of the coefficient
$A_{{\cal R}^c_2}$ (the constant term in the original $\tilde{c}^{rg}_2(\tau)$),
since it allows to propose a refined SPDERG 2nd order UA to the correct 
solution. We find:
\begin{eqnarray}
A_{{\cal R}^c_2}(\tilde{s}^*_0,\tilde{s}^*_1)&=&
-\frac{\textstyle M(\tilde{s}^*_0)^4}{\textstyle ({\tilde{s}^*_0}+M)^7}
+\frac{\textstyle M(\tilde{s}^*_0)^3(9M-11m)}{\textstyle 2({\tilde{s}^*_0}+M)^7}
-\frac{\textstyle M(\tilde{s}^*_0)^2}{\textstyle 2({\tilde{s}^*_0}+M)^7}
(12M^2-27Mm+13m^2)+
\nonumber \\ 
&+&\frac{\textstyle M^2\tilde{s}^*_0(M^2-6mM+6m^2)}
{\textstyle ({\tilde{s}^*_0}+M)^7}+
\frac{\textstyle M\tilde{s}^*_1}{\textstyle ({\tilde{s}^*_0}+M)^5}
[2(\tilde{s}^*_0)^2-\tilde{s}^*_0(5M-6m)+M(1-2m)]+
\nonumber \\ 
&-&\frac{\textstyle M(\tilde{s}^*_1)^2}{\textstyle ({\tilde{s}^*_0}+M)^3},
\label{arcrg2}
\end{eqnarray}
whereas the dependence on $M$ and $m$ of all the coefficients, 
calculated in $\tilde{s}^*_0=1$ and $\tilde{s}^*_1=0$, is given by:
\begin{eqnarray}
A_{{\cal R}^c_2}&=&
\frac{\textstyle M^4-6M^3(m+1)}{\textstyle (1+M)^7}+
\frac{\textstyle 3M^2(4m^2+9m+3)}{\textstyle 2(1+M)^7} 
-\frac{\textstyle M(13m^2+11m+2)}{\textstyle 2(1+M)^7} ; \nonumber \\
B_{{\cal R}^c_2}&=&
-\frac{\textstyle M^4-6M^3(m+1)}{\textstyle (1+M)^7}
-\frac{\textstyle M^2(6m^2+10m+3)}{\textstyle (1+M)^7} 
+\frac{\textstyle M(m-7)-9m^2-m}{\textstyle 4(1+M)^7} ;
\nonumber \\
C_{{\cal R}^c_2}&=&
\frac{\textstyle M^3(2m+1)}{\textstyle (1+M)^6}
-\frac{\textstyle M^2(3m^2+7m+5)}{\textstyle (1+M)^6}
+\frac{\textstyle M(12m^2+15m+5)}{\textstyle 2(1+M)^6}
+\frac{\textstyle 3m^2+3m+2}{\textstyle 2(1+M)^6};
\nonumber \\
D_{{\cal R}^c_2}&=&
-\frac{\textstyle 1}{\textstyle 2(1+M)^5}
[ 2M^3+M^2(m^2-6m-3)+ M(4m^2-m-3)+ 2m^2+3m+1];
 \nonumber \\
E_{{\cal R}^c_2}&=&
\frac{\textstyle (M-m)}{\textstyle 6(1+M)^4}
[M^2+M(2m+3)-3(m+1)]; \hspace{.3in}
F_{{\cal R}^c_2}=
-\frac{\textstyle (M-m)^2}{\textstyle 8(1+M)^3}; \nonumber \\
G_{{\cal R}^c_2}&=&
-\frac{\textstyle M^2(7m+3)}{\textstyle 2(1+M)^7} 
+\frac{\textstyle M(13m^2+11m+6)}{\textstyle 2(1+M)^7} 
+\frac{\textstyle 3m^2+2m+1}{\textstyle (1+M)^7} ; \nonumber \\
H_{{\cal R}^c_2}&=&
\frac{\textstyle 1}{\textstyle (1+M)^6} [M^2+ M(2m^2+m+1)-3m-2]; 
\nonumber \\
I_{{\cal R}^c_2}&=&
-\frac{\textstyle (M-m)}{\textstyle (1+M)^5} (m+1); 
\hspace{.3in}
J_{{\cal R}^c_2}=
-\frac{\textstyle (m+1)}{\textstyle 4(1+M)^7}(M+3m+4).
\label{coeffsolrcrg2}
\end{eqnarray}
One can check that ${\cal R}^c_2(0)=A_{{\cal R}^c_2}+
B_{{\cal R}^c_2}+G_{{\cal R}^c_2}+J_{{\cal R}^c_2}=0$, correctly.
}

\section*{Appendix C}
{\footnotesize 
Let us relabel $T_1(\tilde{s}^*_0)$, $T_2(\tilde{s}^*_0)$, 
$T_3(\tilde{s}^*_0,\tilde{s}^*_1)$ and $T_4(\tilde{s}^*_0)$
the coefficients of the first four terms in $\tilde{c}^{rg}_{div}(\tau)$
at the 2nd order given in (\ref{cdivrgII}). Hence, one has:
\begin{eqnarray}
\tilde{c}^{rg}_{div}(\tau)&=&T_1(\tilde{s}^*_0)
+\varepsilon T_2(\tilde{s}^*_0)
+\varepsilon T_3(\tilde{s}^*_0,\tilde{s}^*_1)
+\varepsilon T_4(\tilde{s}^*_0)(\tau-\tau_0)
+ \nonumber \\ 
&+&\varepsilon^2 A_{{c}^{rg}_{2,div}}({\tilde{s}^*_0}) (\tau-\tau_0)
+\varepsilon^2 B_{{c}^{rg}_{2,div}}({\tilde{s}^*_0},{\tilde{s}^*_1}) (\tau-\tau_0)
+\varepsilon^2 C_{{c}^{rg}_{2,div}}({\tilde{s}^*_0}) (\tau-\tau_0)^2,
\label{cdivrgIIa}
\end{eqnarray}
with:
\begin{eqnarray}
T_1(\tilde{s}^*_0)&=&
\frac{\textstyle \tilde{s}^*_0}
{\textstyle \tilde{s}^*_0+M};
\hspace{.5in}
T_2(\tilde{s}^*_0)=-
\frac{\textstyle M(\tilde{s}^*_0-M+2m)}
{\textstyle \left (\tilde{s}^*_0 +M \right )^4}
\tilde{s}^*_0; \nonumber \\ 
T_3(\tilde{s}^*_0,\tilde{s}^*_1)&=&\frac{\textstyle M\tilde{s}^*_1}
{\textstyle \left (\tilde{s}^*_0 +M \right )^2}; 
\hspace{.3in}
T_4(\tilde{s}^*_0)=-\frac{\textstyle M  (M-m) \tilde{s}^*_0}
{\textstyle \left (\tilde{s}^*_0 +M \right )^3};
\end{eqnarray}
Here, in detail: $T_1$ is the coefficient of the 0th order term
that was already present in $\tilde{c}^{rg}_{0}(\tau)$ given in 
(\ref{solrg0}); $T_2+T_3=B_{{c}^{rg}_1}$ with $B_{{c}^{rg}_1}$ (calculated
in $\tilde{c}^{*}_0=0$) the constant term
in $\tilde{c}^{rg}_{1}(\tau)$ given in (\ref{coeffsolcrg1}); 
$T_4=A_{{c}^{rg}_1}$, with $A_{{c}^{rg}_1}$ the coefficient of the single
1st order secular term in  $\tilde{c}^{rg}_{1}(\tau)$ given again 
in (\ref{coeffsolcrg1}); $A_{{c}^{rg}_{2,div}}$, 
$B_{{c}^{rg}_{2,div}}$ and $C_{{c}^{rg}_{2,div}}$ are the coefficients of the 
three 2nd order secular terms in $\tilde{c}^{rg}_{2}(\tau)$
given in  (\ref{coeffsolcrg2div}), respectively.

When renormalizing the bare constants by $\tilde{s}^*_0 = 
(1+\varepsilon z_{s_0,1}+
\varepsilon^2 z_{s_0,2})\tilde{s}^{rg *}_0$ and 
$\tilde{s}^*_1 = (1+\varepsilon z_{s_0,1})\tilde{s}^{rg *}_1$, respectively
(with the already chosen $z_{s_0,1}$ given in (\ref{zs_s0_1}) and
$z_{s_0,2}$, $z_{s_1,1}$ given in (\ref{zs_II})), one finds (up to order
$\varepsilon^2$):
\begin{eqnarray}
 T_1(\tilde{s}^*_0)- T_1(\tilde{s}^{rg *}_0)\!\hspace{-.1in}&=&\hspace{-.1in}\!
\left [ \frac{d T_1(\tilde{s}^{*}_0)}{d\tilde{s}^{*}_0} \left  
\lvert_{\!\!\shortmid_{\!\shortmid_{\tilde{s}^*_0=\tilde{s}^{rg *}_0}}} \right. \!\! \right ] \!
(\varepsilon z_{s_0,1}+
\varepsilon^2 z_{s_0,2})\tilde{s}^{rg *}_0 
+  \frac{1}{2}  
\left [ \frac{d^2 T_1(\tilde{s}^{*}_0)}{d(\tilde{s}^{*}_0)^2} 
\left  \lvert_{\!\!\shortmid_{\!\shortmid_{\tilde{s}^*_0=\tilde{s}^{rg *}_0}}} \right. \!\!  \right ] 
\! (\varepsilon z_{s_0,1})^2 (\tilde{s}^{rg *}_0)^2= \nonumber \\ 
\!\hspace{-.1in}&=&\hspace{-.1in}\! 
\varepsilon \frac{M(M\!-\!m)\tilde{s}^{rg *}_0}
{  \left ( \tilde{s}^{rg *}_0 \!\!+\!M \right )^3 }(\lambda\!-\!\tau_0)
-\varepsilon^2 Mm\frac{(M\!-\!m)\tilde{s}^{rg *}_0}
{ \left ( \tilde{s}^{rg *}_0 \!\!
+\!M\right )^5 }  (\lambda \!-\!\tau_0)+ 
\nonumber \\ 
\!\hspace{-.1in}&-&\hspace{-.1in}\!
\varepsilon^2 M \frac{(M\!-\!m)^2\tilde{s}^{rg *}_0}
{ 2 \left ( \tilde{s}^{rg *}_0 
\!\!+\!M \right )^5 } (2\tilde{s}^{rg *}_0\!\!-\!M)  (\lambda \!-\!\tau_0)^2= 
\nonumber \\  
\!\hspace{-.1in}&=&\hspace{-.1in}
\!-\varepsilon T_4(\tilde{s}^{rg *}_0) (\lambda\!-\!\tau_0) 
-\varepsilon^2 Mm\frac{(M\!-\!m)\tilde{s}^{rg *}_0}
{ \left ( \tilde{s}^{rg *}_0\!\! +M
\right )^5 }(\lambda\!-\!\tau_0) + 
\varepsilon^2  C_{{c}^{rg}_{2,div}}({\tilde{s}^{rg *}_0}) (\lambda\!-\!\tau_0)^2;
\label{firstterm}
\end{eqnarray}
\begin{eqnarray}
\varepsilon \left [T_2(\tilde{s}^*_0)- T_2(\tilde{s}^{rg *}_0)\right ]
\!\hspace{-.1in}&=&\hspace{-.1in}\!
\varepsilon \left [ \frac{d T_2(\tilde{s}^{*}_0)}{d\tilde{s}^{*}_0} \left  
\lvert_{\!\!\shortmid_{\!\shortmid_{\tilde{s}^*_0=\tilde{s}^{rg *}_0}}} \right. \!\! \right ] \!
(\varepsilon z_{s_0,1})\tilde{s}^{rg *}_0= \nonumber \\ 
\!\hspace{-.1in}&=&\hspace{-.1in}\!\varepsilon^2 
\frac{M(M\!-\!m)\tilde{s}^{rg *}_0 \!\!}{ \left  ( \tilde{s}^{rg *}_0 \!\!+\!M
\right )^6 }  \left  [2(\tilde{s}^{rg *}_0)^2\!-\!(5M\!-\!6m)\tilde{s}^{rg *}_0+
 M(M\!-\!2m)\right ](\lambda\!-\!\tau_0)=
\nonumber \\ 
\!\hspace{-.1in}&=&\hspace{-.1in}\! 
-\varepsilon^2 A_{{c}^{rg}_{2,div}}({\tilde{s}^{rg *}_0}) (\lambda\!-\!\tau_0)
+\varepsilon^2 Mm\frac{(M-m)\tilde{s}^{rg *}_0}{ \left ( \tilde{s}^{rg *}_0 +M
\right )^5 } (\lambda\!-\!\tau_0);
\label{secondterm}
\end{eqnarray}
\begin{eqnarray}
\varepsilon \left [T_3(\tilde{s}^*_0,\tilde{s}^*_1)\!-\! 
T_3(\tilde{s}^{rg *}_0\!\!,\tilde{s}^{rg *}_1)\right ]
\!\hspace{-.1in}&=&\hspace{-.1in}\!
\varepsilon\! \left [ \frac{\partial T_3(\tilde{s}^{*}_0,\tilde{s}^{*}_1)}
{\partial \tilde{s}^{*}_0} \left  
\lvert_{\!\!\shortmid_{\!\shortmid^{{\tilde{s}^*_0=\tilde{s}^{rg *}_0}}_{\tilde{s}^*_1=\tilde{s}^{rg *}_1}}} 
\right. \!\! \right ] \!\!
(\varepsilon z_{s_0,1})\tilde{s}^{rg *}_0 \!\!+\!
\varepsilon\! \left [ \frac{\partial T_3(\tilde{s}^{*}_0,\tilde{s}^{*}_1)}
{\partial \tilde{s}^{*}_1} \left  
\lvert_{\!\!\shortmid_{\!\shortmid^{{\tilde{s}^*_0=\tilde{s}^{rg *}_0}}_{\tilde{s}^*_1=\tilde{s}^{rg *}_1}}} 
\right. \!\! \right ] \!\!
(\varepsilon z_{s_1,1})\tilde{s}^{rg *}_1\!\!=
\nonumber \\ 
\!\hspace{-.1in}&=&\hspace{-.1in}\!
\varepsilon^2 \left [ - \frac{2M(M-m)\tilde{s}^{rg *}_1}
{ \left ( \tilde{s}^{rg *}_0 +M \right )^4 } \tilde{s}^{rg *}_0 
  + \frac{M^2(M-m)}
{ \left ( \tilde{s}^{rg *}_0 +M\right )^4 } \tilde{s}^{rg *}_1 
\right ](\lambda\!-\!\tau_0)=
\nonumber \\ 
\!\hspace{-.1in}&=&\hspace{-.1in}\!-\varepsilon^2 
B_{{c}^{rg}_{2,div}}({\tilde{s}^{rg *}_0},{\tilde{s}^{rg *}_1}) (\lambda\!-\!\tau_0);
\label{thirdterm}
\end{eqnarray}
\begin{eqnarray}
\varepsilon \left [T_4(\tilde{s}^*_0)- T_4(\tilde{s}^{rg *}_0)\right ]
\!\hspace{-.1in}&=&\hspace{-.1in}\!
\varepsilon \left [ \frac{d T_4(\tilde{s}^{*}_0)}{d\tilde{s}^{*}_0} \left  
\lvert_{\!\!\shortmid_{\!\shortmid_{\tilde{s}^*_0=\tilde{s}^{rg *}_0}}} \right. \!\! \right ] \!
(\varepsilon z_{s_0,1})\tilde{s}^{rg *}_0=\varepsilon^2 
\frac{M(M\!-\!m)^2\tilde{s}^{rg *}_0}{ \left ( \tilde{s}^{rg *}_0 \!\!+\!M
\right )^4 }(2\tilde{s}^{rg *}_0\!\!-\!M)(\lambda\!-\!\tau_0)=
\nonumber \\ 
\!\hspace{-.1in}&=&\hspace{-.1in}\!
-2 \varepsilon^2 C_{{c}^{rg}_{2,div}}({\tilde{s}^{rg *}_0})(\lambda\!-\!\tau_0).
\label{fourthterm}
\end{eqnarray}
Therefore, when moreover writing 
$(\tau-\tau_0)=(\tau-\lambda)+(\lambda-\tau_0)$,
since obviously 
$(\tau-\tau_0)^2=(\tau-\lambda)^2+2(\tau-\lambda)(\lambda-\tau_0)+
(\lambda-\tau_0)^2$, one gets exactly the same form of 
$\tilde{c}^{rg}_{div}(\tau,\lambda)$ given in (\ref{cdivrgII}), with
$\tau_0 \rightarrow \lambda$, $\tilde{s}^{*}_0\rightarrow 
\tilde{s}^{rg *}_0(\lambda)$
and $\tilde{s}^{*}_1\rightarrow \tilde{s}^{rg *}_1(\lambda)$.

Let us now also remind that $d{\tilde{s}^{rg *}_0}/{d\lambda}=-\varepsilon
z_{s_0,1}\tilde{s}^{rg *}_0/(\lambda-\tau_0)$. Correspondingly, one can
make partially use again of the previous formulas in the study of the 
derivative with respect to $\lambda$ of $\tilde{c}^{rg}_{div}(\tau,\lambda)$:
\begin{eqnarray}
\frac{d\tilde{c}^{rg}_{div}(\tau,\lambda)}{d\lambda}\!\!&=&\!\!
\frac{dT_1({\tilde{s}^{rg *}_0})}{d{\tilde{s}^{rg *}_0}}
\frac{d{\tilde{s}^{rg *}_0}}{d\lambda}
+\varepsilon \frac{dT_2({\tilde{s}^{rg *}_0})}{d{\tilde{s}^{rg *}_0}}
\frac{d{\tilde{s}^{rg *}_0}}{d\lambda}+
\varepsilon \frac{\partial T_3({\tilde{s}^{rg *}_0},{\tilde{s}^{rg *}_1})}
{\partial {\tilde{s}^{rg *}_0}}
\frac{d{\tilde{s}^{rg *}_0}}{d\lambda}+
\nonumber \\ 
\!\!&+&\!\!\varepsilon \frac{\partial T_3({\tilde{s}^{rg *}_0},
{\tilde{s}^{rg *}_1})}
{\partial {\tilde{s}^{rg *}_1}}
\frac{d{\tilde{s}^{rg *}_1}}{d\lambda}
+\varepsilon \frac{dT_4({\tilde{s}^{rg *}_0})}{d{\tilde{s}^{rg *}_0}}
\frac{d{\tilde{s}^{rg *}_0}}{d\lambda}(\tau-\lambda)-
\varepsilon T_4({\tilde{s}^{rg *}_0})+
\nonumber \\ 
\!\!&-&\!\!\varepsilon^2 A_{{c}^{rg}_{2,div}}({\tilde{s}^{rg *}_0})
-\varepsilon^2 B_{{c}^{rg}_{2,div}}({\tilde{s}^{rg *}_0},{\tilde{s}^{rg *}_1})
-2\varepsilon^2 C_{{c}^{rg}_{2,div}}({\tilde{s}^{rg *}_0})(\tau-\lambda).
\end{eqnarray}
In detail, one can notice first of all that, to satisfy 
$d\tilde{c}^{rg}_{div}(\tau,\lambda)/d\lambda=0$, the 1st order known result on 
$d{\tilde{s}^{rg *}_0}/{d\lambda}$ given in Eqs.~(\ref{scalingcondition1}) 
needs to be once again satisfied. From this point of view, see in particular
Eqs.~(\ref{firstterm}). Indeed, the contribution that comes from
the term ${dT_1({\tilde{s}^{rg *}_0})}/{d{\tilde{s}^{rg *}_0}}$ in the
same way as the first term in that formula and the one that is equal to 
$-\varepsilon T_4({\tilde{s}^{rg *}_0})$ are the only ones to be proportional 
to $\varepsilon$. Thus, they obviously need to cancel each other.
Then, the term involving the derivative of $T_2$ is partially cancelled by
the one proportional to $A_{{c}^{rg}_{2,div}}$, leaving a contribution equal to 
$-\varepsilon^2Mm{(M-m)\tilde{s}^{rg *}_0}/{(\tilde{s}^{rg *}_0 +M )^5 }$, 
as can be seen from Eqs.~(\ref{secondterm}). Moreover, the two terms 
proportional to $(\tau-\lambda)$ cancel  each other, too, as can be seen from
Eqs.~(\ref{fourthterm}). 

Therefore, by using $d{\tilde{s}^{rg *}_0(\lambda)}/{d\lambda}=
-\varepsilon(M-m)\tilde{s}^{rg *}_0/(\tilde{s}^{rg *}_0+M)$, 
by imposing the scaling condition $d\tilde{c}^{rg}_{div}(\tau,\lambda)
/d\lambda=0$ at the 2nd order, and by making partially use of 
Eqs.~(\ref{thirdterm}), we end up with the equation:
\begin{equation}
-\varepsilon\frac{ Mm(M\!-\!m)\tilde{s}^{rg *}_0}
{ ( \tilde{s}^{rg *}_0 \!\!+\!M )^5 }+\varepsilon \frac{2M(M\!-\!m)
\tilde{s}^{rg *}_0
\tilde{s}^{rg *}_1 \!\!}{ \left ( \tilde{s}^{rg *}_0\!\! +\!M \right )^4 } +
\frac{M}{({\tilde{s}^{rg *}_0\!\!+\!M)^2}}
\frac{d{\tilde{s}^{rg *}_1}}{d\lambda}
-\varepsilon \frac{M(M\!-\!m)\tilde{s}^{rg *}_1\!\!}
{ ( \tilde{s}^{rg *}_0\!\! +\!M )^4 }(2\tilde{s}^{rg *}_0\!\!-\!M)\!=\!0.
\end{equation}
Hence, we get the expected result:
\begin{equation}
\frac{d{\tilde{s}^{rg *}_1}}{d\lambda}=
\varepsilon \left [ \frac{m(M-m)\tilde{s}^{rg *}_0}
{ ( \tilde{s}^{rg *}_0 +M )^3 }-
\frac{M(M-m)\tilde{s}^{rg *}_1}
{ ( \tilde{s}^{rg *}_0 +M )^2 } \right ],
\end{equation}
that completes the present verification. Indeed, this is the same 
ODE to be obeyed by $\tilde{s}^{rg *}_1(\lambda)$ that we previously 
obtained in the study of the substrate, and that is reported in 
Eqs.~(\ref{scalingcondition2}).}

\end{document}